\documentclass[aps, prx,twocolumn,longbibliography,superscriptaddress,amsmath,amssymb,floatfix]{revtex4}
\usepackage[latin9]{inputenc}
\usepackage{amssymb}
\usepackage{graphicx}
\usepackage{subfigure}
\usepackage{amsmath}
\usepackage{color}
\usepackage{mathrsfs}
\usepackage{float}
\usepackage{indentfirst}
\usepackage{mathrsfs}
\usepackage{float}
\usepackage{indentfirst}
\usepackage{textcomp}
\usepackage{comment}
\usepackage{mathtools}
\usepackage{natbib,hyperref}
\usepackage{soul}

\begin{document}

\title{Half-filled stripe to N$\acute e$el antiferromagnetism transition in the $t'$-Hubbard model on honeycomb lattice}

\author{Yang Shen}
\affiliation{Key Laboratory of Artificial Structures and Quantum Control, School of
Physics and Astronomy, Shanghai Jiao Tong University, Shanghai 200240, China}

\author{Mingpu Qin}\thanks{qinmingpu@sjtu.edu.cn}
\affiliation{Key Laboratory of Artificial Structures and Quantum Control, School of
Physics and Astronomy, Shanghai Jiao Tong University, Shanghai 200240, China}

\begin{abstract}
We study the ground state of the doped Hubbard model on honeycomb lattice with both nearest ($t$) and next-nearest neighboring hoppings ($t'$) in the small doping and strongly interacting region. Previous study on the model without $t'$ showed the ground state is a half-filled stripe. We employ density matrix renormalization group and extrapolate the results with truncation errors in the converged region. In the $t' < 0$ side, we find the half-filled stripe phase at $t' = 0$ is stable against the frustration of $t'$ until a critical point $-0.4 < t'_c < -0.3$, beyond which the ground state switches to anti-ferromagnetic N$\acute e$el phase with charge modulation. With further increase of $t'$ to $-0.7$, the ground state becomes paramagnetic. In the $t' > 0$ side, the half-filled stripe stretches to $t' \approx 0.7$. We don't find obvious enhancement of pairing for the range of $t'$ studied. We study width-4 cylinders in this work but the results for spin, charge, and pairing correlation agree qualitatively for periodic and anti-periodic boundary conditions in the half-filled stripe and anti-ferromagnetic N$\acute e$el phases, suggesting the results are likely to be representative for true two-dimensional systems. The half-filled stripe to anti-ferromagnetic N$\acute e$el phase transition can be realized on real materials or ultra-cold atom platform. 

\end{abstract}

\maketitle
Understanding the physics of doped Mott insulators is crucial to reveal the microscopic mechanism of high-Tc superconductivity \cite{RevModPhys.78.17}.
Hubbard model and its descendant, i.e., $t-J$ model, are the prototype models to study Mott-related physics \cite{Hubbard,RevModPhys.84.1383,qin2022hubbard,annurev-conmatphys-031620-102024}.
With a simple form, Hubbard model can host many exotic quantum states \cite{Hubbard,RevModPhys.84.1383,qin2022hubbard,annurev-conmatphys-031620-102024}.  

Recently, progresses have been made in the study of the doped Hubbard model on the square lattice \cite{PhysRevX.5.041041}. Collaboration of state-of-the-art numerical methods \cite{doi:10.1126/science.aam7127} established the filled-stripe phase in the doped Hubbard model. It was also found that the ``pure" Hubbard model (with only nearest neighboring hoppings) on square lattice doesn't host superconductivity in its ground state \cite{PhysRevX.10.031016}.  

The Hubbard model on honeycomb lattice \cite{RevModPhys.81.109} has also been extensively studied, partly due to the interests in the correlation-driven metal-insulator transition and the connection to graphene \cite{PhysRevX.6.011029,PhysRevX.3.031010}.
Similar to the square lattice, honeycomb lattice is bipartite, and the half-filled Hubbard model on honeycomb lattice is also a Mott insulator with antiferromagnetic (AF) N$\acute e$el order at strong interactions \cite{PhysRevX.3.031010, sorella1992semi}. 
These similarities make honeycomb lattice another playground to study Mott-related physics and to explore possible mechanisms of high-Tc superconductivity.
Recent calculation showed the ground state of the slightly doped Hubbard model on honeycomb lattice also has stripe order similar as the square lattice case \cite{PhysRevB.103.155110, PhysRevB.105.035111}. The difference between them is that the filling of stripe on honeycomb lattice is $1/2$ \cite{PhysRevB.103.155110, PhysRevB.105.035111}. This finding was also confirmed by followed calculations \cite{xu2022competing,peng2023superconductivity}.

Given that superconductivity is absent in the pure Hubbard model on the square lattice \cite{PhysRevX.10.031016}, other terms need to be included in the Hamiltonian to search for the possible superconducting ground state. The simplest term is the next-nearest neighboring hoppings $t'$. To account for the particle-hole asymmetry of the phase diagram of cuprates \cite{RevModPhys.84.1383}, a nonzero $t'$ term is needed in the framework of single band Hubbard model. 
Recent calculation for the $t'$-Hubbard model on square lattice showed superconductivity indeed emerges with the inclusion of $t'$ \cite{xu2023coexistence}. It is argued that $t'$ frustrates the stripe phase and helps the pairs across the stripes to build coherence, so the long-range superconducting order emerges \cite{xu2023coexistence}.    
 

\begin{figure}[t]
	\includegraphics[width=0.5\textwidth]{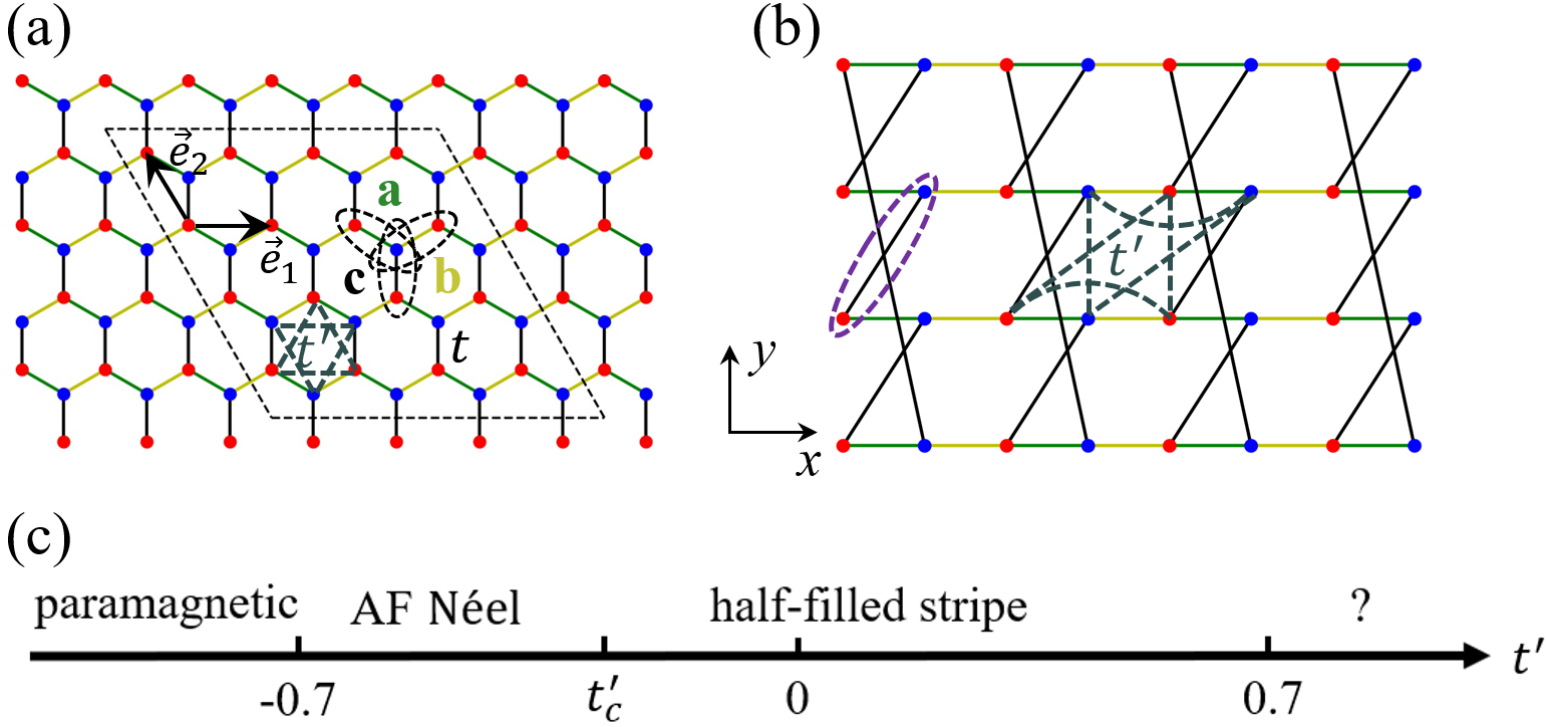}
	\caption{ 
		(a) Sketch of the honeycomb lattice which is rearranged into a square lattice in (b).
		Blue and red dots represent two inequivalent sublattices, respectively.  Bonds along different directions are distinguished with colors and are labeled as A, B and C.
		The two arrows in (a) are the primitive vectors of the Bravais lattice. The c bond in the dashed oval in (b) is the reference bond when calculating the pair-pair correlation function. We also vary the position of the reference c bond when calculating pair-pair correlations to analyze the boundary effect (details can be found in the Supplementary Materials). Gray dashed lines in (a) and (b) represent the next-nearest neighboring hopping $t'$ on honeycomb and the corresponding square lattices. Periodic/anti-periodic (open) boundary conditions are imposed for the vertical (horizontal) direction. (c) shows a rough phase diagram based on the DMRG results.}
	\label{honeycomb_geometry}
\end{figure}

\begin{figure*}[t]
	\includegraphics[width=0.45\textwidth]{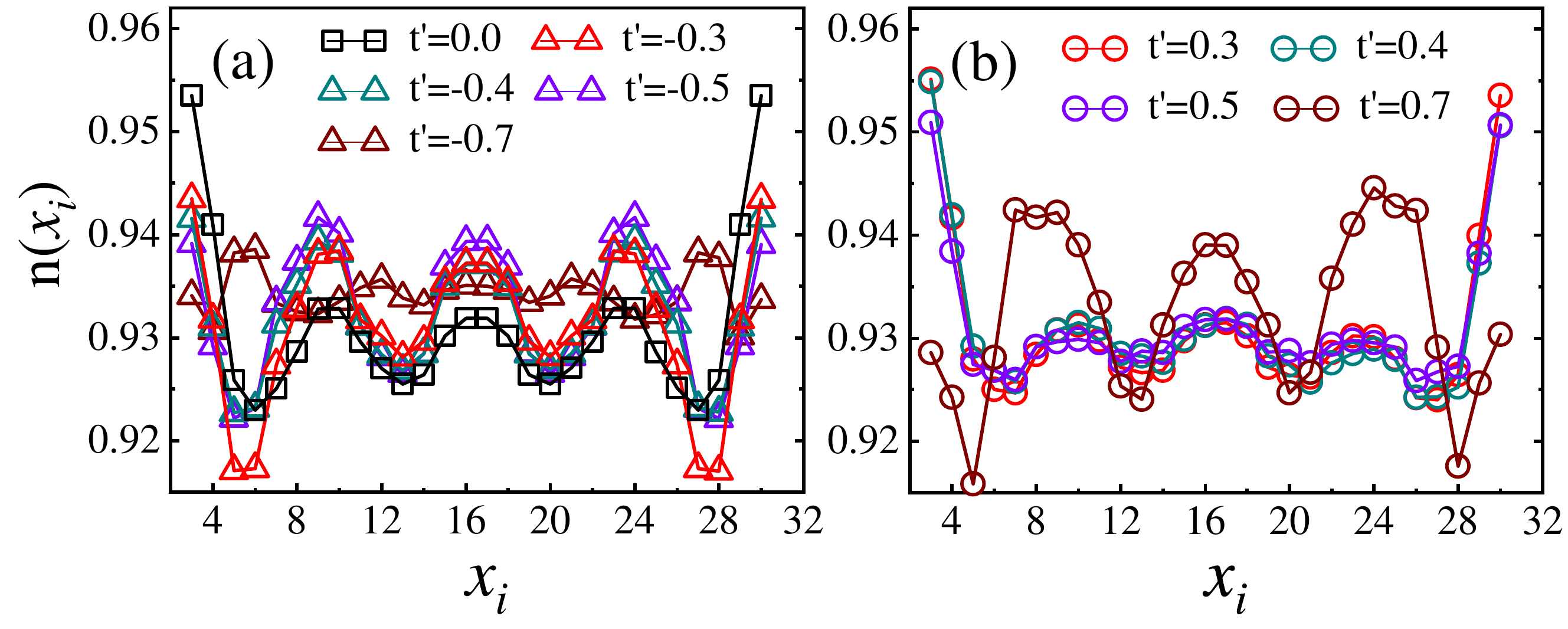}
	\includegraphics[width=0.45\textwidth]{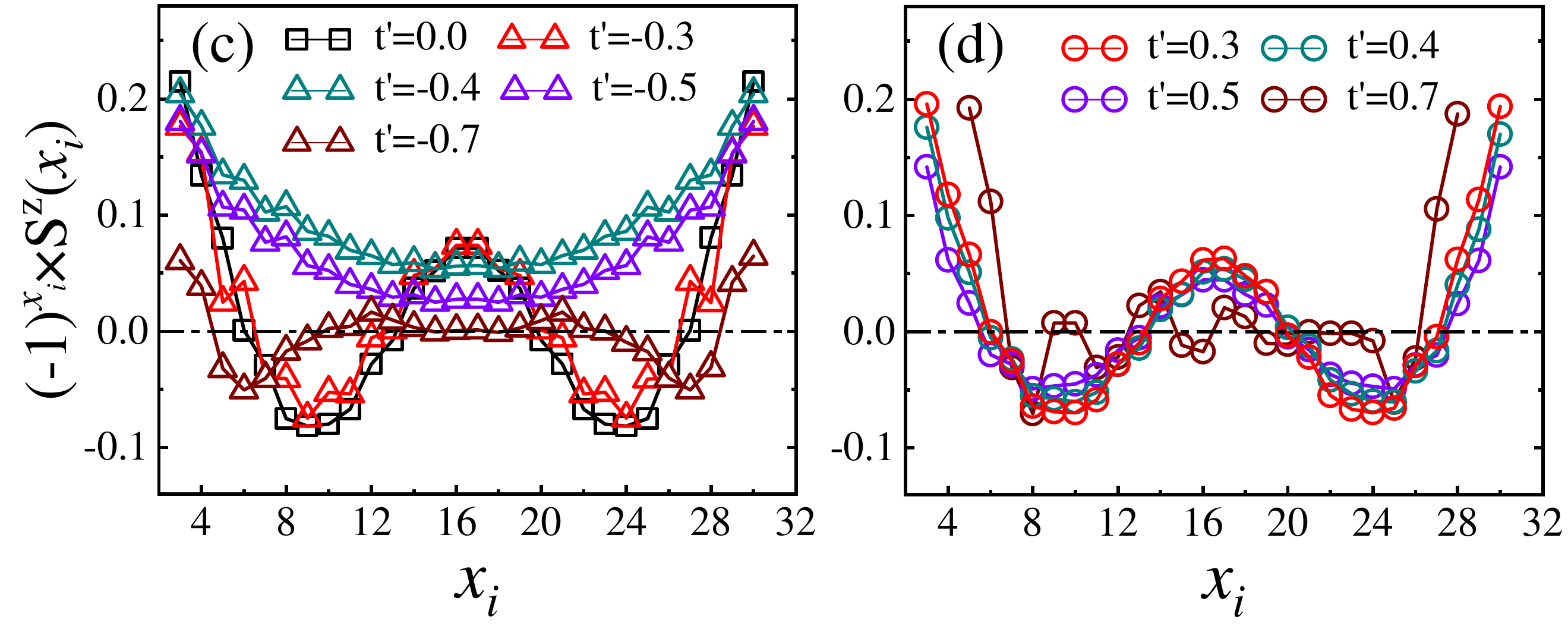}
	\includegraphics[width=0.45\textwidth]{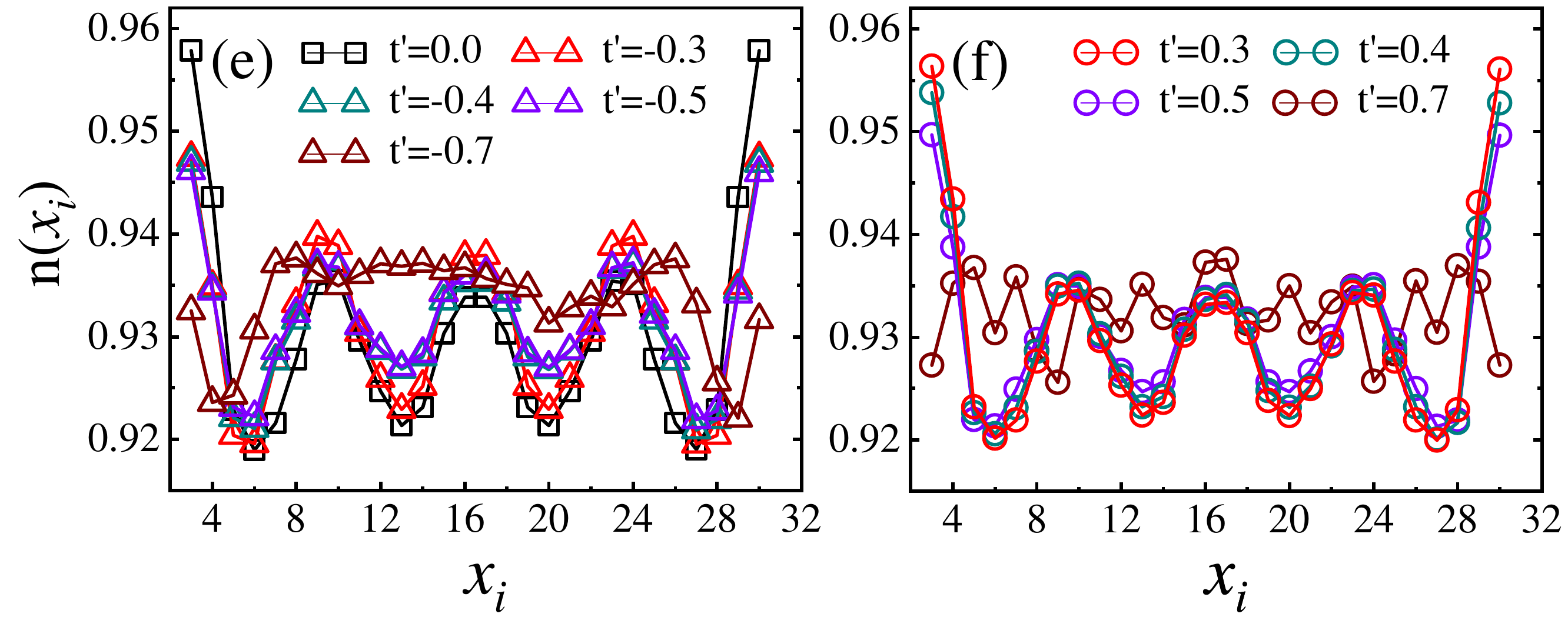}
	\includegraphics[width=0.45\textwidth]{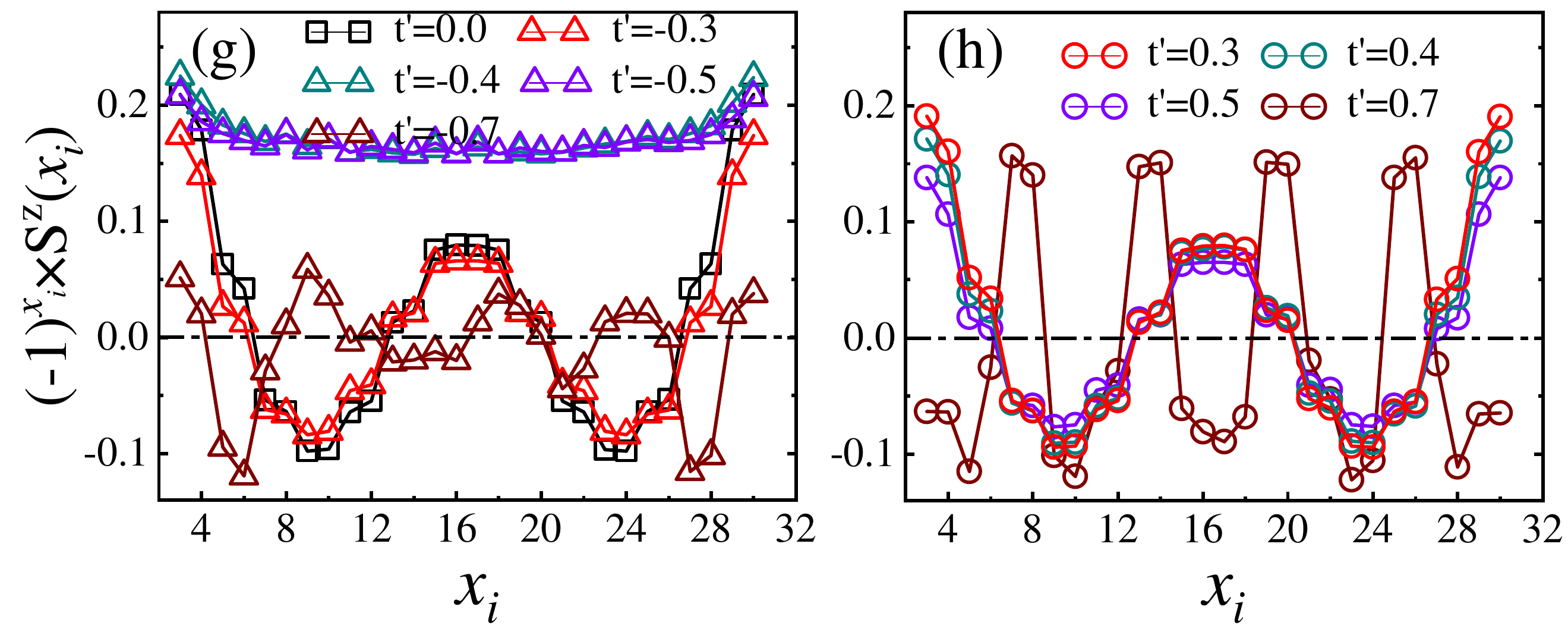}
	
	\caption{Electron density and staggered spin density for positive (b, d) and negative (a, c) $t'$ under PBC (upper panel) and APBC (lower panel). The dashed horizontal lines in (c), (d), (g), and (h) represent zero. We can find a phase transition from half-filled stripe order to AF N$\acute e$el order at critical $t'_c$ between $-0.3$ and $-0.4$. The results for spin and charge density agree qualitatively between PBC and APBC in the half-filled stripe and AF N$\acute e$el phases. The results are from extrapolation of truncation errors in DMRG except the $\pm 0.7$ case and $t'$ close to the phase transition point, for which results for largest bond dimension is shown. See the Supplementary Materials for details.
	}
	\label{stagSzElec}
\end{figure*}

The inclusion of $t'$ term in the Hubbard model on honeycomb lattice also frustrates the N$\acute e$el order. 
Actually, the strong-coupling limit of the Hubbard model with $t'$ at half-filling, i.e., $J_1-J_2$ Heisenberg model on honeycomb lattice, has a rich phase diagram \cite{PhysRevB.96.104401,PhysRevB.88.165138}. 
It was shown that there is an AF N$\acute e$el phase on the small $J_2$ side ($J_2$ $\lessapprox $ $0.2J_1$) \cite{PhysRevB.85.060402, bishop2012frustrated}, and a staggered valence-bond  solid phase on the large $J_2$ side ($J_2$ $\gtrapprox$ $0.4J_1$) \cite{PhysRevB.84.014417,PhysRevLett.107.087204}.
In the intermediate region, the frustrated exchange melts the AF N$\acute e$el state and results in competing phases including various quantum spin liquids and plaquette valence bond solid state \cite{PhysRevB.96.104401,PhysRevB.88.165138}. Based on these results, it is natural to ask whether superconductivity can be also enhanced by $t'$ for the Hubbard model on the honeycomb lattice.

In this work, we systematically study the ground state properties of the doped Hubbard model with both $t$ and $t'$ on honeycomb lattice in the strong interacting and slight doping region. We employ density matrix renormalization group (DMRG) \cite{PhysRevLett.69.2863,PhysRevB.48.10345} which can provide accurate results for narrow cylinders \cite{RevModPhys.77.259, stoudenmire2012studying}. We study the evolution of the ground state with $t'$ in both positive and negative sides. At the negative $t'$ side, we find the half-filled stripe state, previously found in the absence of $t'$ \cite{PhysRevB.105.035111,PhysRevB.103.155110}, is stable against the frustration of $t'$ until a critical point $t'_c$ between $-0.3$ and $-0.4$, at which a phase transition from half-filled stripe to AF N$\acute e$el phase with charge modulation occurs. With the further increase of $t'$ to about $-0.7$, the ground state becomes paramagnetic and the charge modulation disappears. At the positive $t'$ side, the half-filled stripe stretches to $t' \approx 0.7$. We also analyze the long-range behavior of the pair-pair correlation functions. We don't find obvious enhancement of pairing for the range of $t'$ studied. A rough phase diagram is shown in Fig.~\ref{honeycomb_geometry} (c).

The ground state of the $t'$-Hubbard model on square lattice is found to be very sensitive to the boundary conditions and widths of the studied systems, indicating the presence of large finite size effect \cite{xu2023coexistence}. 
To evaluate the finite size effect in the width-4 systems studied, we compare the results with periodic (PBC) and anti-periodic (APBC) boundary conditions. The spin, charge and pairing properties turn out to be insensitive to the boundary conditions in the half-filled stripe and AF N$\acute e$el phases, indicating the results we obtained in this work are likely to be representative of true two dimensional systems.

\begin{figure*}[t]
\includegraphics[width=0.32\textwidth]{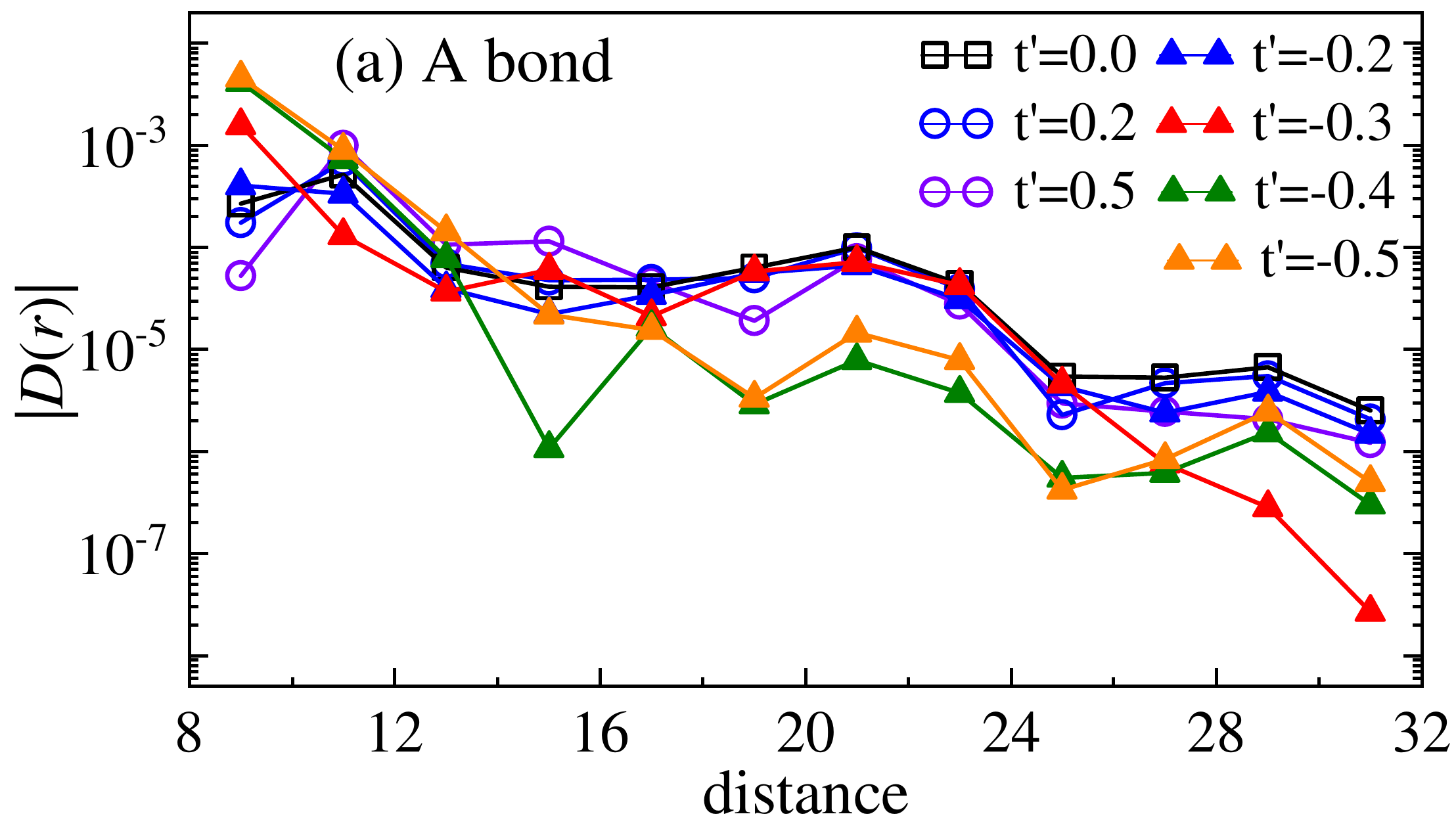}
\includegraphics[width=0.32\textwidth]{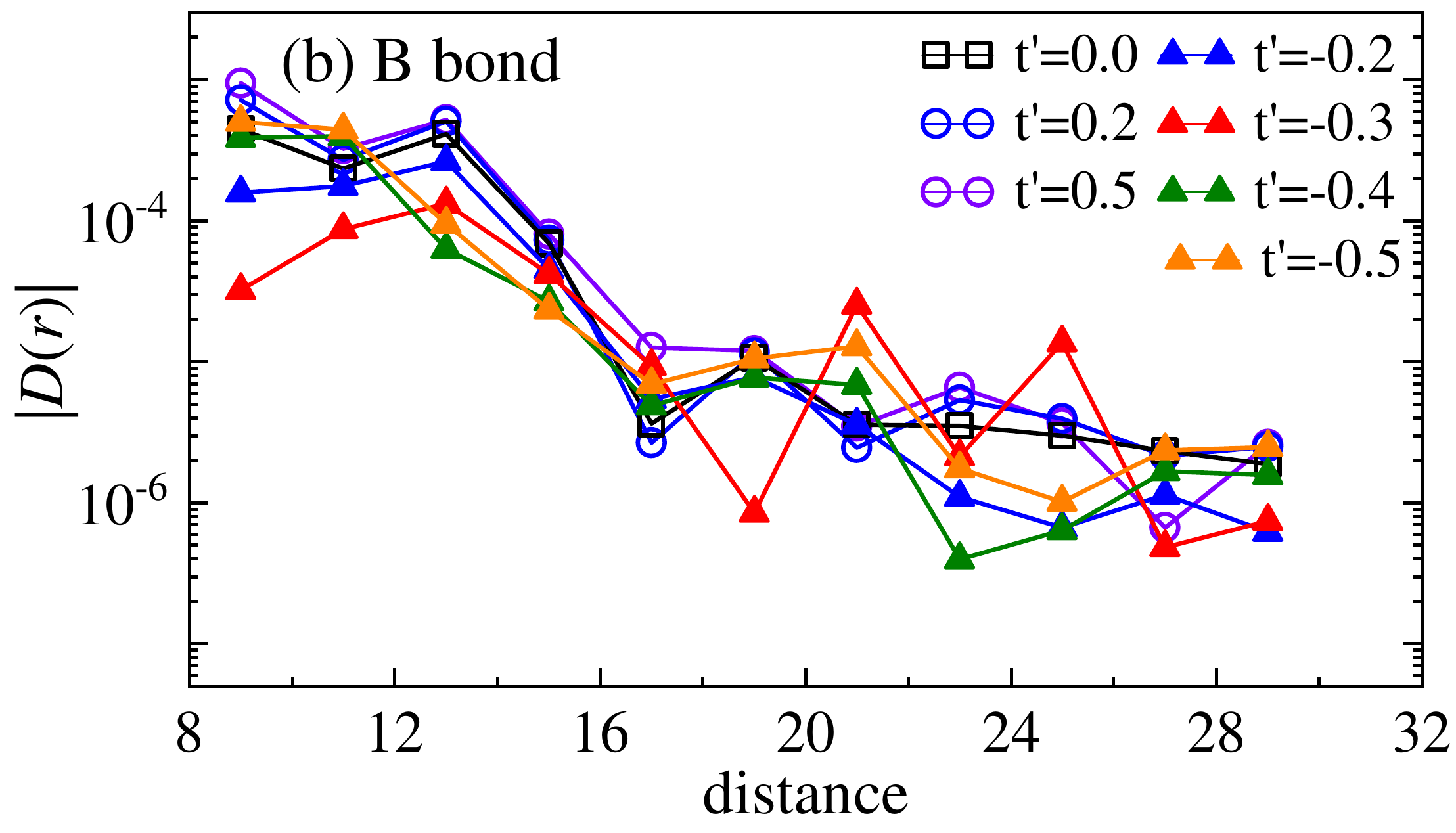}
\includegraphics[width=0.32\textwidth]{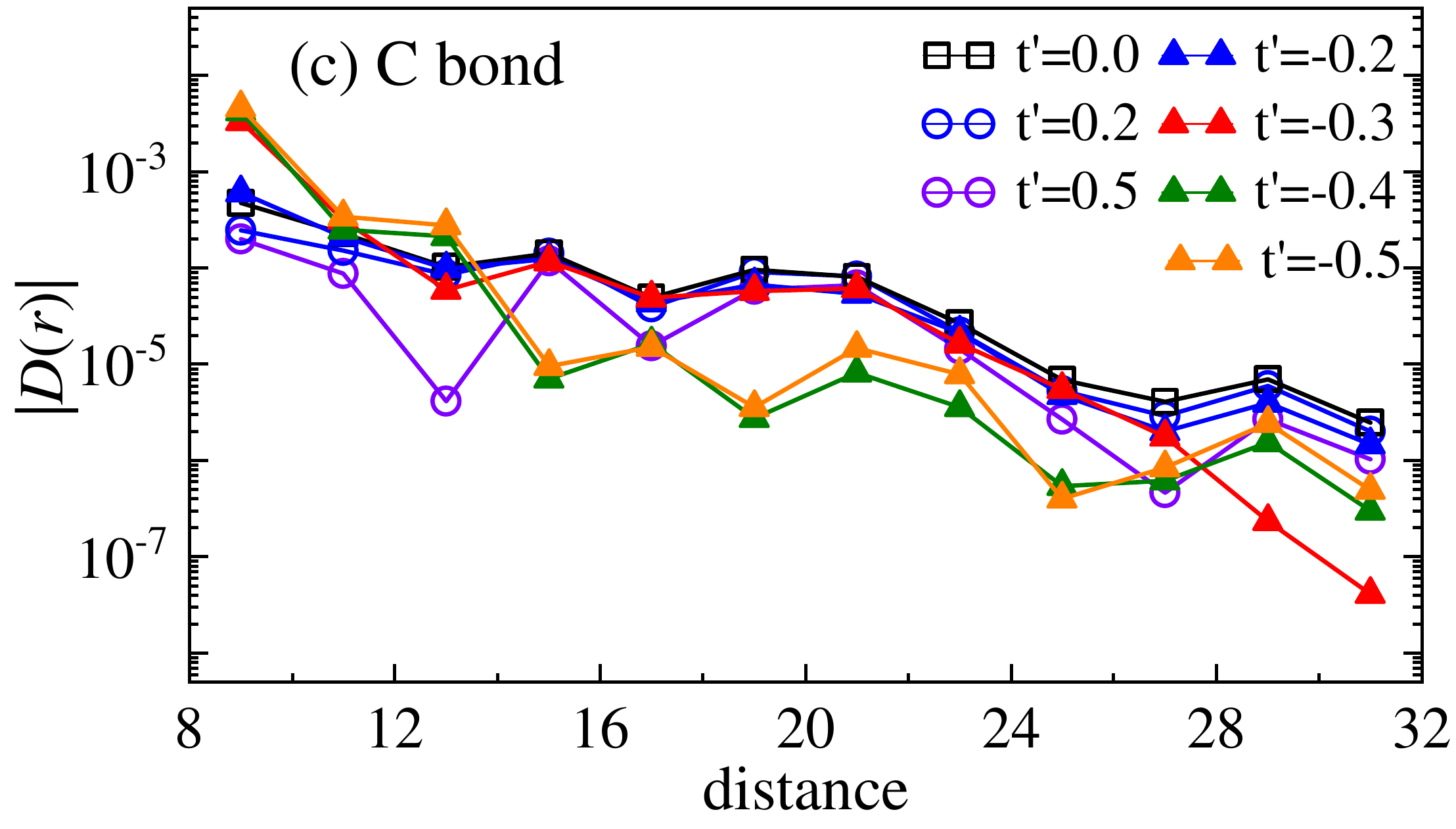}
\includegraphics[width=0.32\textwidth]{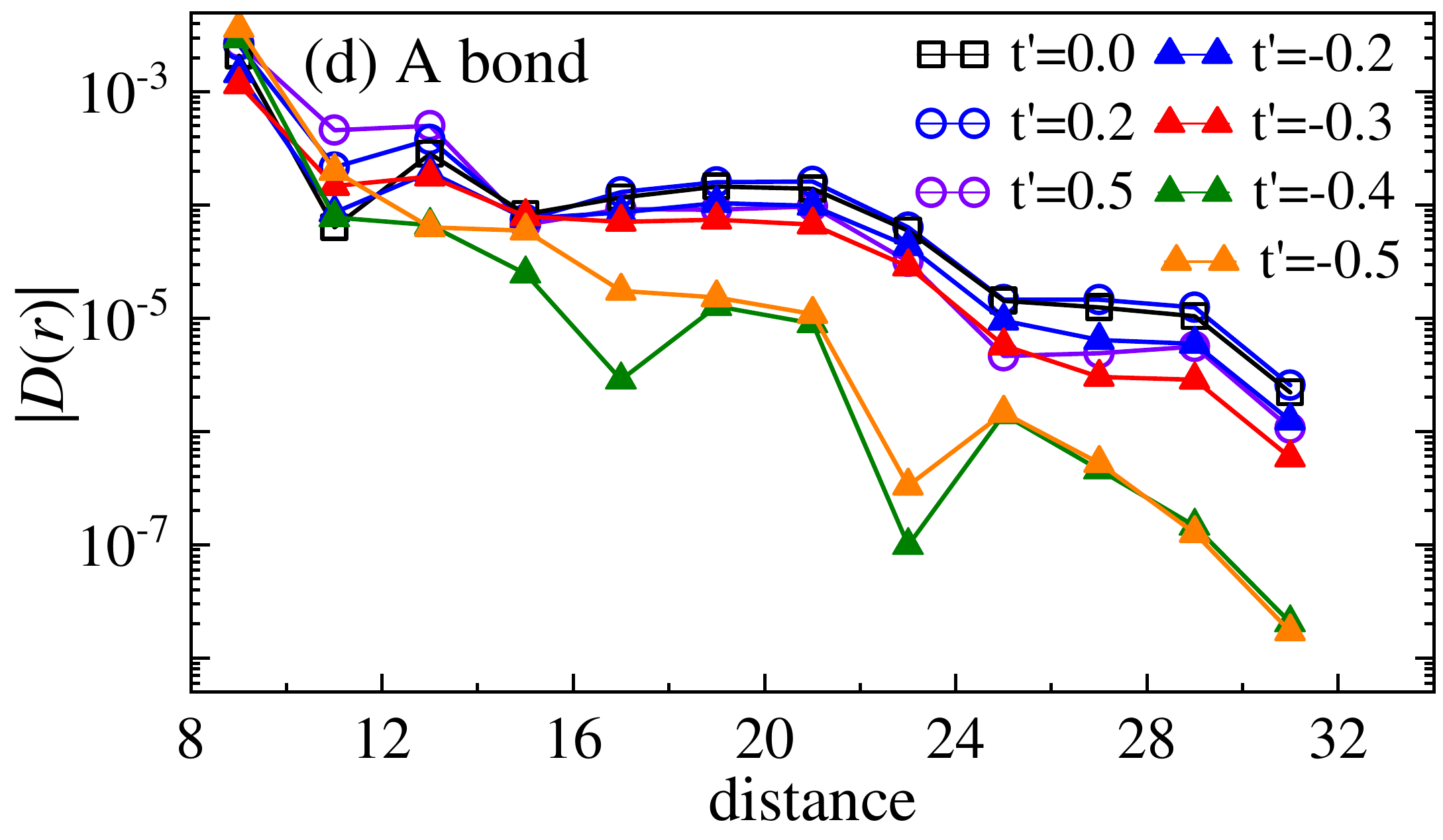}
\includegraphics[width=0.32\textwidth]{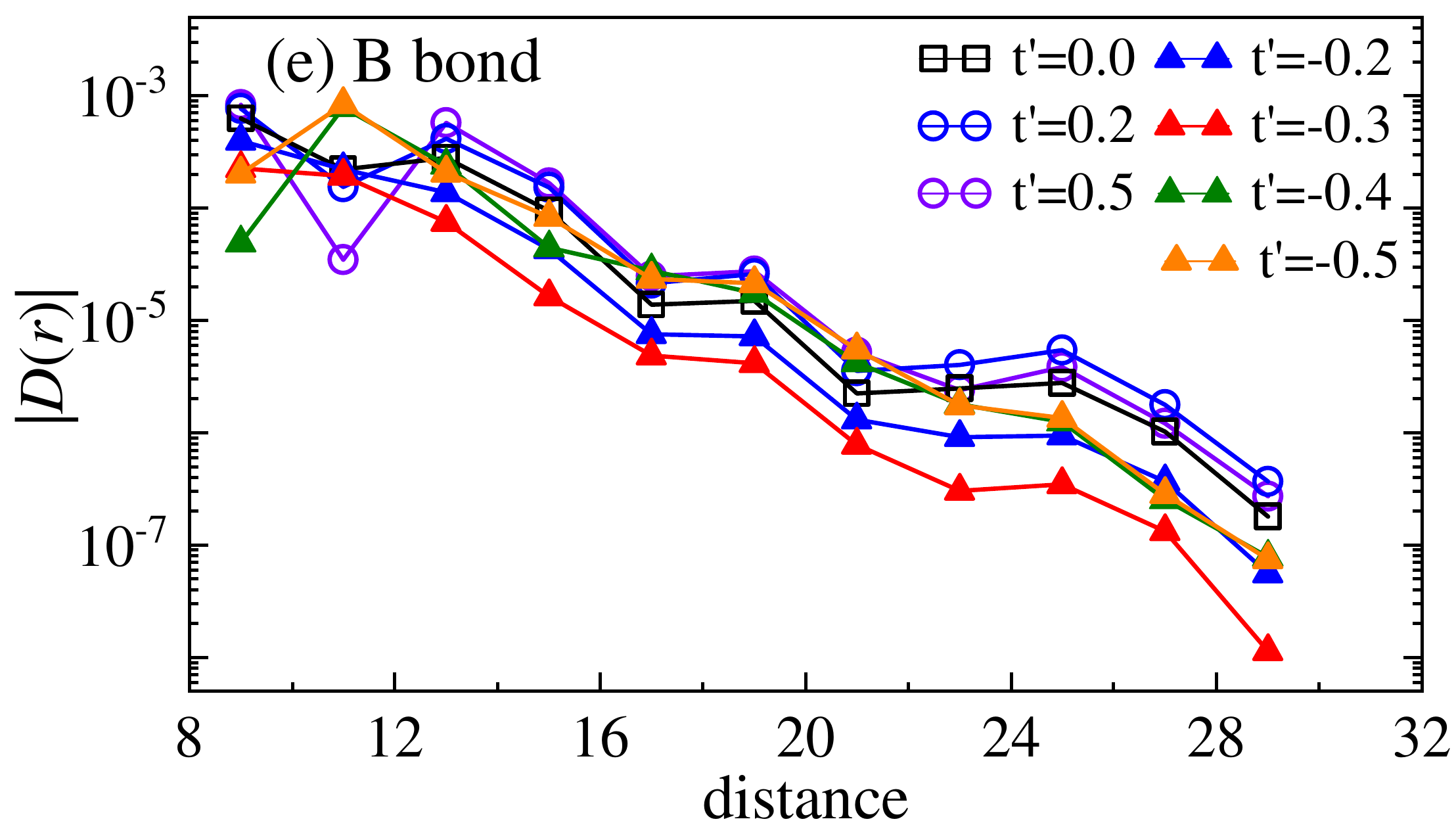}
\includegraphics[width=0.32\textwidth]{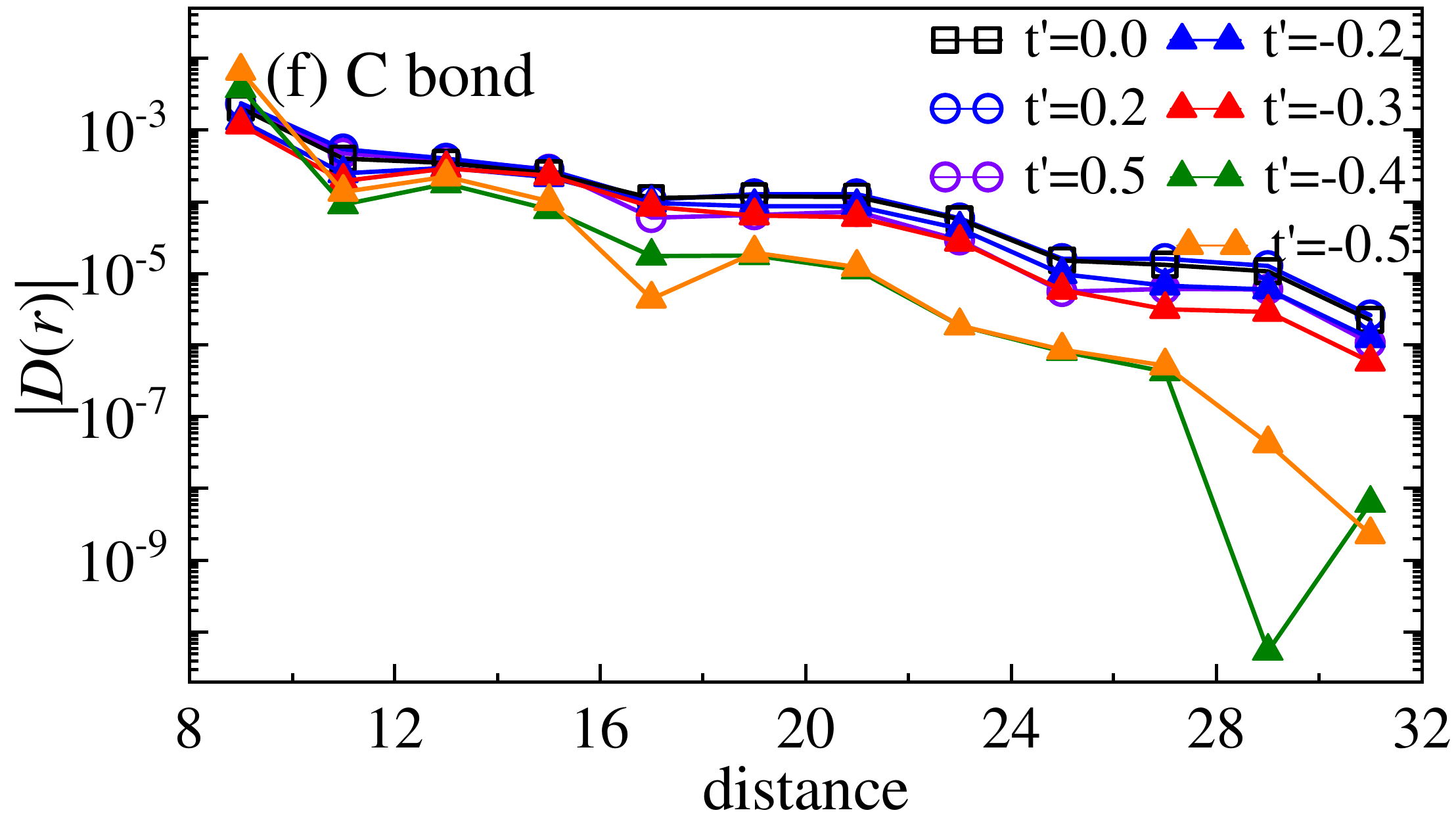}
\caption{Absolute value of pair-pair correlation functions of A (a, d), B (b, e) and C (c, f) bonds for different $t'$. The upper (lower) panel denotes results with PBC (APBC). The reference bond is placed at the bond between site (8, 2) and (9, 3) to remove the boundary effect \cite{2023arXiv230316487S}. The long-distance behaviors of pair-pair correlation don't change qualitatively within the range of $t'$ studied and under different boundary conditions.}
\label{paircorr}
\end{figure*}

\emph{Model and Method --} 
The Hamiltonian of the Hubbard model is

\begin{equation}
\hat{H} =-\sum_{ (i,j), \sigma} t_{i j}\left(\hat{c}_{i \sigma}^{\dagger} \hat{c}_{j \sigma}+h.c.\right)+U \sum_{i } \hat{n}_{i \uparrow} \hat{n}_{i \downarrow}
\label{Ham}
\end{equation}
where $\hat{c}_{i \sigma}^{\dagger}(\hat{c}_{i \sigma})$ creates (annihilates) an electron on site $i=(x_i,y_i)$ with spin $\sigma$, and $U$ represents the on-site Coulomb interaction. We consider electron hopping terms up to next-nearest neighbors ($t'$) and set the nearest neighboring hopping $t$ as energy unit. We focus on $U = 8$ in this work. We scan a range of $t'$ from $-0.7$ to $0.7$. We only study the hole-doped case. The electron doped case can be transformed to the hole doped case by reversing the sign of $t'$ through a particle-hole transformation.
The averaged hole concentration away from half-filling is defined as $\delta = N_h/N$ with $N_h=\sum_i(1-n_i)$.  An illustration of the honeycomb lattice is sketched in Fig.~\ref{honeycomb_geometry}(a). We rearrange it into a square lattice as shown in Fig.~\ref{honeycomb_geometry}(b) to index the sites more conveniently. After the rearrangement, we study the square system with cylinder geometry, i.e., with periodic/anti-periodic (open) boundary conditions along $y$ ($x$) direction. There are totally $N=L_x \times L_y$ lattice sites. We compare the results with PBC and APBC in our calculations to test the finite size effect.

We employ DMRG \cite{PhysRevLett.69.2863,PhysRevB.48.10345} in this work, which can provide accurate results of narrow cylinders \cite{RevModPhys.77.259, stoudenmire2012studying}. We focus on 1/16 hole doping regime in which half-filled stripe was found in the model without $t'$ before \cite{PhysRevB.105.035111,PhysRevB.103.155110}. We calculate systems with width $L_y = 4$ and length $L_x = 32$. We apply AF pinning fields with strength $h_m = 0.5$ on the open edges of the width-4 cylinder, which allows us to detect spin order by measuring the local magnetization $\langle \hat{S}_{i}^z \rangle$ instead of the more demanding spin-spin correlation functions and makes the DMRG calculations easier to converge \cite{PhysRevLett.99.127004}. To characterize the charge and spin order, we define rung charge density as $n(x)=\sum_{y=1}^{L_{y}}\left\langle\hat{n}_{i}(x,y)\right\rangle / L_{y}$ along the width direction of the cylinders and staggered spin density as $(-1)^i \langle \hat{S}_i^z \rangle$. We also measure the singlet pair-pair correlation function defined as $D(r) = \langle \hat{\Delta}_i^{\dagger}\hat{\Delta}_{i+r}\rangle$ with $\hat{\Delta}_i^{\dagger}=\hat{c}_{(i,1),\uparrow}^{\dagger}\hat{c}_{(i,2),\downarrow}^{\dagger}-\hat{c}_{(i,1),\downarrow}^{\dagger}\hat{c}_{(i,2),\uparrow}^{\dagger}$. The state kept in the DMRG calculation is as large as $m = 11000$ with a typical truncation error $\epsilon \approx 1.5 \times 10^{-5}$ and careful extrapolations of DMRG results with truncation errors are performed in the converged region.

\begin{figure*}[t]
\includegraphics[width=0.24\textwidth]{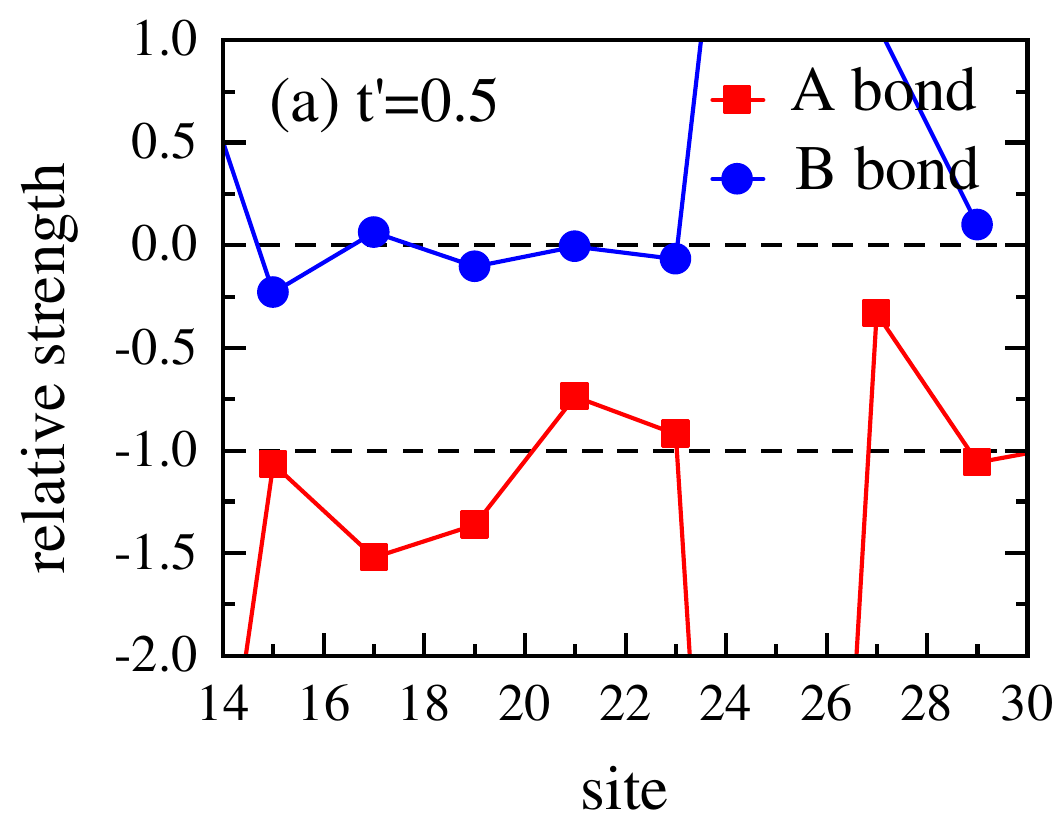}
\includegraphics[width=0.24\textwidth]{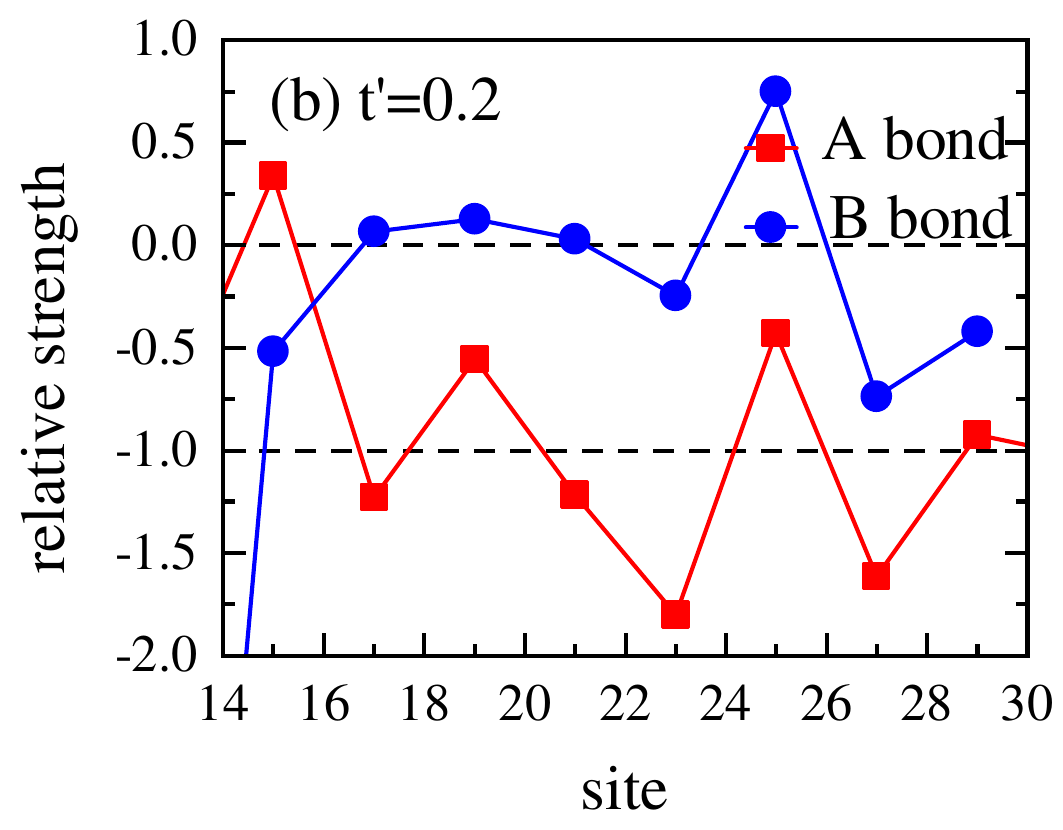}
\includegraphics[width=0.24\textwidth]{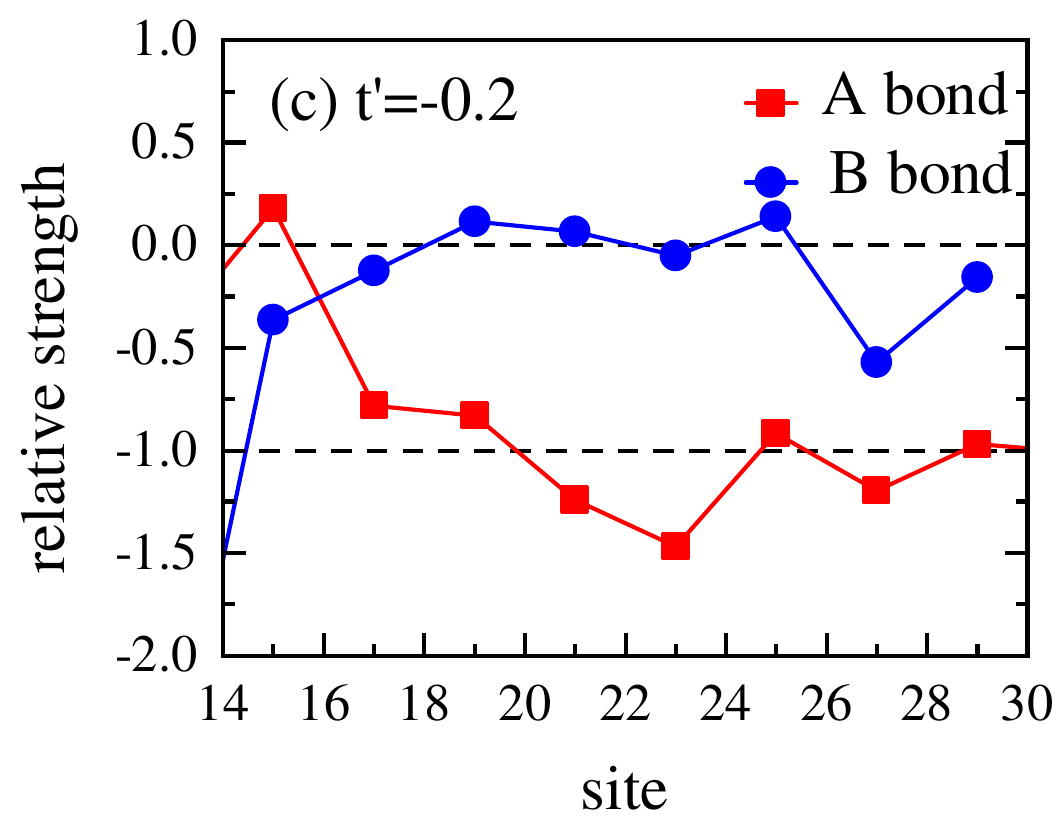}
\includegraphics[width=0.24\textwidth]{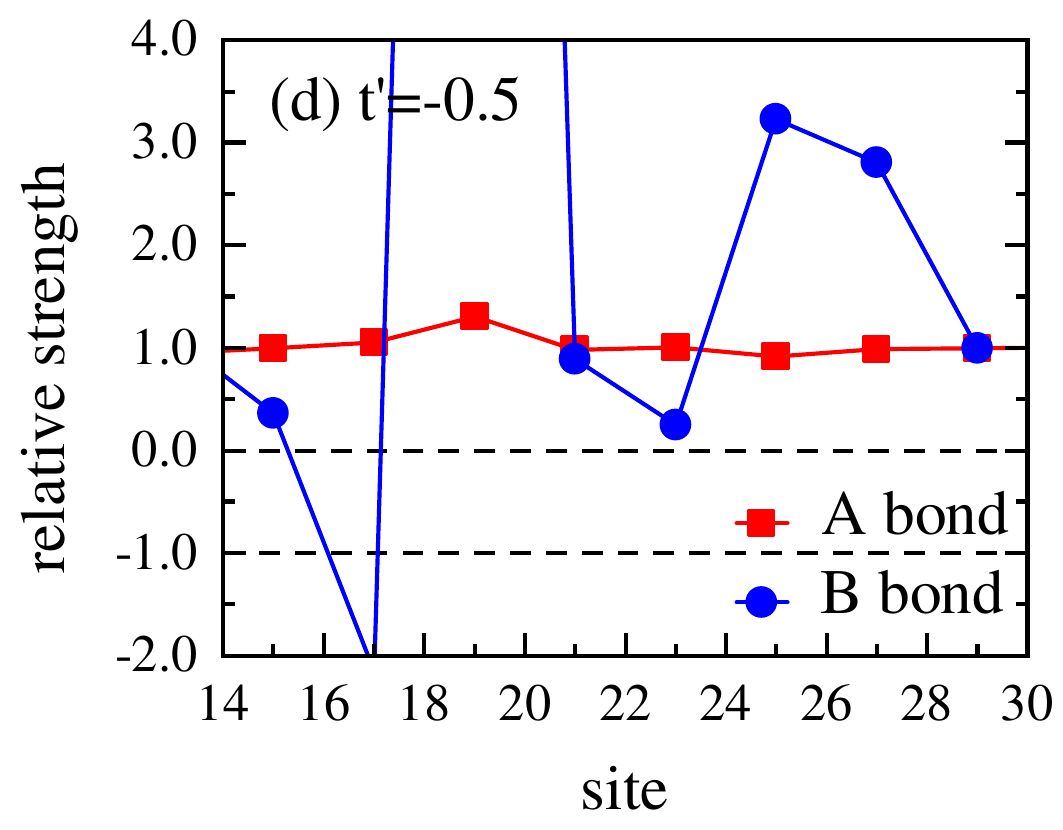}
\includegraphics[width=0.24\textwidth]{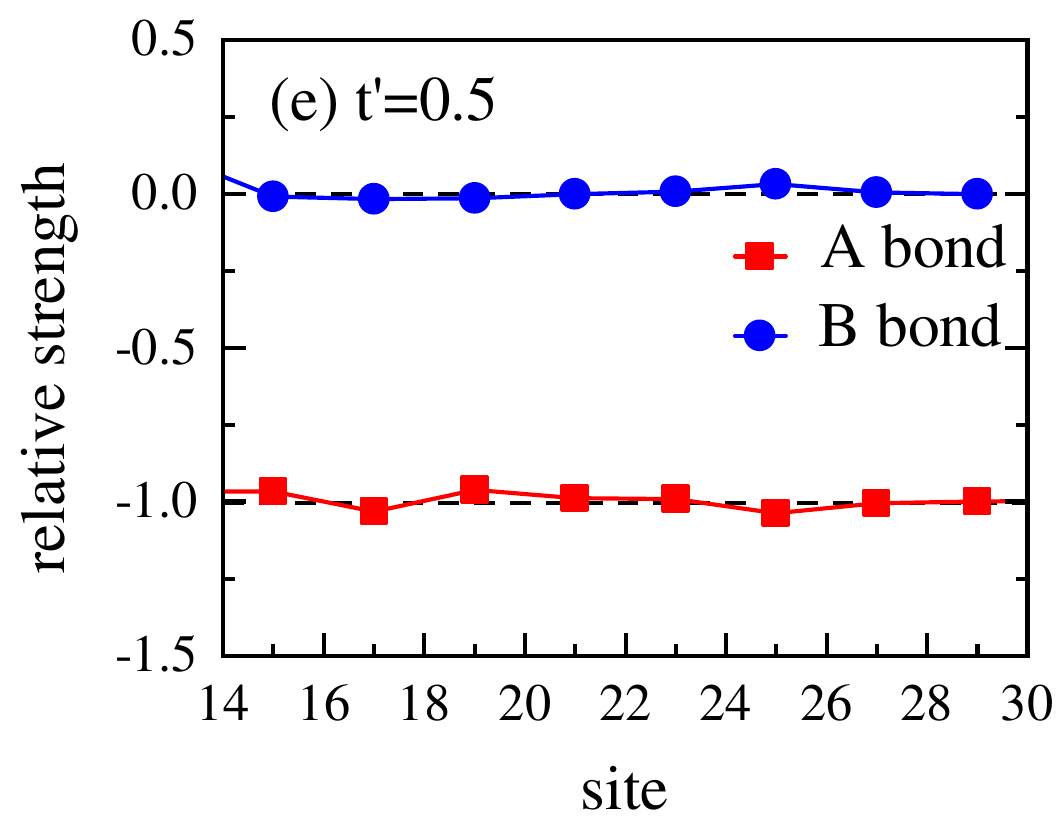}
\includegraphics[width=0.24\textwidth]{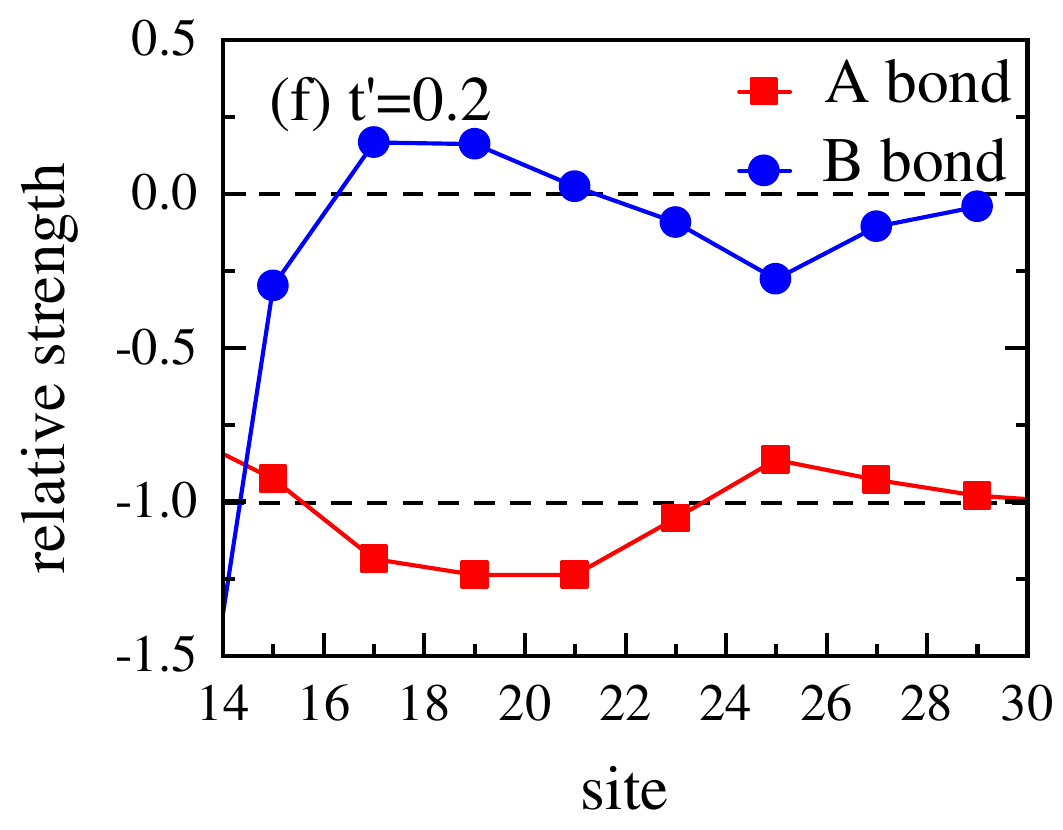}
\includegraphics[width=0.24\textwidth]{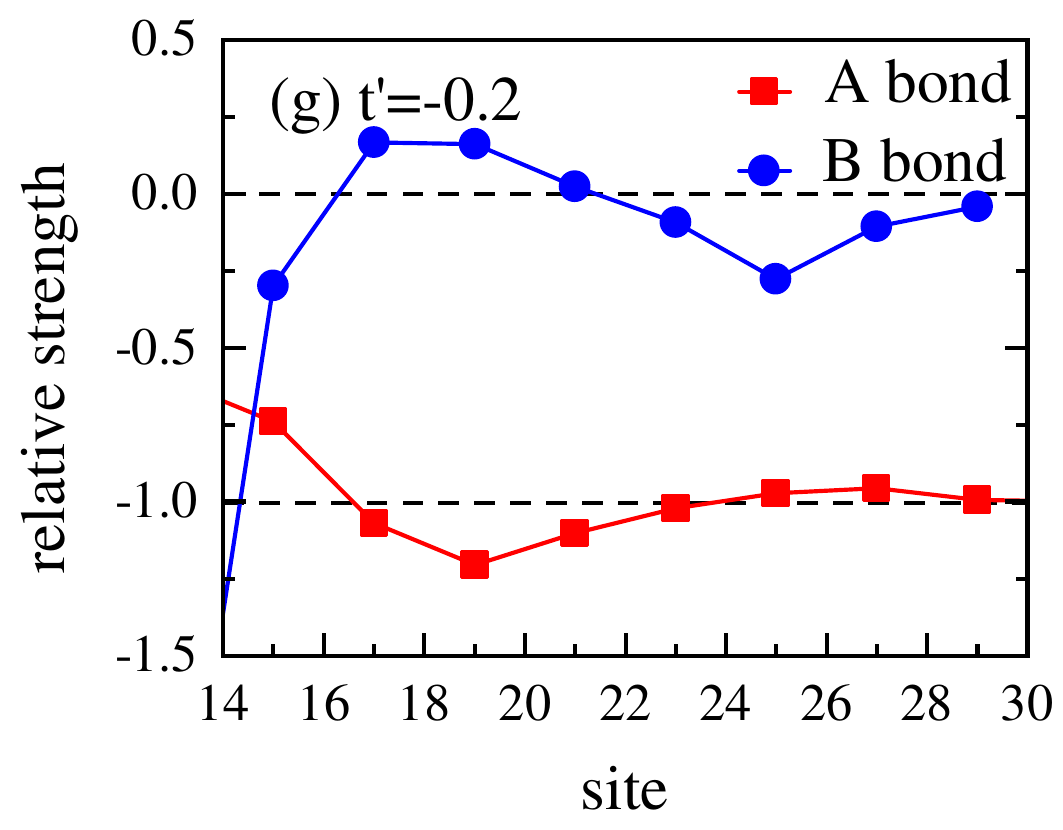}
\includegraphics[width=0.24\textwidth]{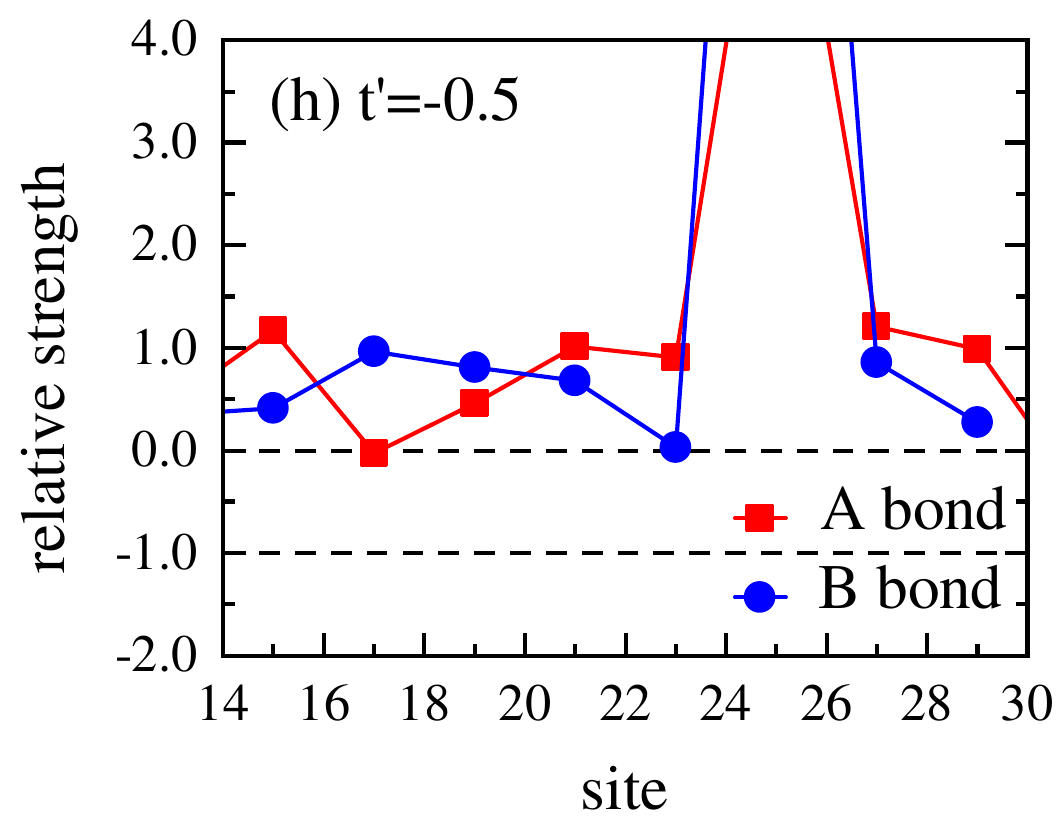}
\caption{The relative strengths of the pair-pair correlations for bonds connected by the same site. The upper (lower) panel denotes results under PBC (APBC) with different $t'$. The reference bond is placed at the bond between sites (8, 2) and (9, 3). The dashed horizontal lines represent 0 and -1. In the half-filled stripe phase ((a), (b), (c), (e), (f), and (g)), we can find a (-1, 0, 1) sign structure at long distance, which is the $d_{xy}$ representation of the $D$6h symmetry of the honeycomb lattice. In the AF N$\acute e$el phase ((d) and (h)), we can't find a clear pattern for the relative strength of the pairing correlations.}
\label{relative_stren}
\end{figure*}

\emph{Charge and spin densities --} In Fig.~\ref{stagSzElec}, we show the results for charge and staggered spin density profiles for various values of $t'$ under PBC (upper panel) and APBC (lower panel). As mentioned above, we only study the hole doped cases and the electron doped case can be mapped to the hole doped case by reversing the sign of $t'$ with a particle-hole transformation.

As shown in Fig.~\ref{stagSzElec} (b) and (f), in the positive $t'$ side, when $t'$ is small, the charge distribution $n(x)$ displays an oscillation with period $\lambda_\rho = 1/2\delta = 8$. And the spatial modulation of staggered spin density has a wavelength twice that of the charge density with a $\pi$ phase flip at the hole concentrated sites (see Fig.~\ref{stagSzElec} (d) and (h)). The spin and charge modulations are consistent with the half-filled stripe state found in the $t' = 0$ case \cite{PhysRevB.105.035111,PhysRevB.103.155110} (also shown in Fig.~\ref{stagSzElec} (a)). The half-filled stripe phase terminates at about $t' = 0.7$. When $t' \ge 0.7$, there is large discrepancy for the states obtained for PBC and APBC. The characterization of the phase in this region needs further investigation on wider systems. 

In the negative $t'$ side, the half-filled stripe phase at $t' = 0$ stretches to $t'_c \approx -0.3$, beyond which a phase transition occurs. When $t'<t'_c$, the system switches to an AF N$\acute e$el phase with charge modulation as shown in Fig.~\ref{stagSzElec}(a), (e), (c), and (g). The charge-modulated N$\acute e$el phase terminates at about $t' = -0.7$, beyond which the ground state is paramagnetic with nearly uniform density. 

The results for spin and charge densities agree qualitatively with PBC and APBC in the half-filled stripe and AF N$\acute e$el phases, suggesting the results are insensitive to the width of studied systems and are likely to be representative for the true two-dimensional system.

A rough phase diagram based on these results can be found in Fig.~\ref{honeycomb_geometry} (c).

\emph{Pairing correlation --} The singlet pair-pair correlation function $D(r)$ is shown in Fig.~\ref{paircorr} for a series of $t'$ values. We plot the absolute value of $D(r)$ for A, B and C bonds (see Fig.~\ref*{honeycomb_geometry} for the definitions) under PBC and APBC in the upper and lower panes, respectively. The reference bond is the C bond placed between sites (8, 2) and (9, 3) (see Fig.~\ref{honeycomb_geometry}(b)). 
We find that the long-distance behaviors of pair-pair correlation don't change qualitatively within the range of $t'$. They don't change qualitatively either under different boundary conditions. There is a tiny oscillation in the decay of the pair-pair correlations induced by the charge modulation. 

On the square lattice, long-range pairing order emerges with the inclusion of $t'$ \cite{xu2023coexistence}. But as shown in Fig.~\ref{paircorr}, we don't find obvious enhancement of pairing for the range of $t'$ studied. Near the transition point $-0.4 < t' < -0.3$ where half-filled stripe switches to AF N$\acute e$el order with charge modulation, the strength of $D(r)$ is actually slightly suppressed. 
We have also calculated the spin-triplet pair-pair correlations (see the results in the Supplement Materials) but found they are much weaker than the singlet pair-pair correlations.

In Fig.~\ref{relative_stren} we show the relative strengths of the pair-pair correlation for PBC and APBC with different $t'$ values, by dividing the correlation on A and B bonds with the value on C bonds which are connected by the same site. In the stripe phase (i.e., $|t'|<|t'_c|$), we can find that the relative strength of pair-pair correlations on A bonds is nearly equal to C bonds in magnitude but with opposite sign. And the pair-pair correlations on B bonds are very tiny comparing to the values on C bonds. Therefore, the pairing order have an approximate (-1,0,1) sign structure locally, which is one of the degenerate $d$-wave representations $d_{xy}$ of $D$6h symmetry of the honeycomb lattice \cite{PhysRevB.75.134512}. In the AF N$\acute e$el phase ($t' = -0.5$ results in Fig.~\ref{relative_stren} (d) and (h)), we can't find a clear pattern for the relative strength of the pairing correlations, which could be resulted from the finite size effect.

\emph{Summary and perspectives --} We systematically study the evolution of the half-filled stripe order with next-nearest neighboring hopping $t'$ on the slightly doped Hubbard model on honeycomb lattice in the strongly interacting region. We employ DMRG and study width-4 cylinders. In the negative $t'$ side, the half-filled stripe phase is stable against the frustration of $t'$ until a critical point $-0.4 < t'_c < -0.3$, at which a phase transition from half-filled stripe to AF N$\acute e$el phase with charge modulation occurs. The charge-modulated AF N$\acute e$el phase terminates at about $t' = -0.7$. In the positive $t'$ side, the half-filled stripe state stretches to about $t' = 0.7$. Different from the square lattice case \cite{xu2023coexistence}, we don't find obvious enhancement of pairing for the range of $t'$ studied. The results with PBC and APBC agree qualitatively in the half-filled stripe and AF N$\acute e$el phases, indicating the results in this work are likely to be representative for real two-dimensional system. 
It will be interesting to try to realize the phase transition from half-filled stripe to AF N$\acute e$el phase in real materials or in ultra-cold atom platforms.



\emph{Acknowledgments --} Y. Shen and M. P. Qin thank Weidong Luo for his generosity to provide computational resources for this work. M. P. Qin acknowledges the support from the National Key Research and Development Program of MOST of China (2022YFA1405400), the National Natural Science Foundation of China (Grant No. 12274290) and the sponsorship from Yangyang Development Fund.

\bibliography{Main_tpHub_DMRG}

\end{document}


\title{Supplementary Materials for ``Half-filled stripe to N$\acute e$el antiferromagnetism transition in the $t'$-Hubbard model on honeycomb lattice"}

\author{Yang Shen}
\affiliation{Key Laboratory of Artificial Structures and Quantum Control, School of
Physics and Astronomy, Shanghai Jiao Tong University, Shanghai 200240, China}

\author{Mingpu Qin}\thanks{qinmingpu@sjtu.edu.cn}
\affiliation{Key Laboratory of Artificial Structures and Quantum Control, School of
Physics and Astronomy, Shanghai Jiao Tong University, Shanghai 200240, China}

\maketitle

In the Supplementary Materials, we provide more details for the results presented in the main text. In Sec. A and B, we show both the finite bond dimension and the extrapolated results for charge and staggered spin density. In Sec. C, we show the details of the extrapolation of charge and staggered spin density with truncation errors.  In Sec. D, we show both the finite bond dimension and the extrapolated results for the singlet pair-pair correlation function. The triplet pair-pair correlation function are shown in Sec. E. For all the quantities, we show the results for both periodic and anti-periodic boundary conditions. 

\subsection{A. Charge density}

\begin{figure*}[t]
   \includegraphics[width=0.45\textwidth]{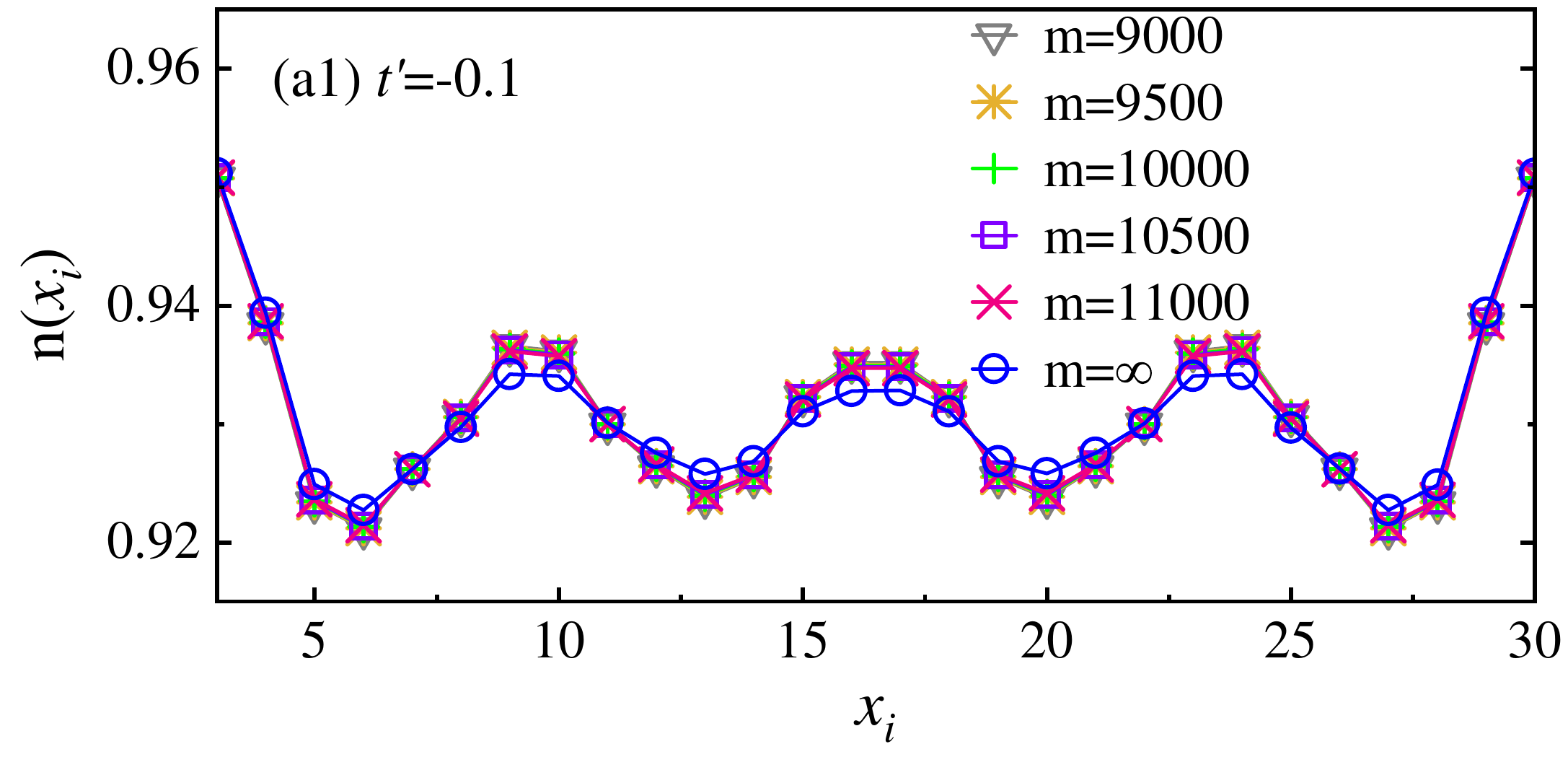}
   \includegraphics[width=0.45\textwidth]{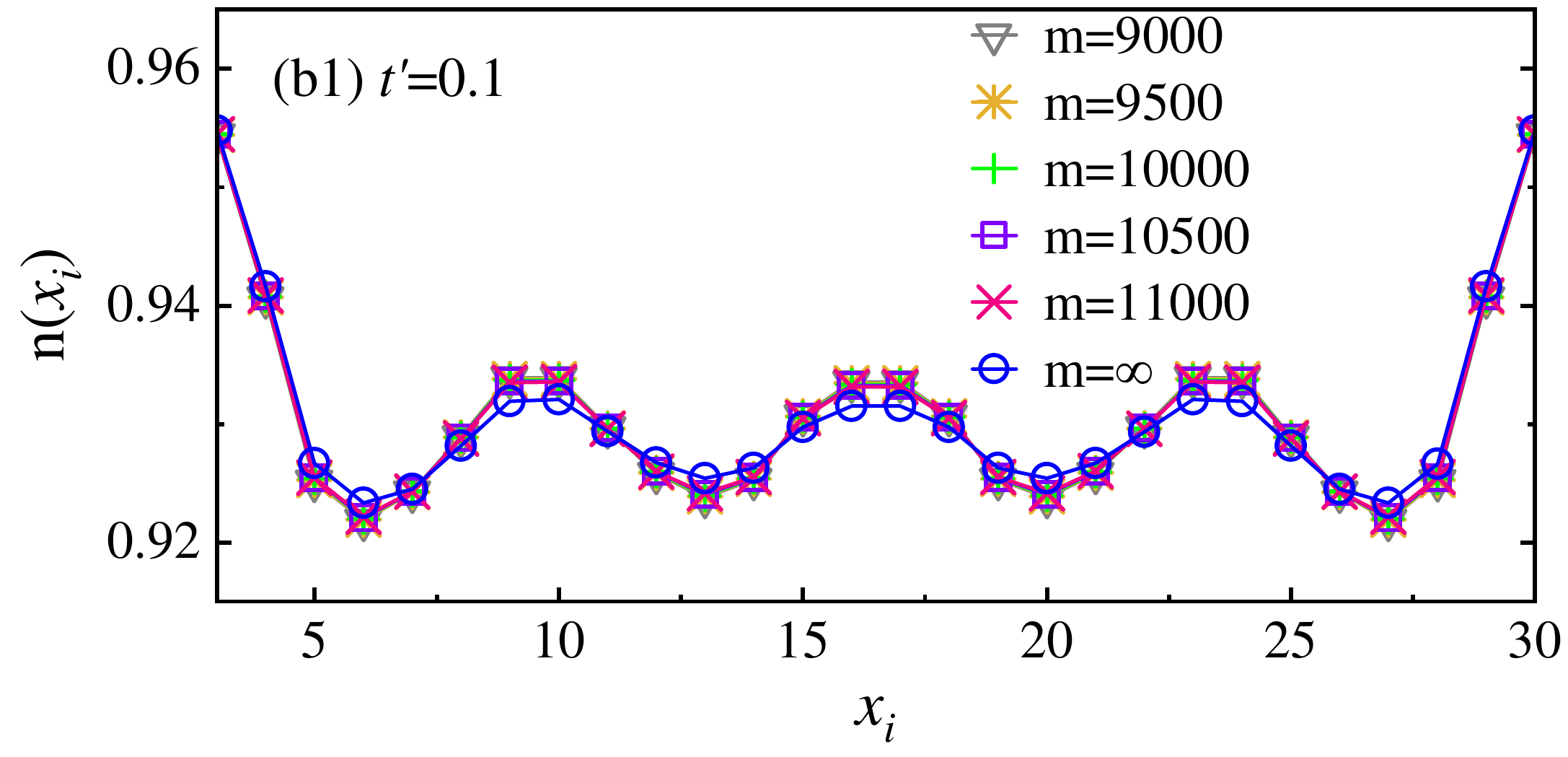}
   \includegraphics[width=0.45\textwidth]{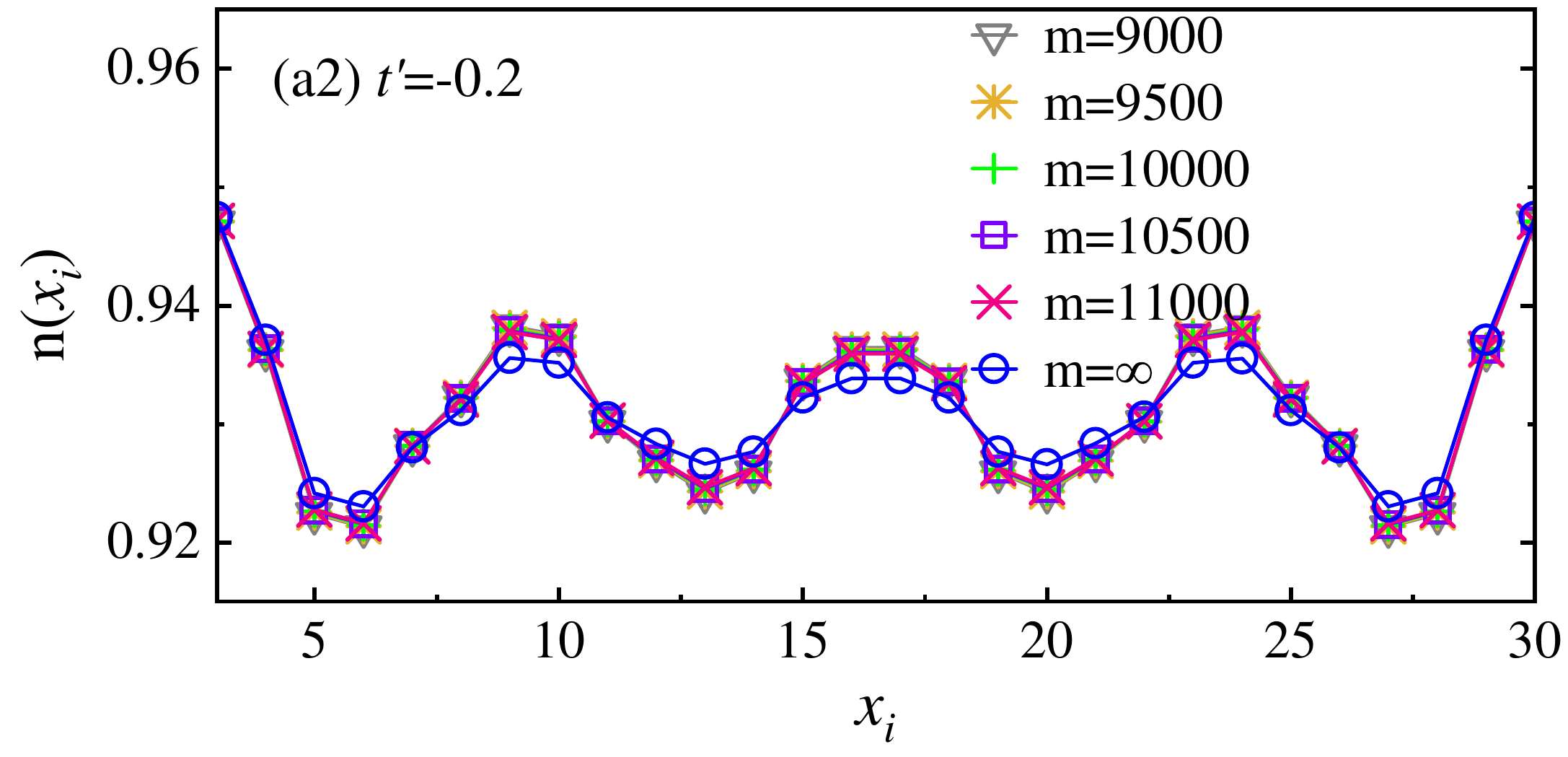}
   \includegraphics[width=0.45\textwidth]{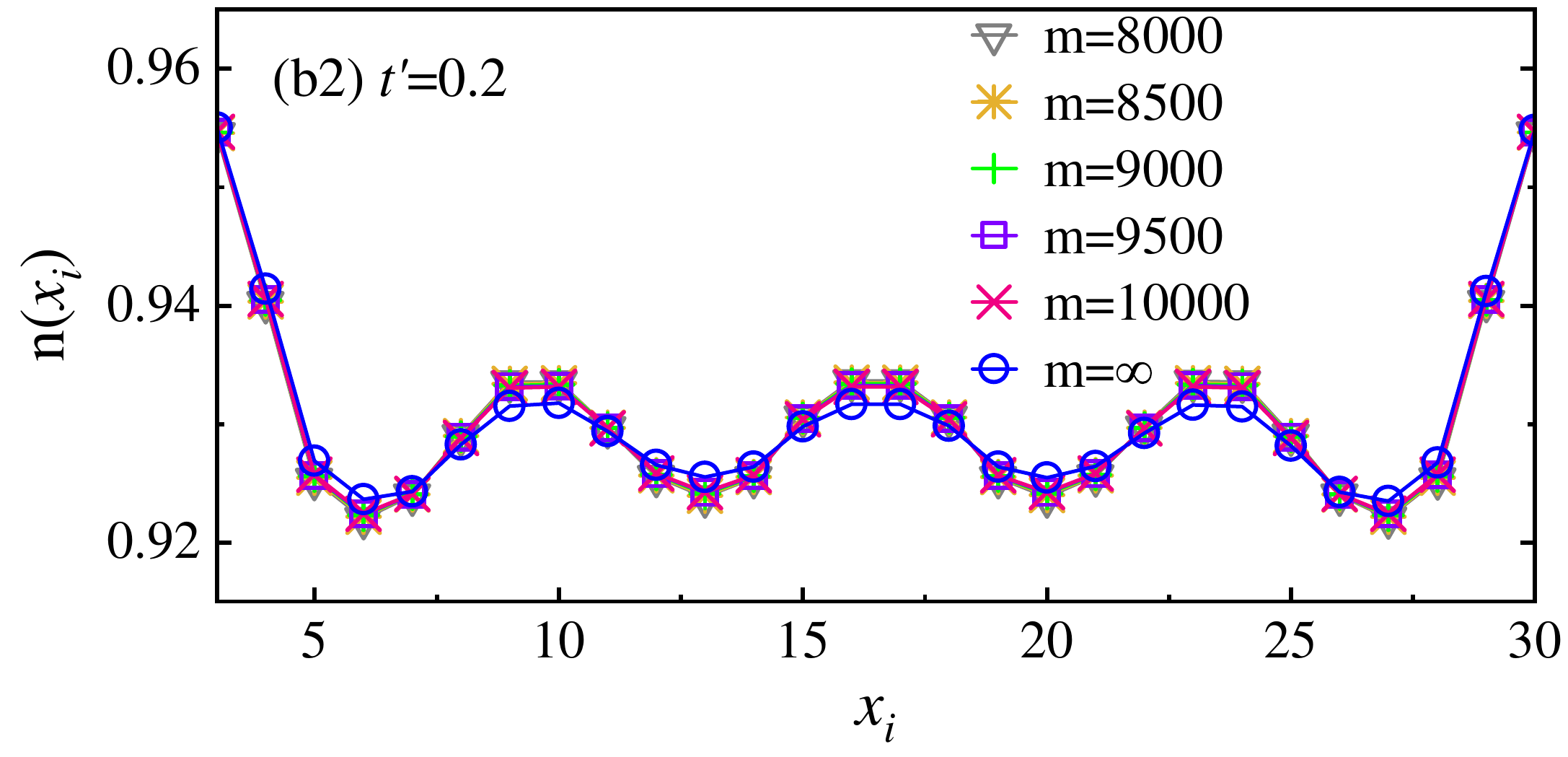}
   \includegraphics[width=0.45\textwidth]{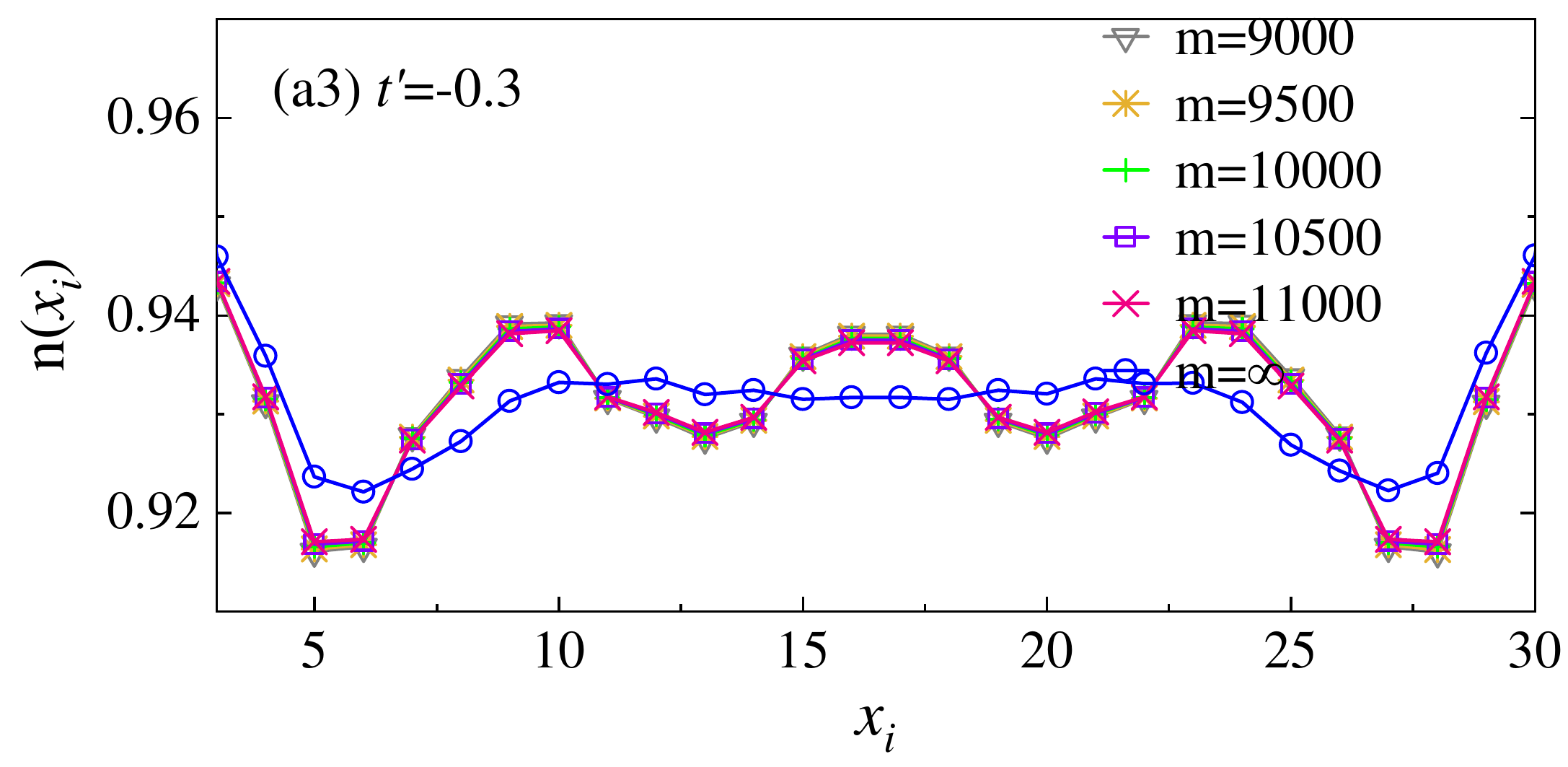}
   \includegraphics[width=0.45\textwidth]{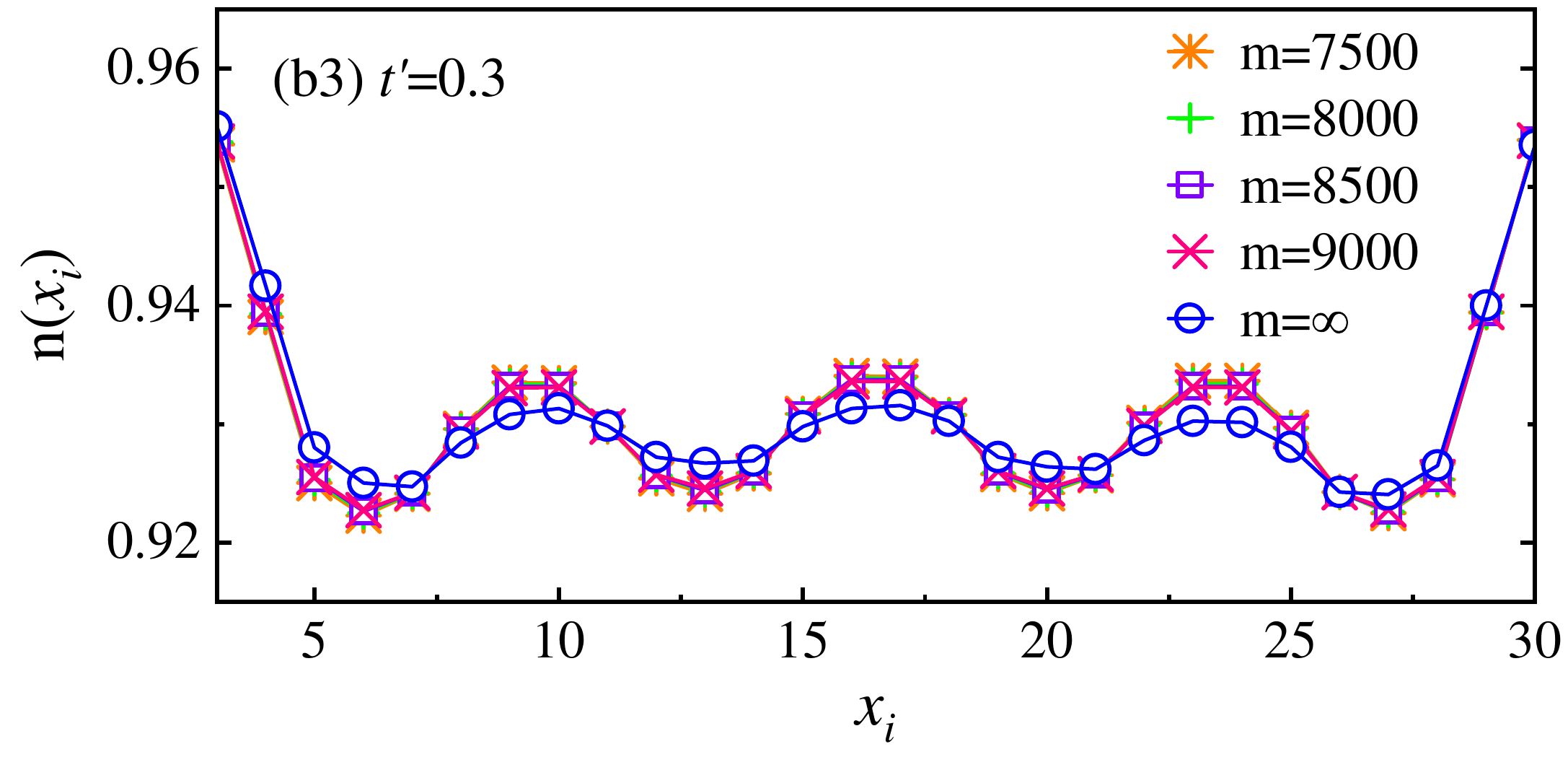}
   \includegraphics[width=0.45\textwidth]{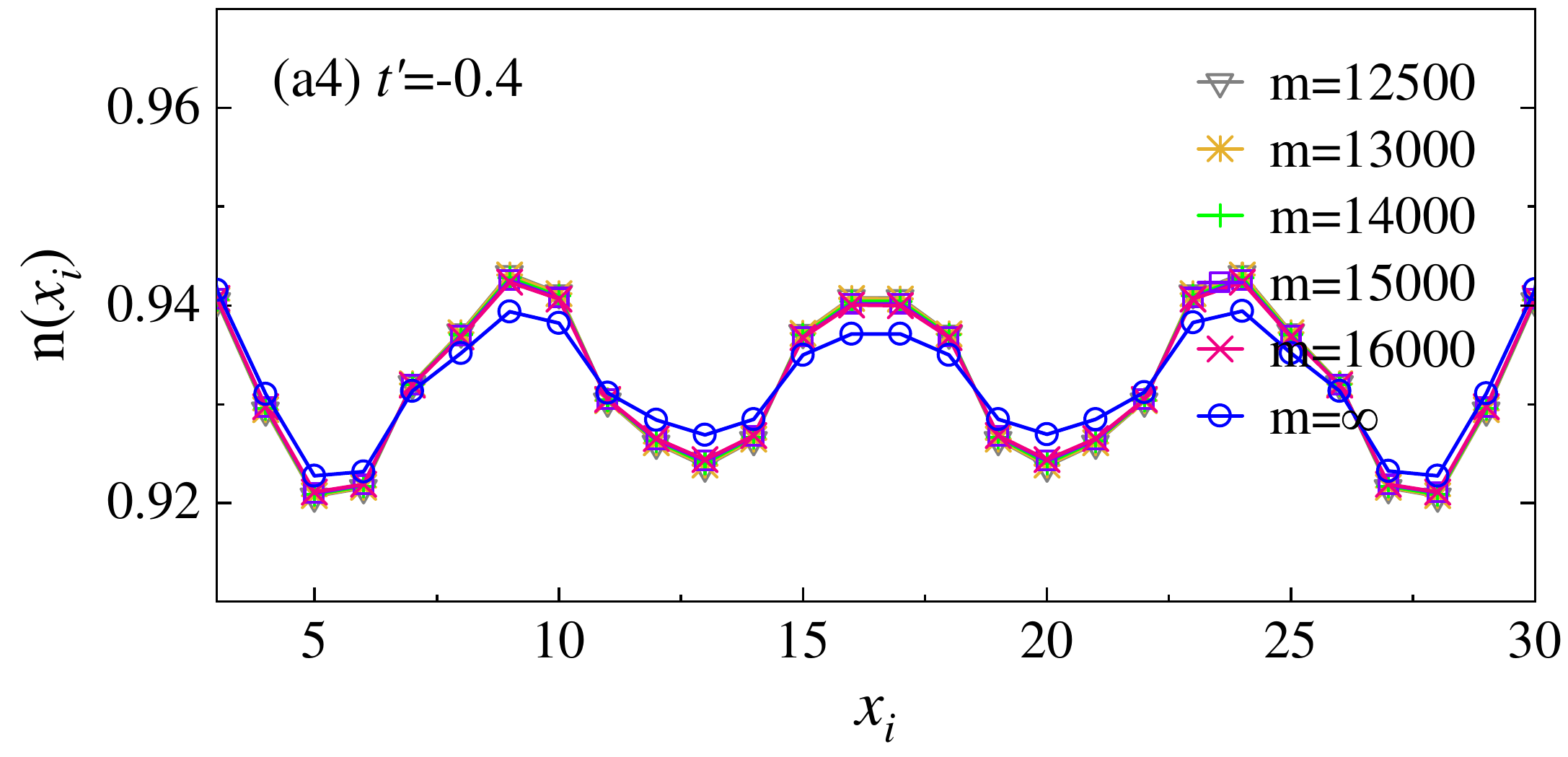}
   \includegraphics[width=0.45\textwidth]{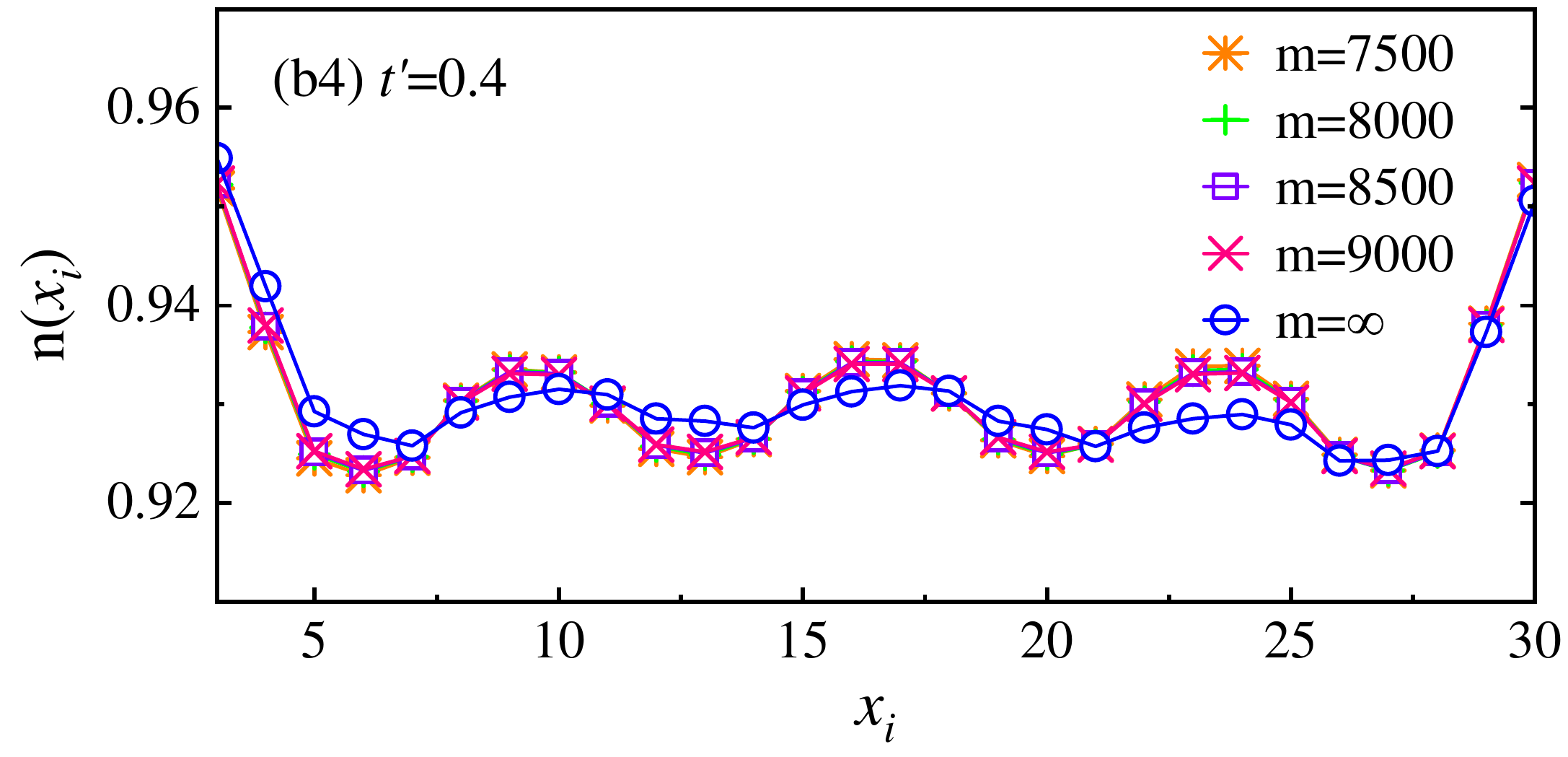}
   \includegraphics[width=0.45\textwidth]{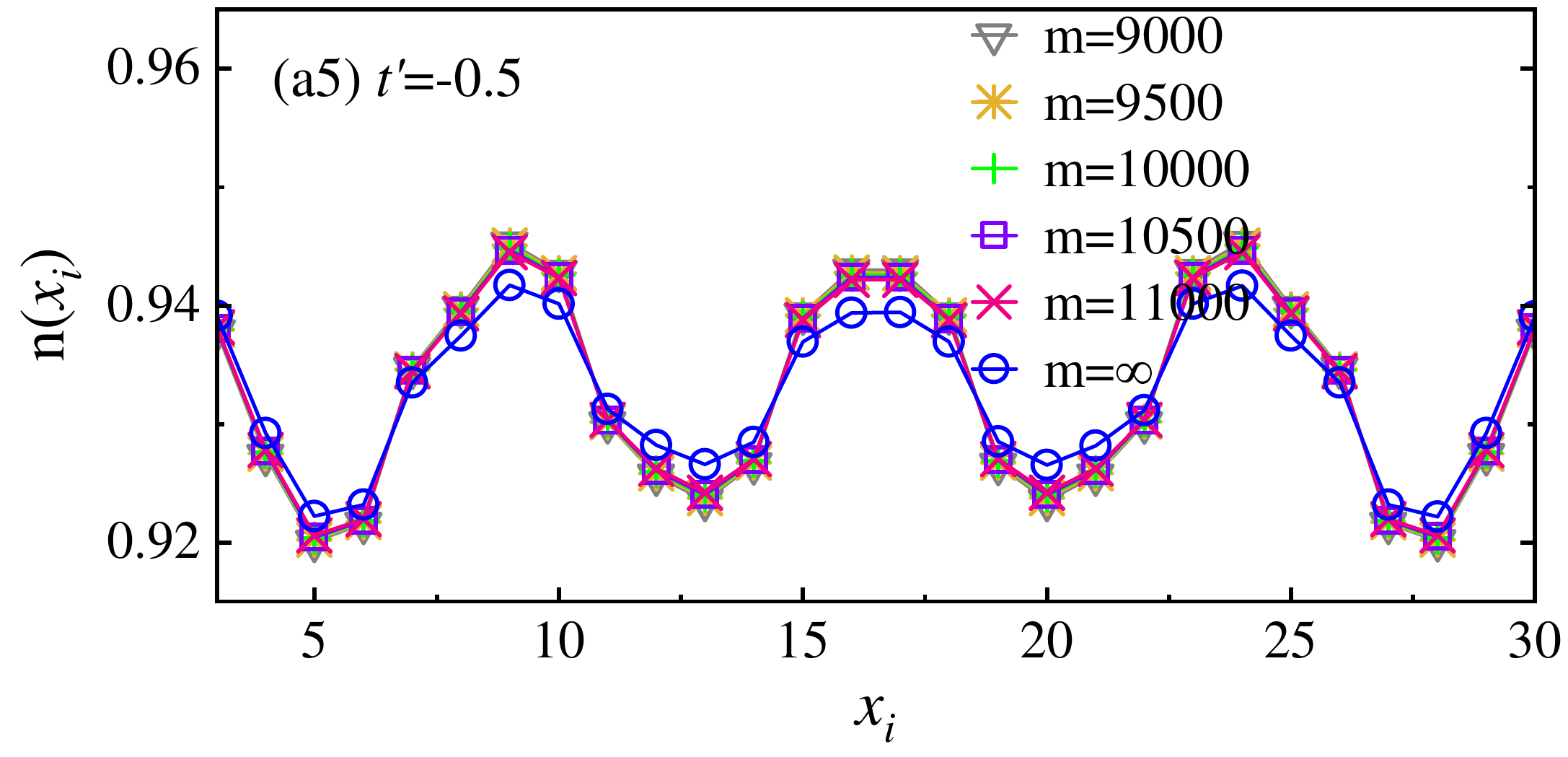}
   \includegraphics[width=0.45\textwidth]{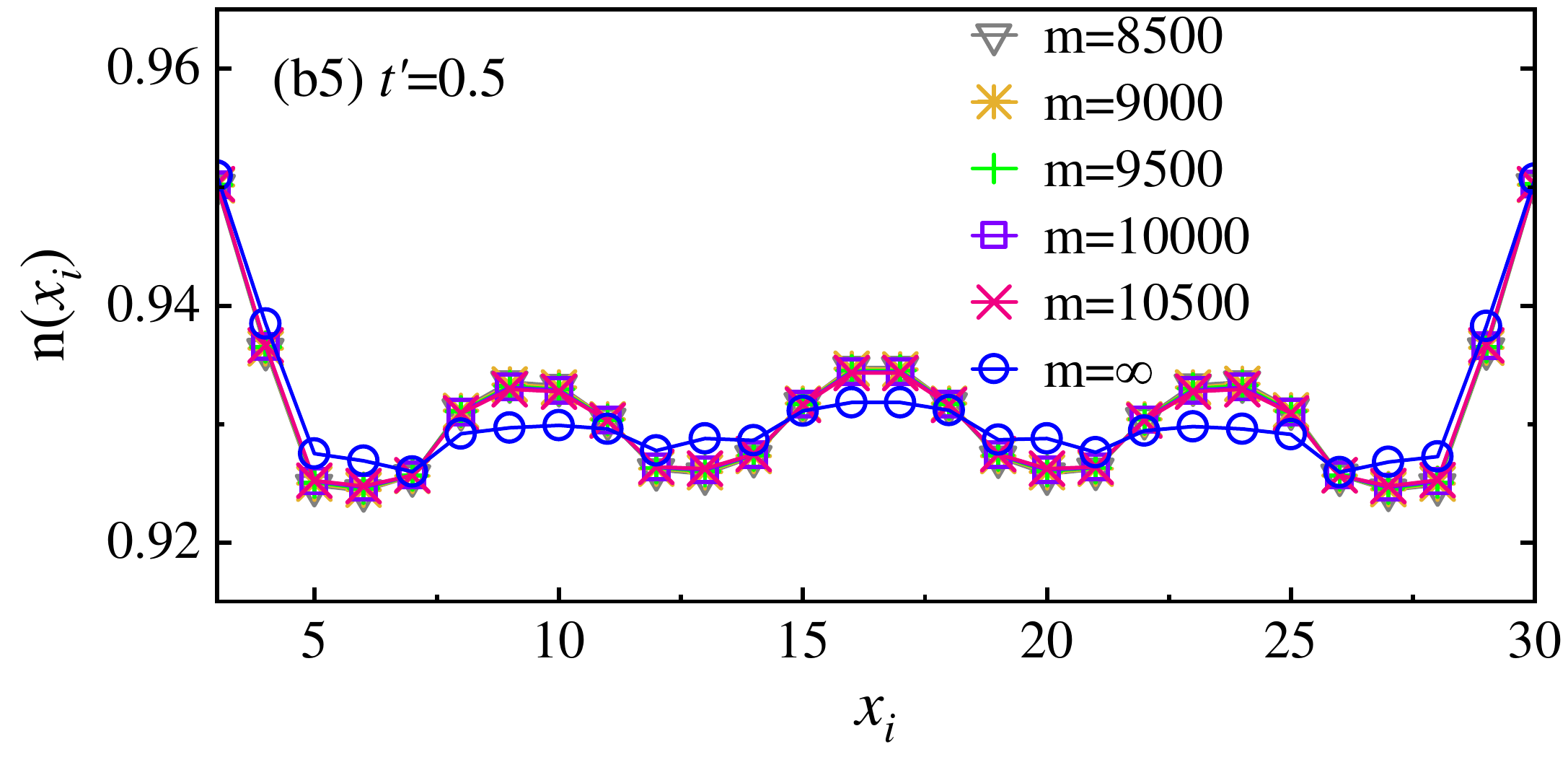}
   \caption{Charge density for a varies of $t'$ values in the stripe and N$\acute e$el antiferromagnetic phases under periodic boundary conditions. Both the finite bond dimension and the extrapolated results are shown.}
   \label{charge_PBC}
\end{figure*}

\begin{figure*}[t]
   \includegraphics[width=0.45\textwidth]{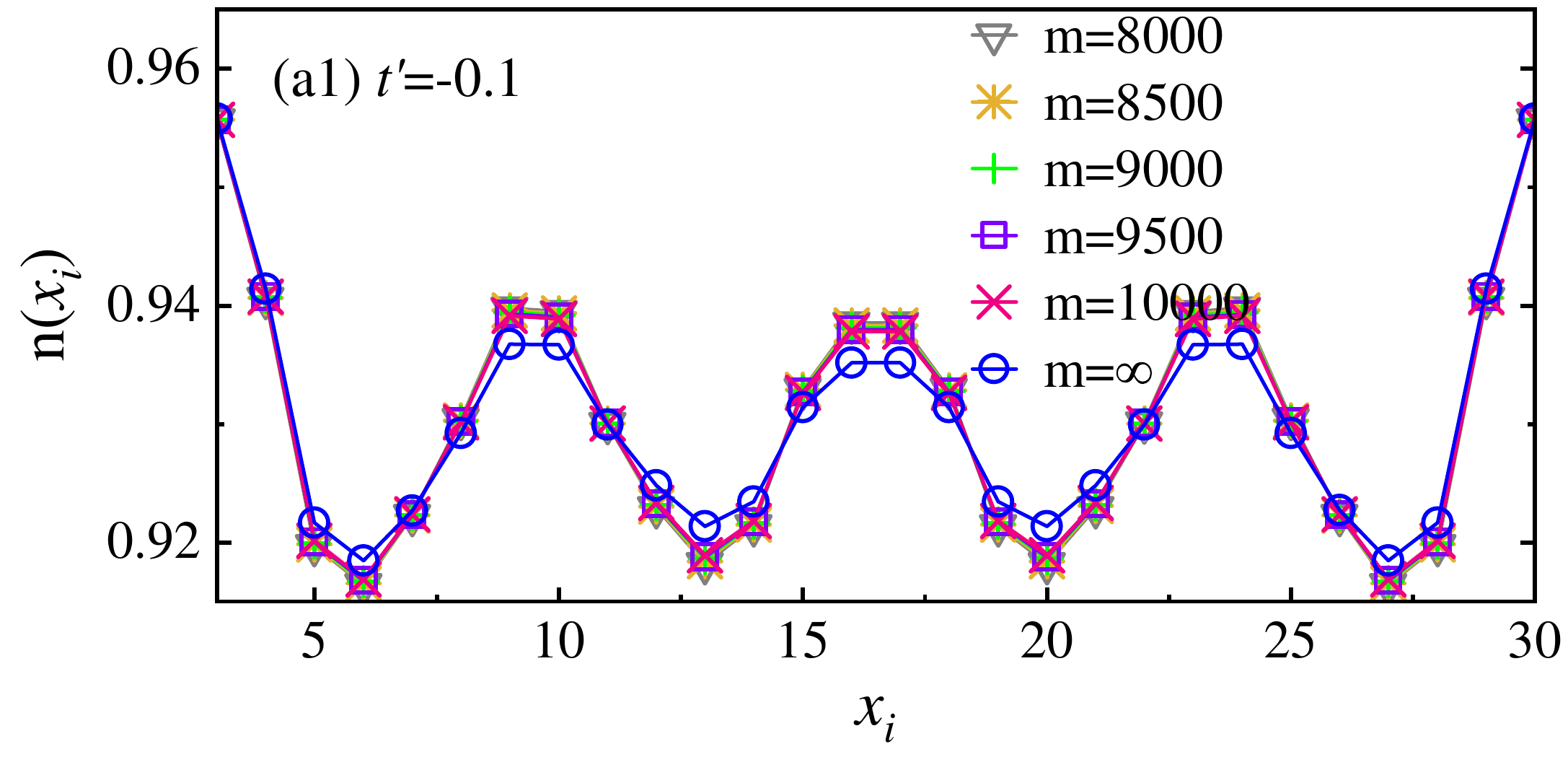}
   \includegraphics[width=0.45\textwidth]{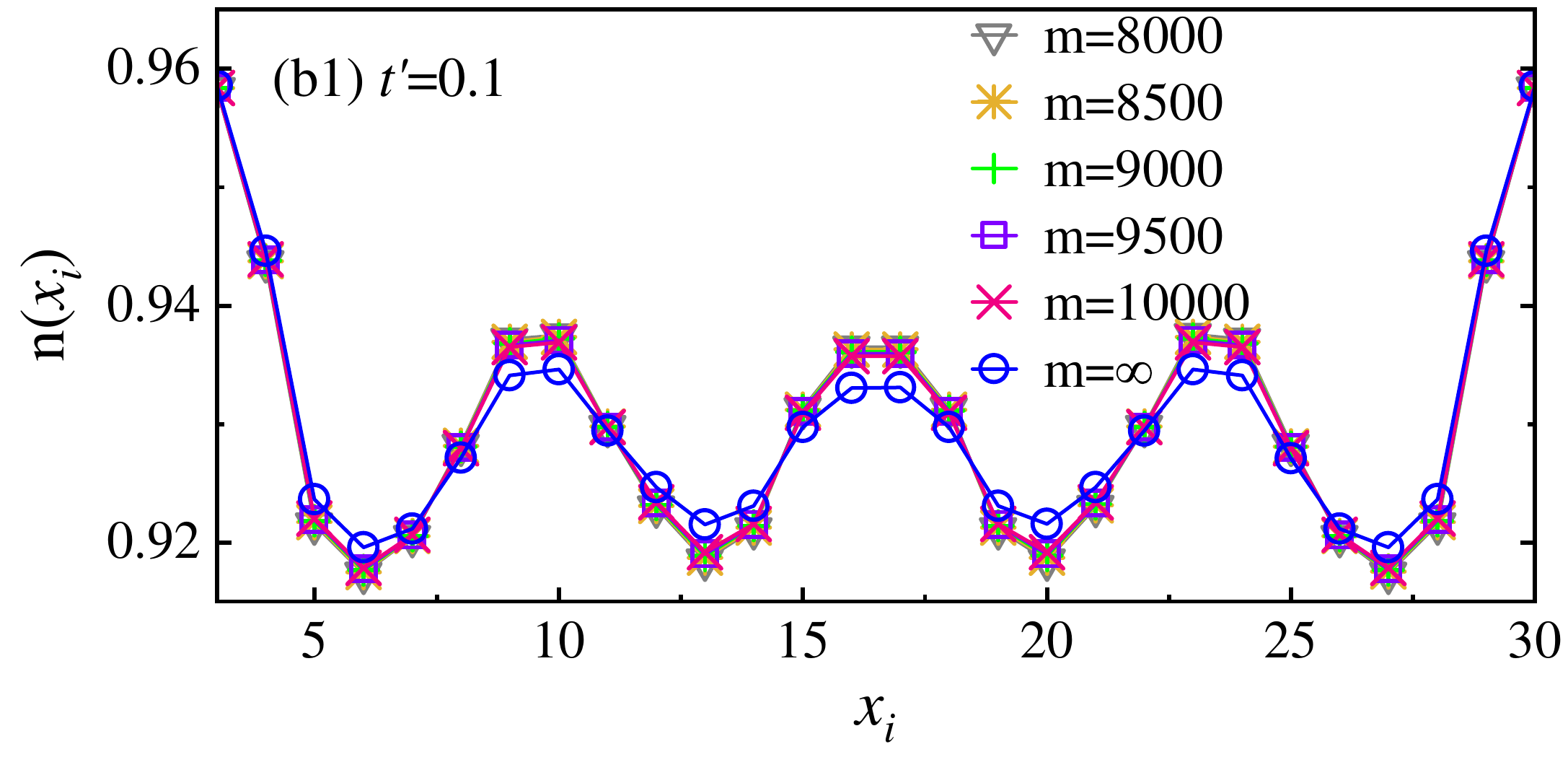}
   \includegraphics[width=0.45\textwidth]{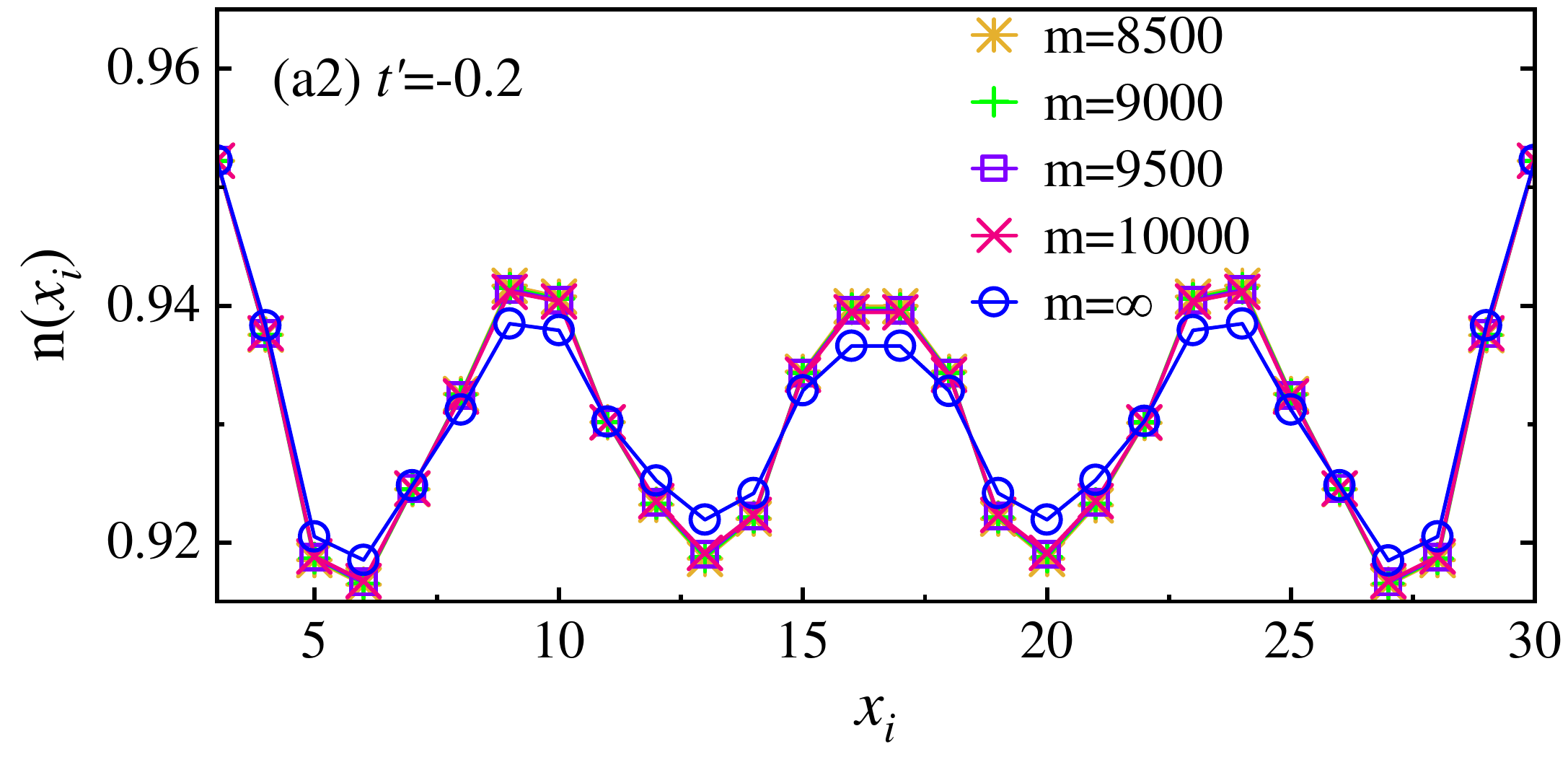}
   \includegraphics[width=0.45\textwidth]{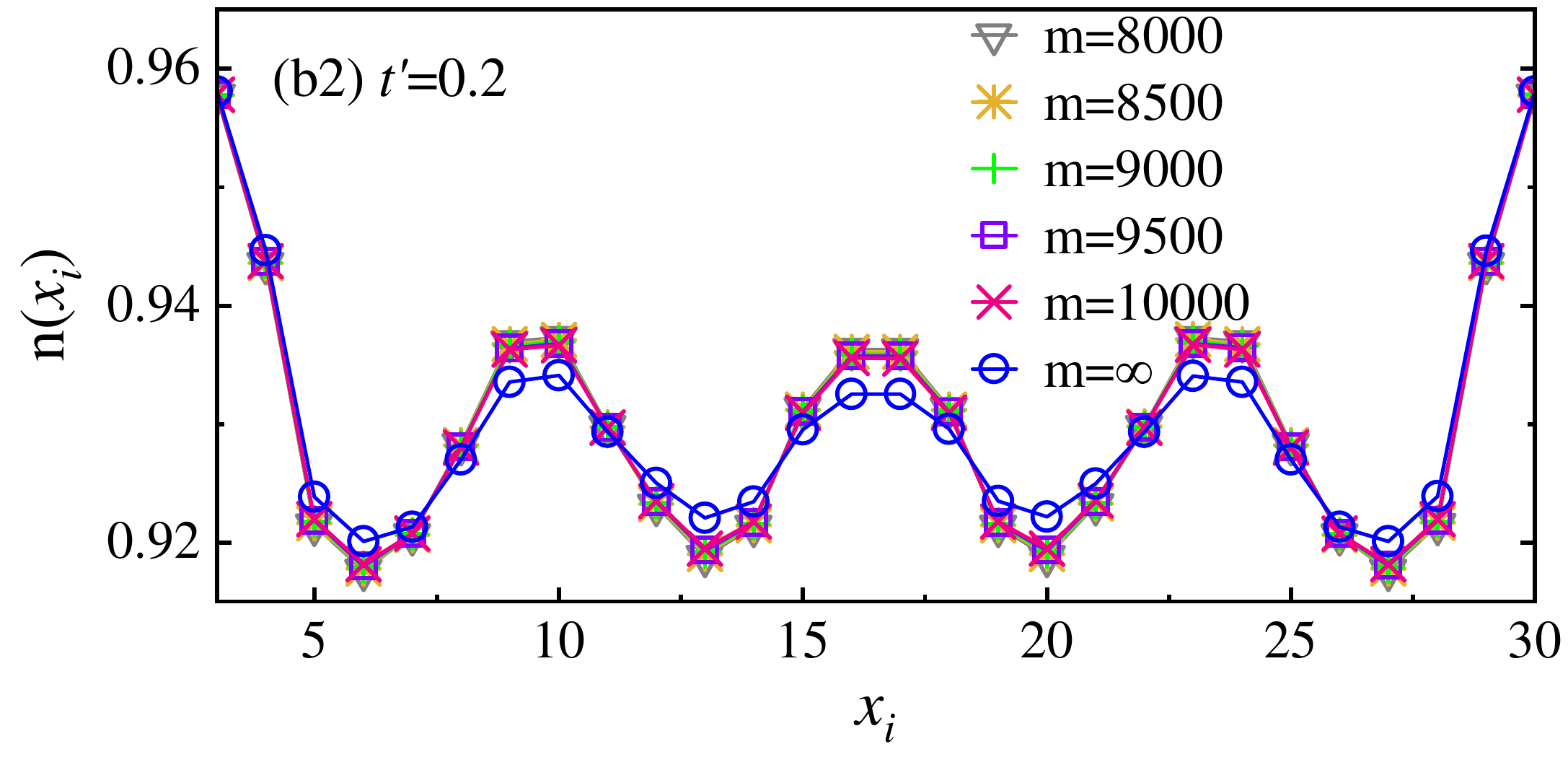}
   \includegraphics[width=0.45\textwidth]{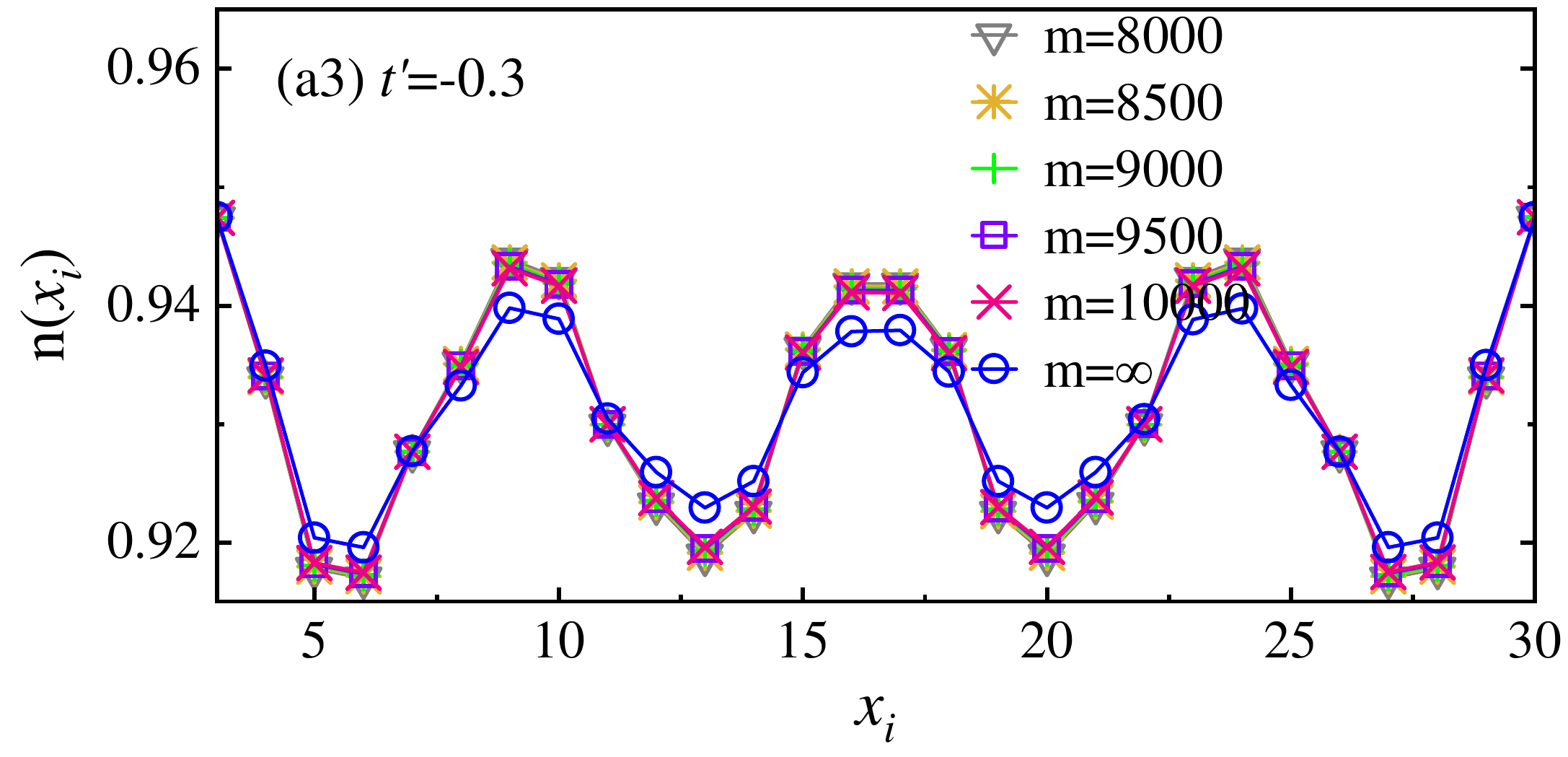}
   \includegraphics[width=0.45\textwidth]{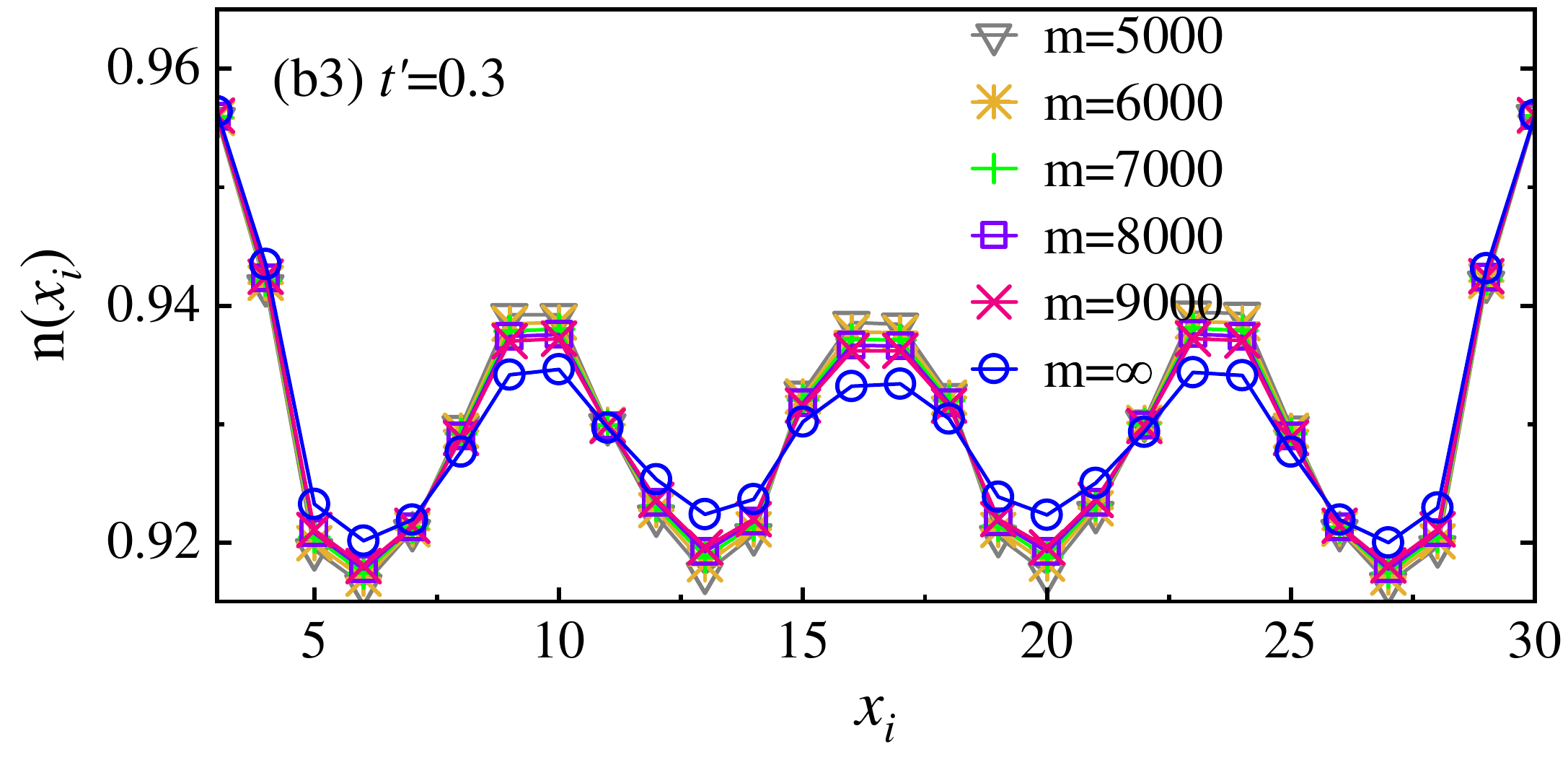}
   \includegraphics[width=0.45\textwidth]{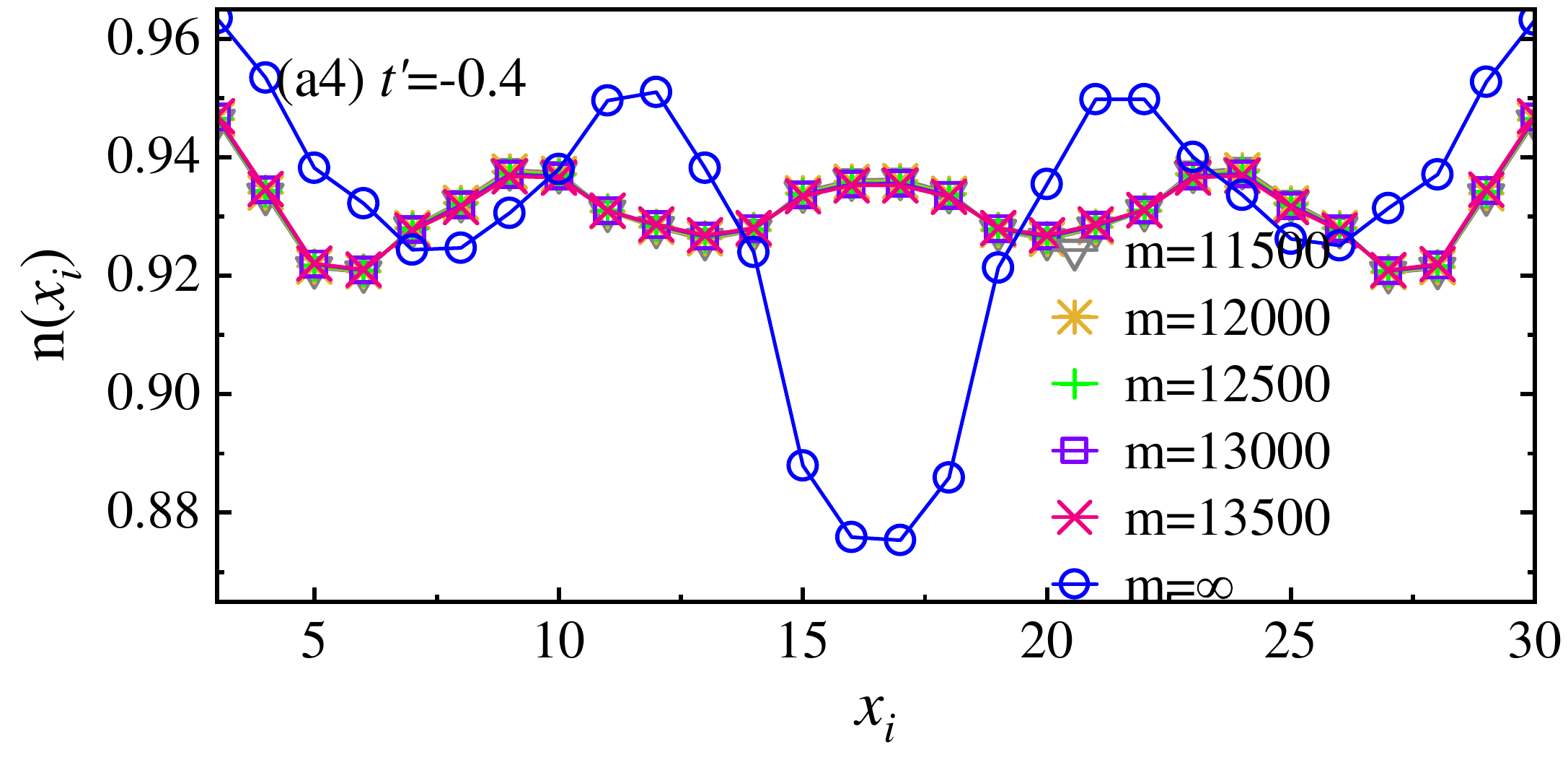}
   \includegraphics[width=0.45\textwidth]{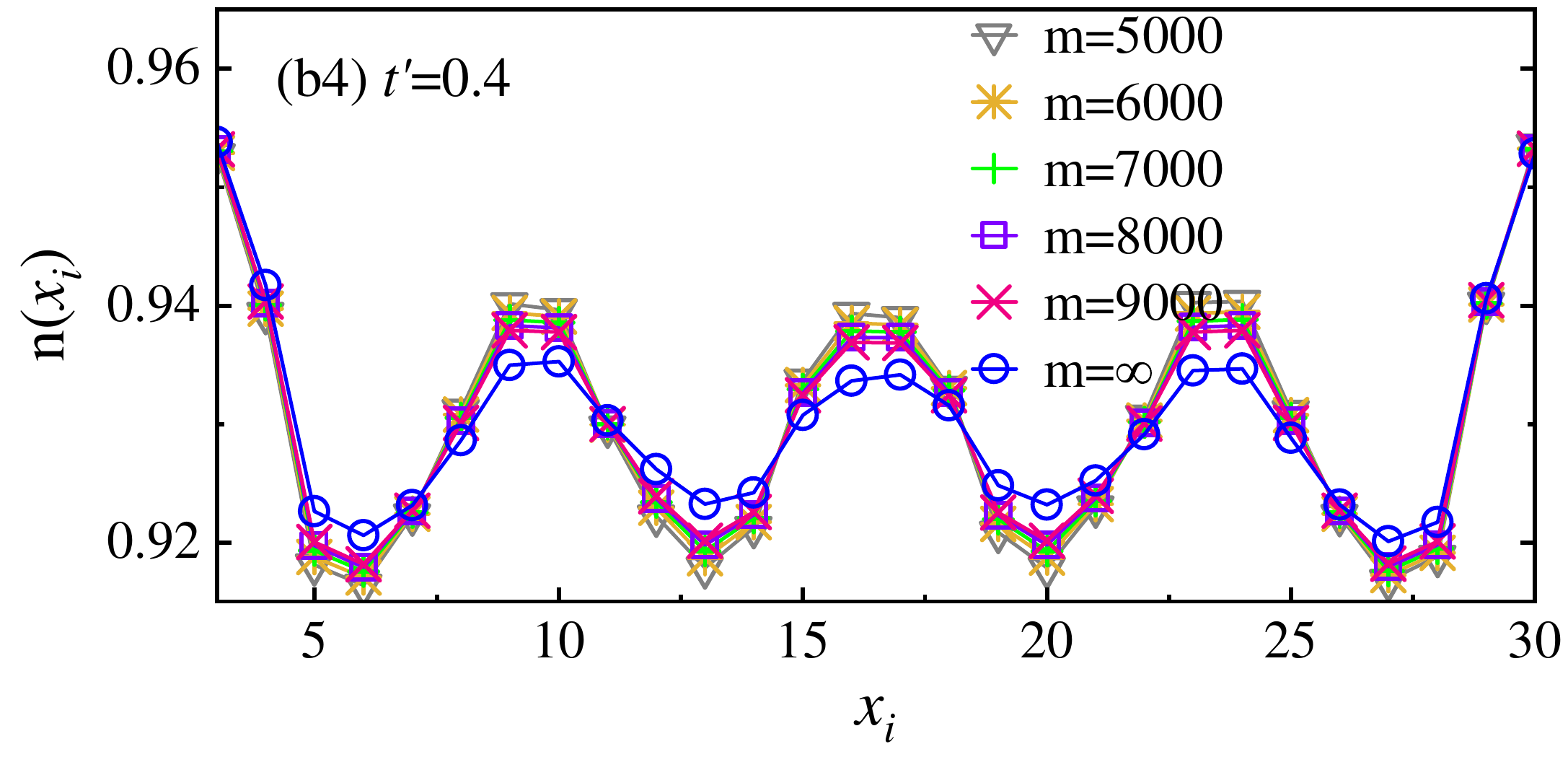}
   \includegraphics[width=0.45\textwidth]{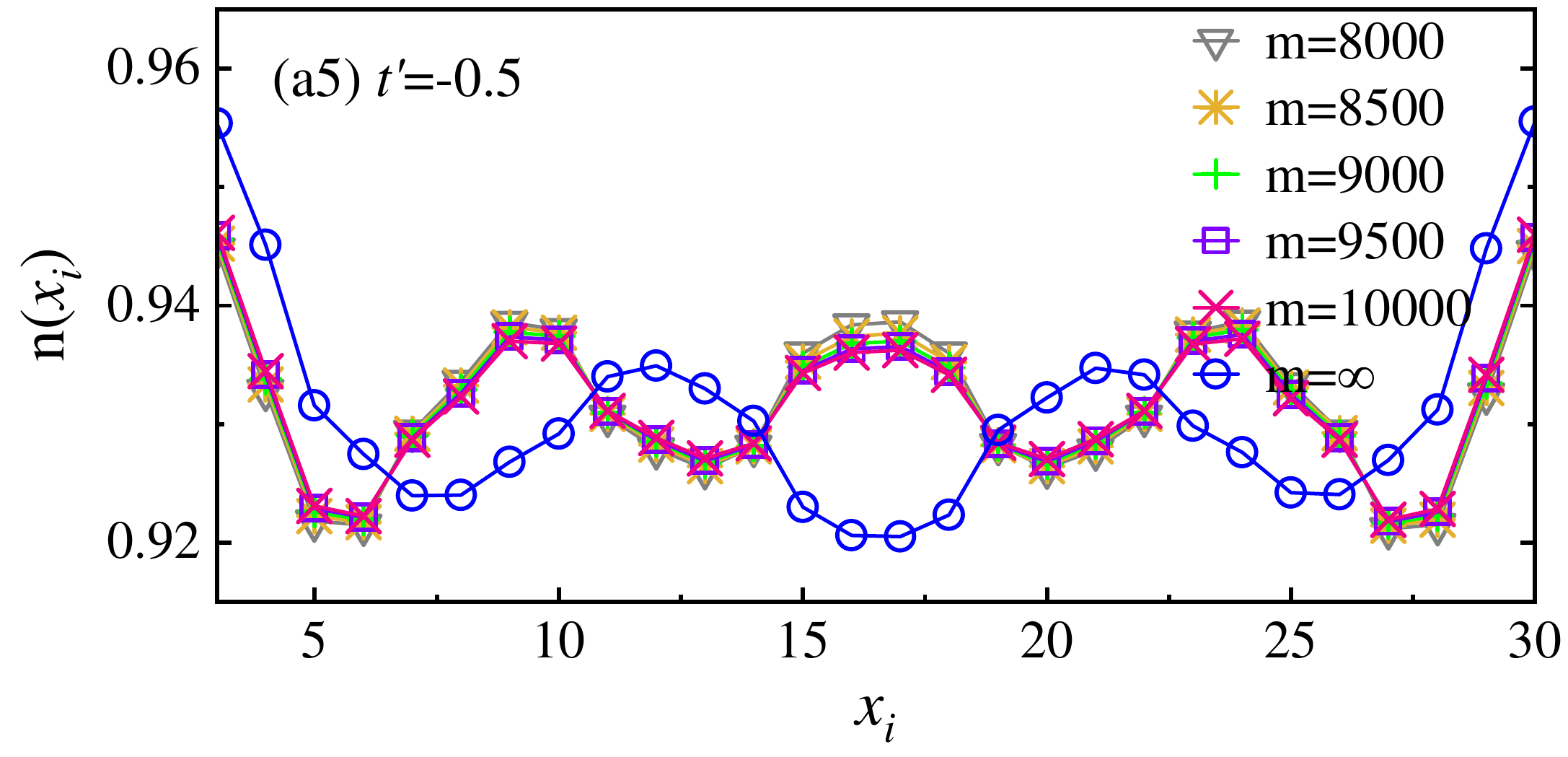}
   \includegraphics[width=0.45\textwidth]{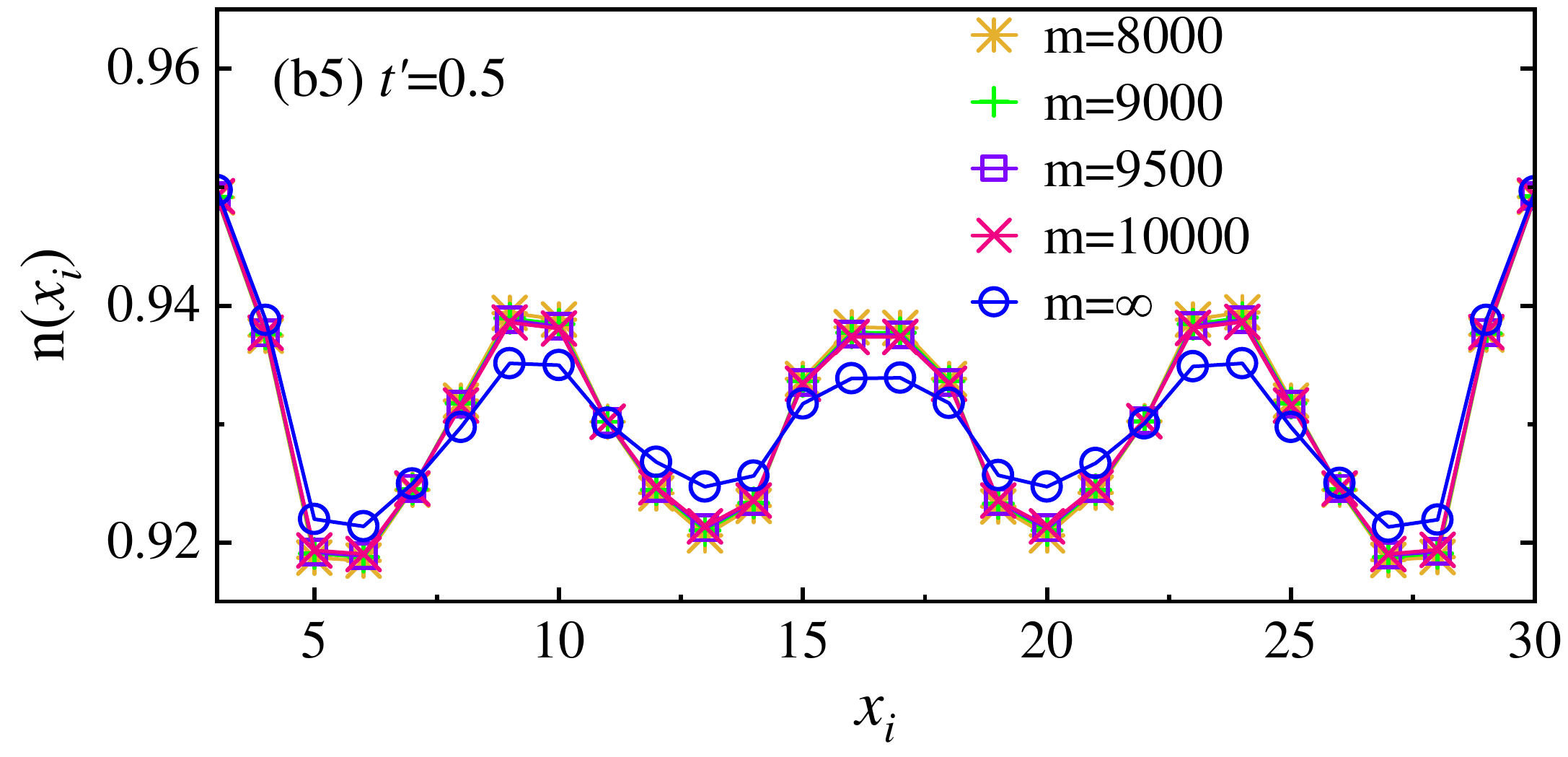}
   \caption{Charge density for a varies of $t'$ values in the stripe and N$\acute e$el antiferromagnetic phases under anti-periodic boundary conditions. Both the finite bond dimension and the extrapolated results are shown.}
   \label{charge_APBC}
\end{figure*}

Figs.~\ref{charge_PBC} and ~\ref{charge_APBC} show the charge density profiles for a varies of $t'$ values in the stripe and N$\acute e$el antiferromagnetic phases under periodic and anti-periodic boundary conditions, respectively. When $t'$ is away from the phase transition point with $-0.4 < t'_c < -0.3$, the local charge density is well converged. Near the phase transition point, the extrapolations are not well-behaved, indicating the requirement for larger bond dimension to enter the converged region in these difficult cases.

\subsection{B. Staggered spin density}

\begin{figure*}[t]
   \includegraphics[width=0.45\textwidth]{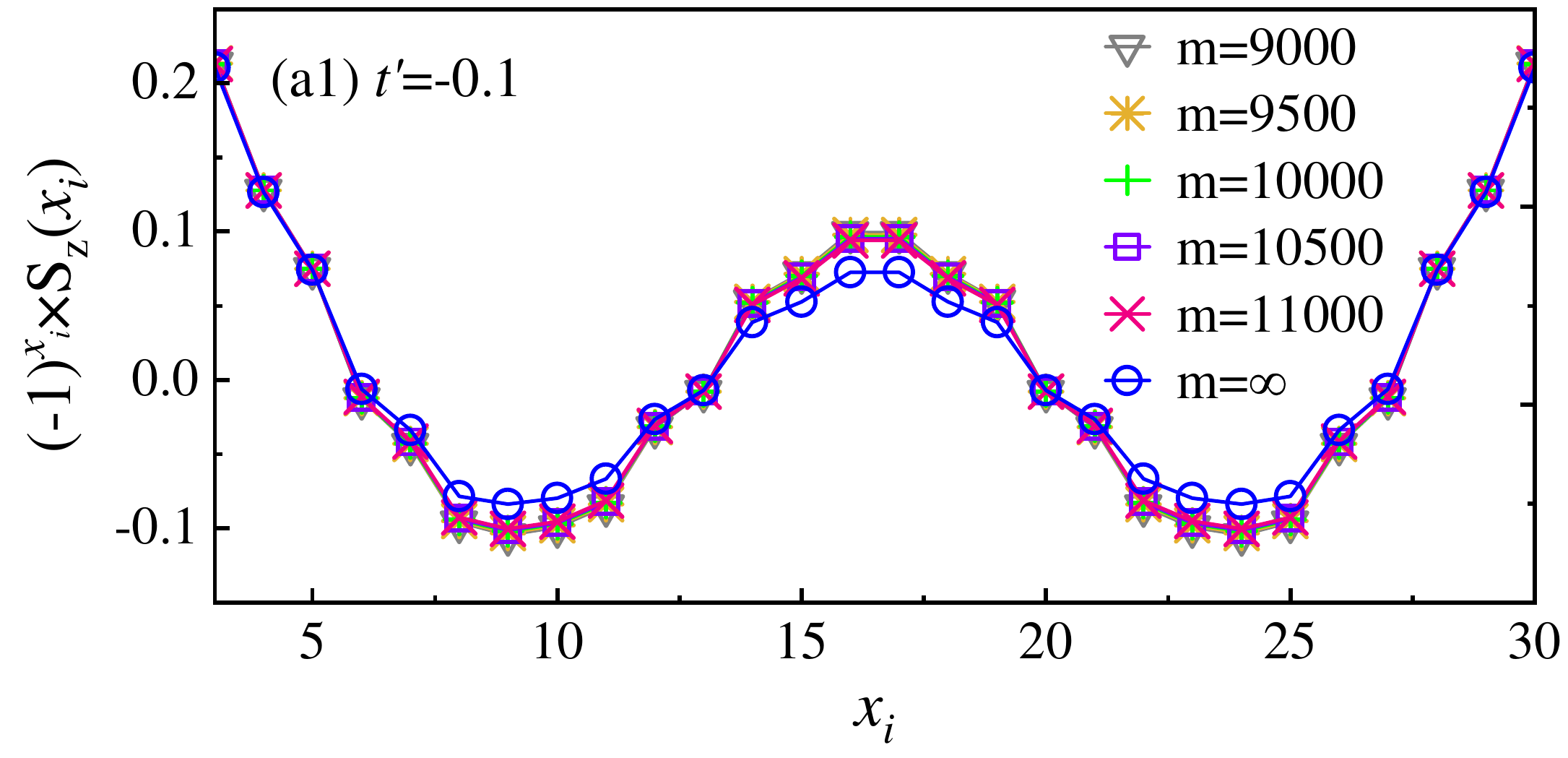}
   \includegraphics[width=0.45\textwidth]{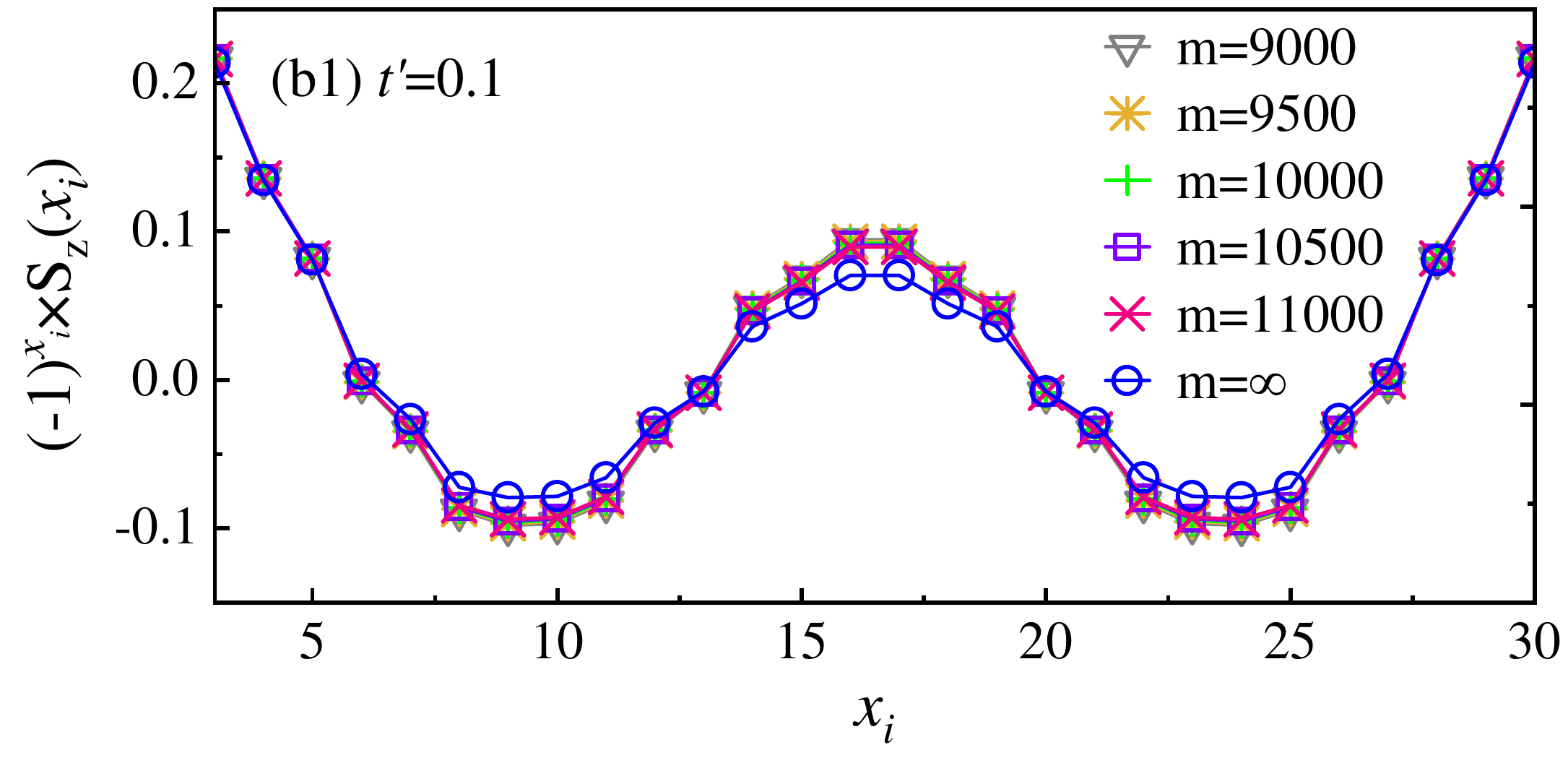}
   \includegraphics[width=0.45\textwidth]{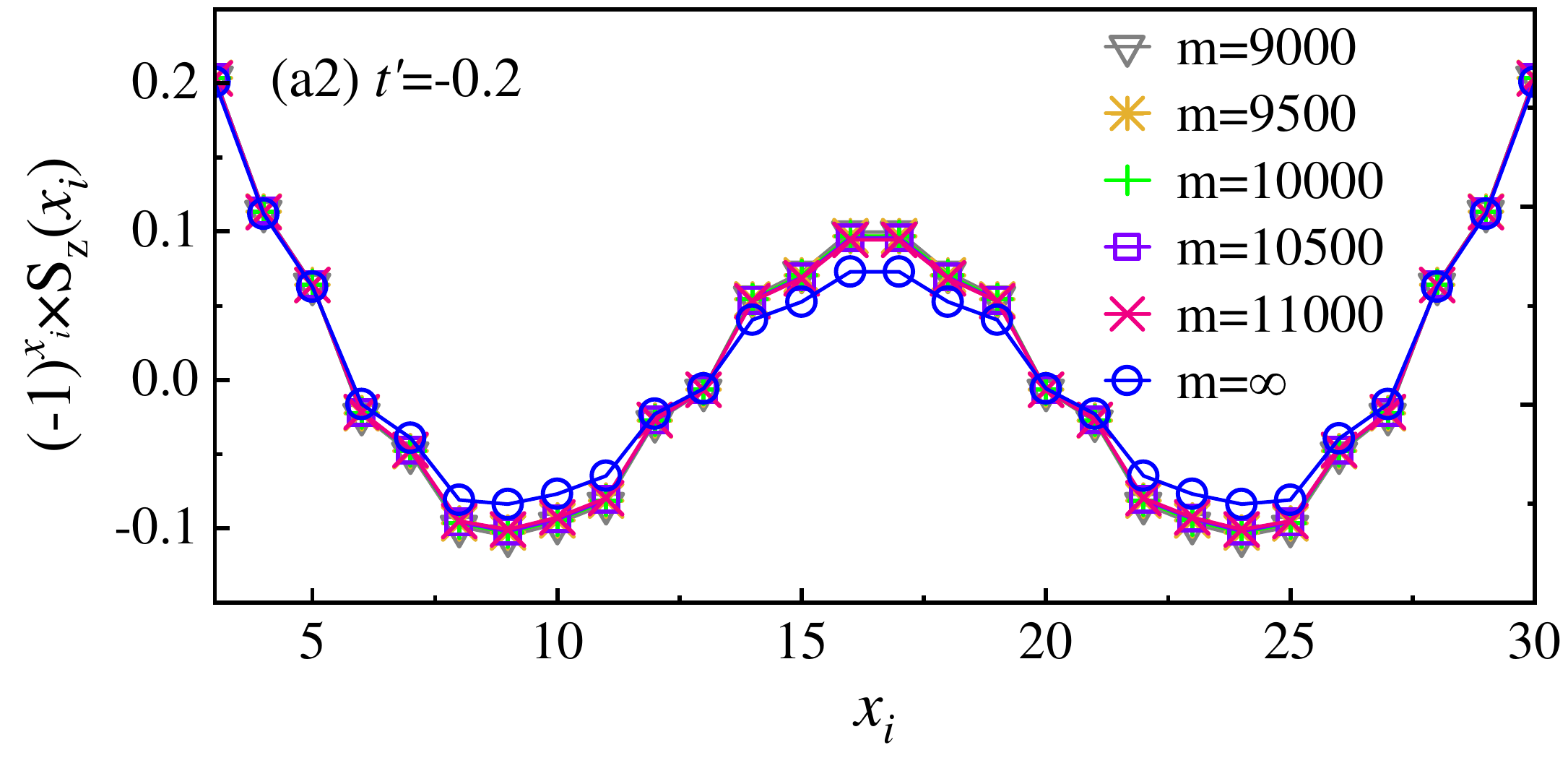}
   \includegraphics[width=0.45\textwidth]{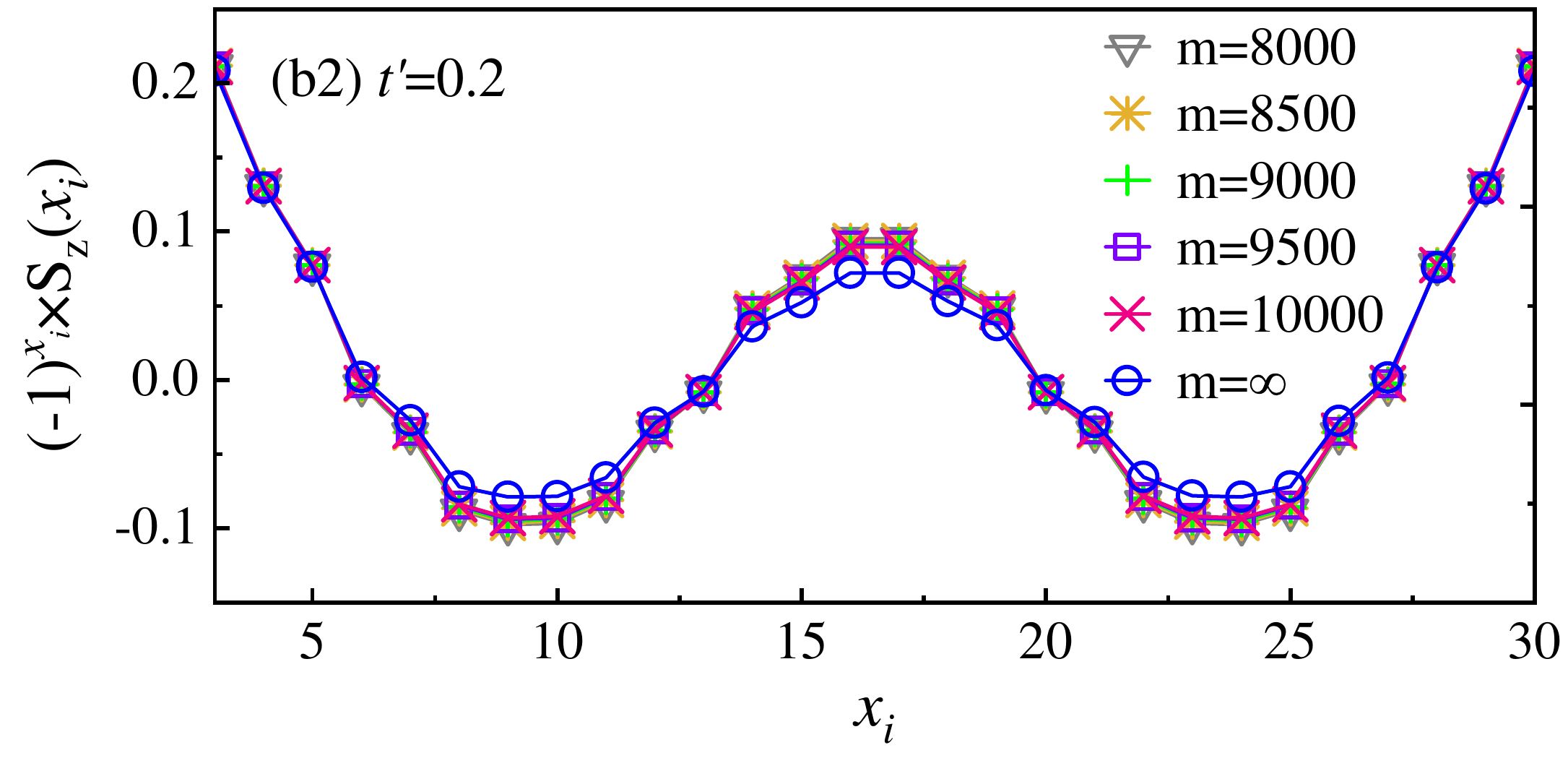}
   \includegraphics[width=0.45\textwidth]{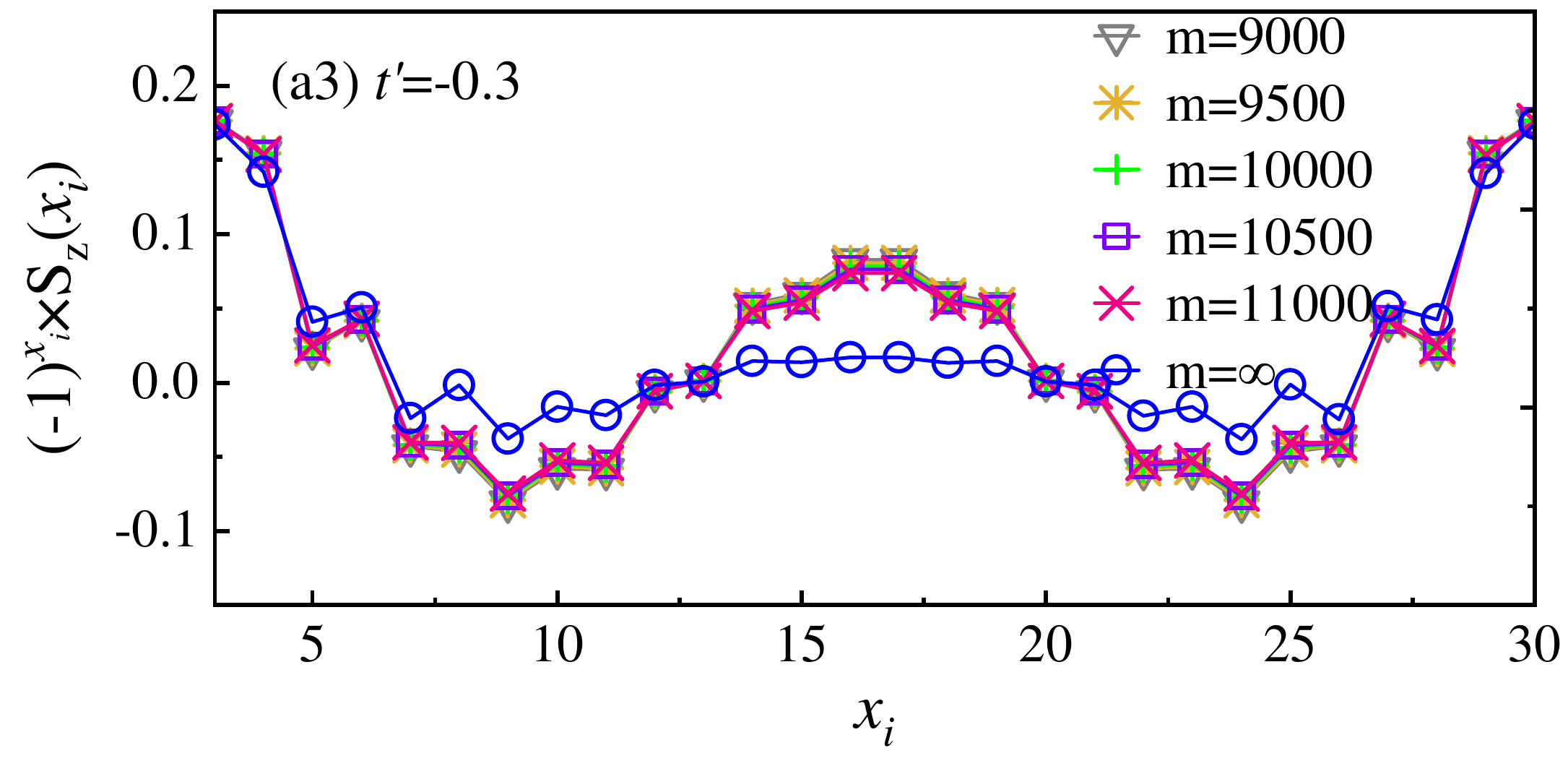}
   \includegraphics[width=0.45\textwidth]{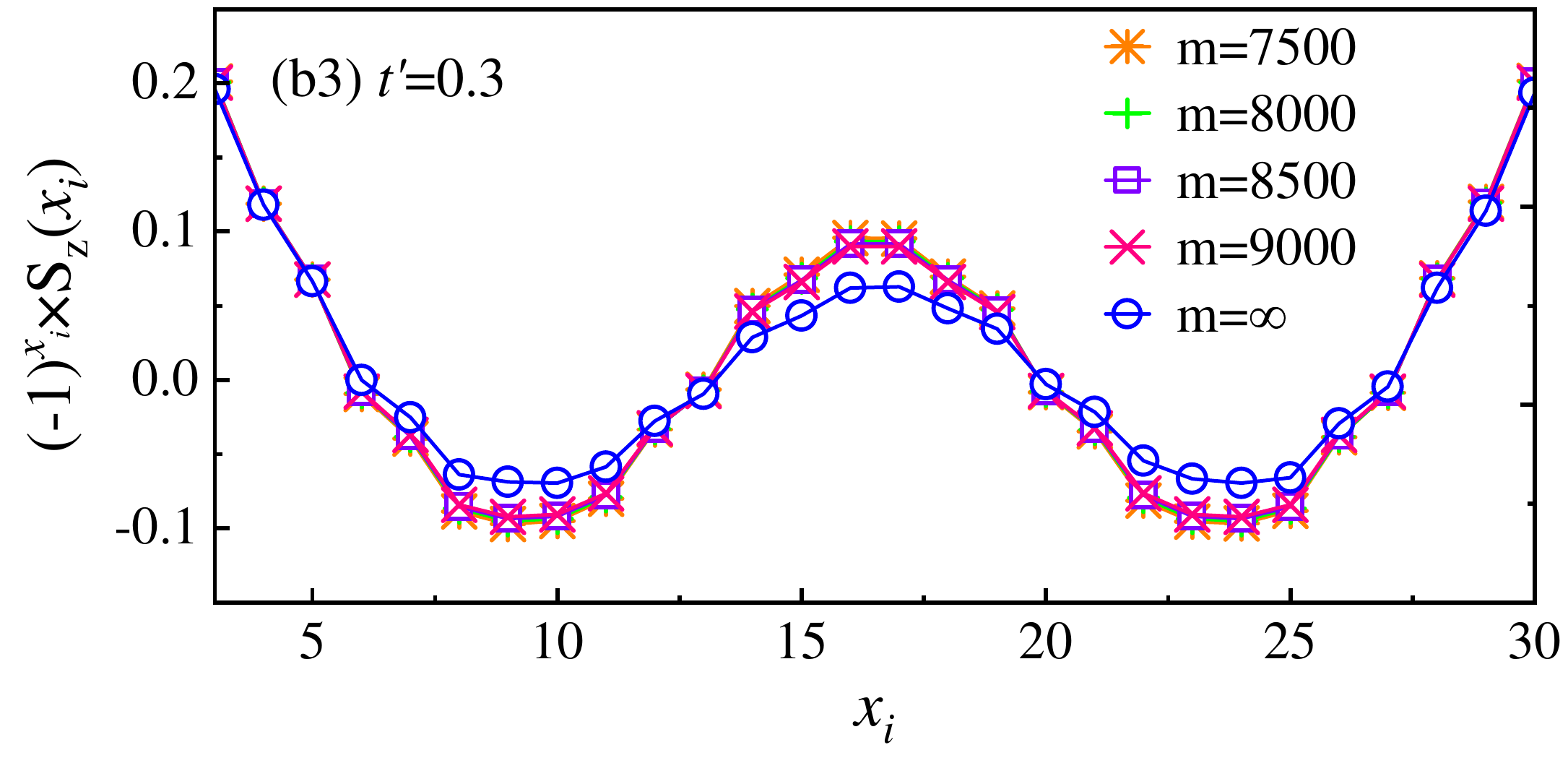}
   \includegraphics[width=0.45\textwidth]{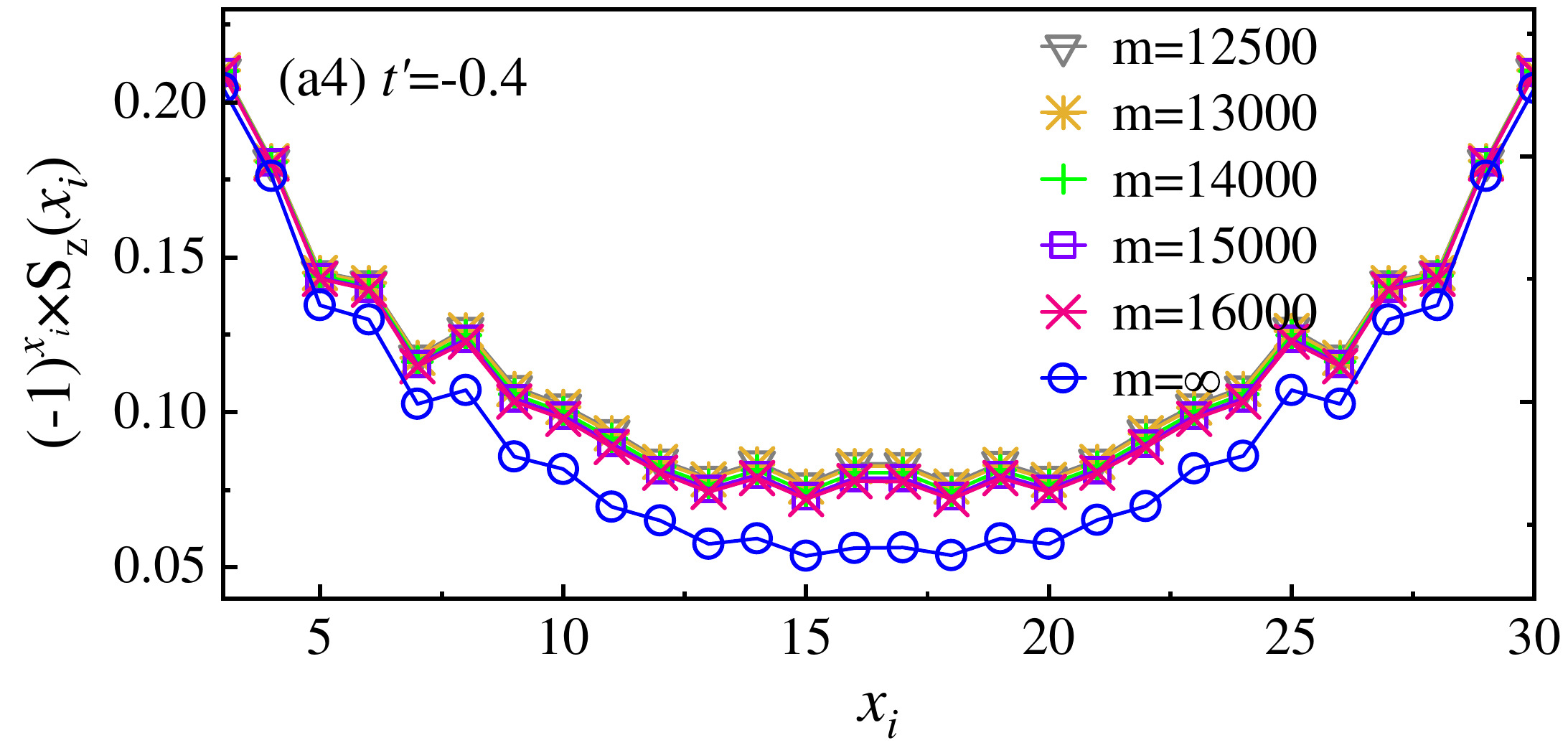}
   \includegraphics[width=0.45\textwidth]{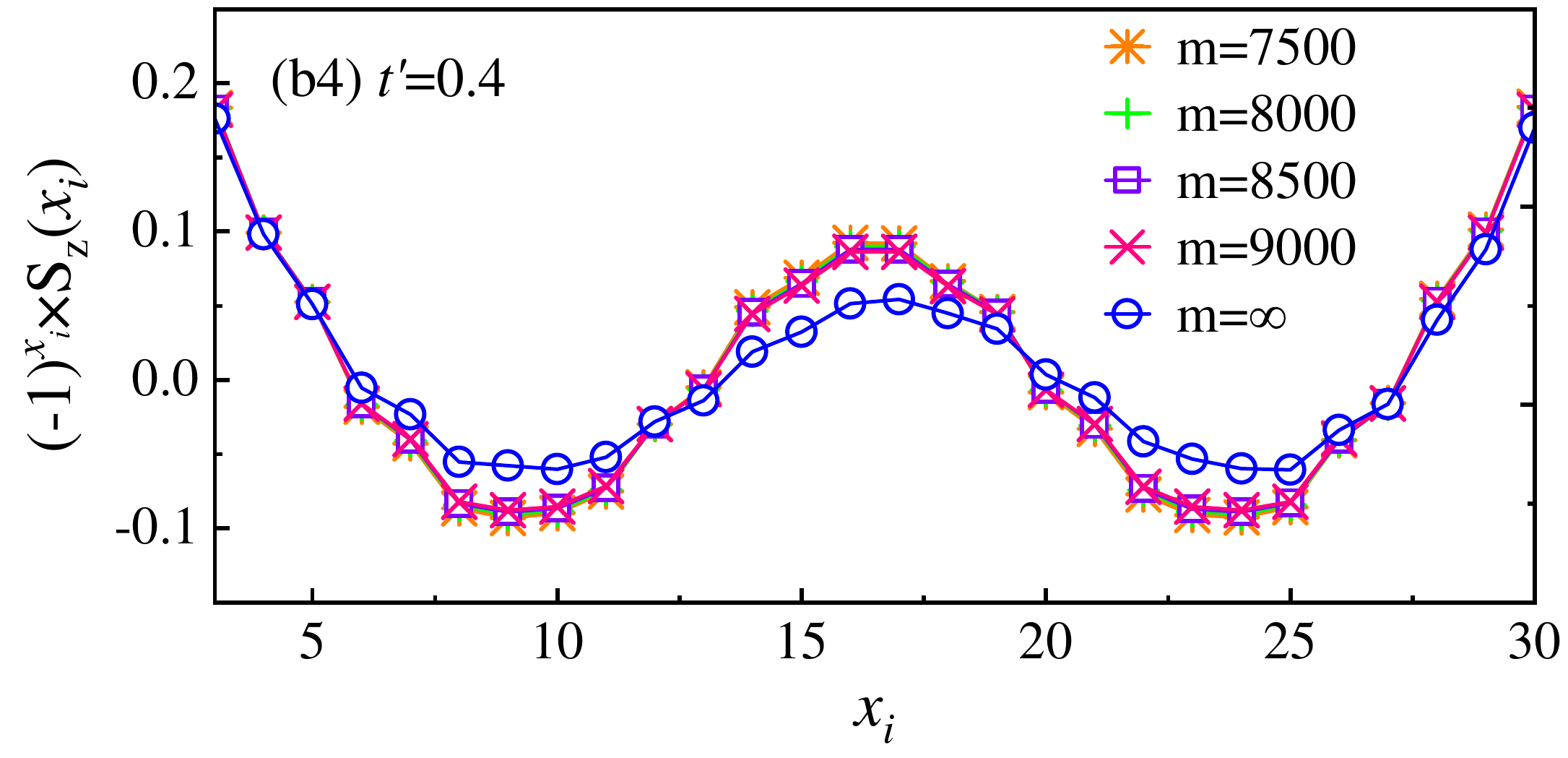}
   \includegraphics[width=0.45\textwidth]{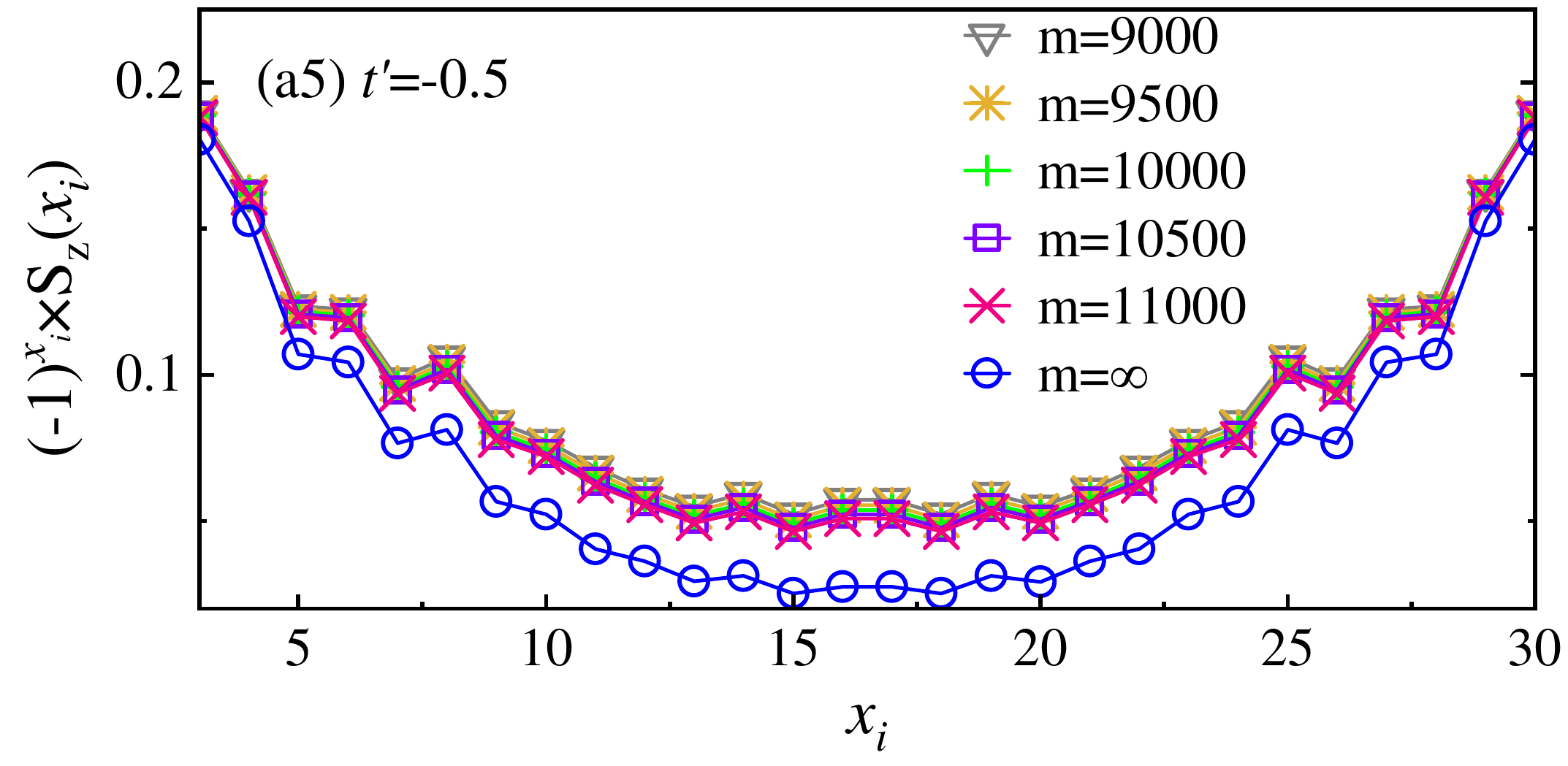}
   \includegraphics[width=0.45\textwidth]{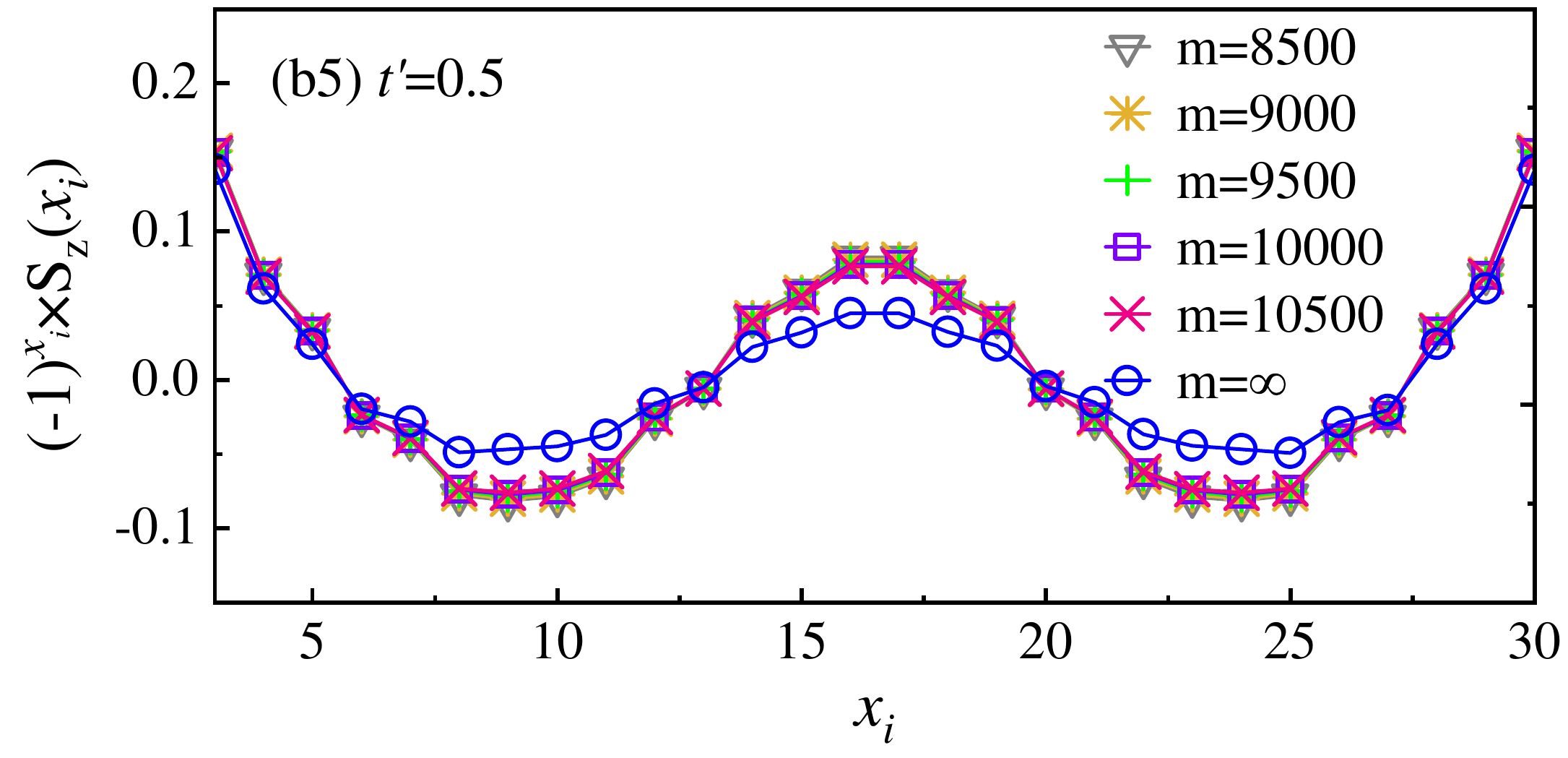}
   \caption{Staggered spin density for a varies of $t'$ values in the stripe and N$\acute e$el antiferromagnetic phases under periodic boundary conditions. Both the finite bond dimension and the extrapolated results are shown.}
   \label{stagSz_PBC}
\end{figure*}

\begin{figure*}[t]
   \includegraphics[width=0.45\textwidth]{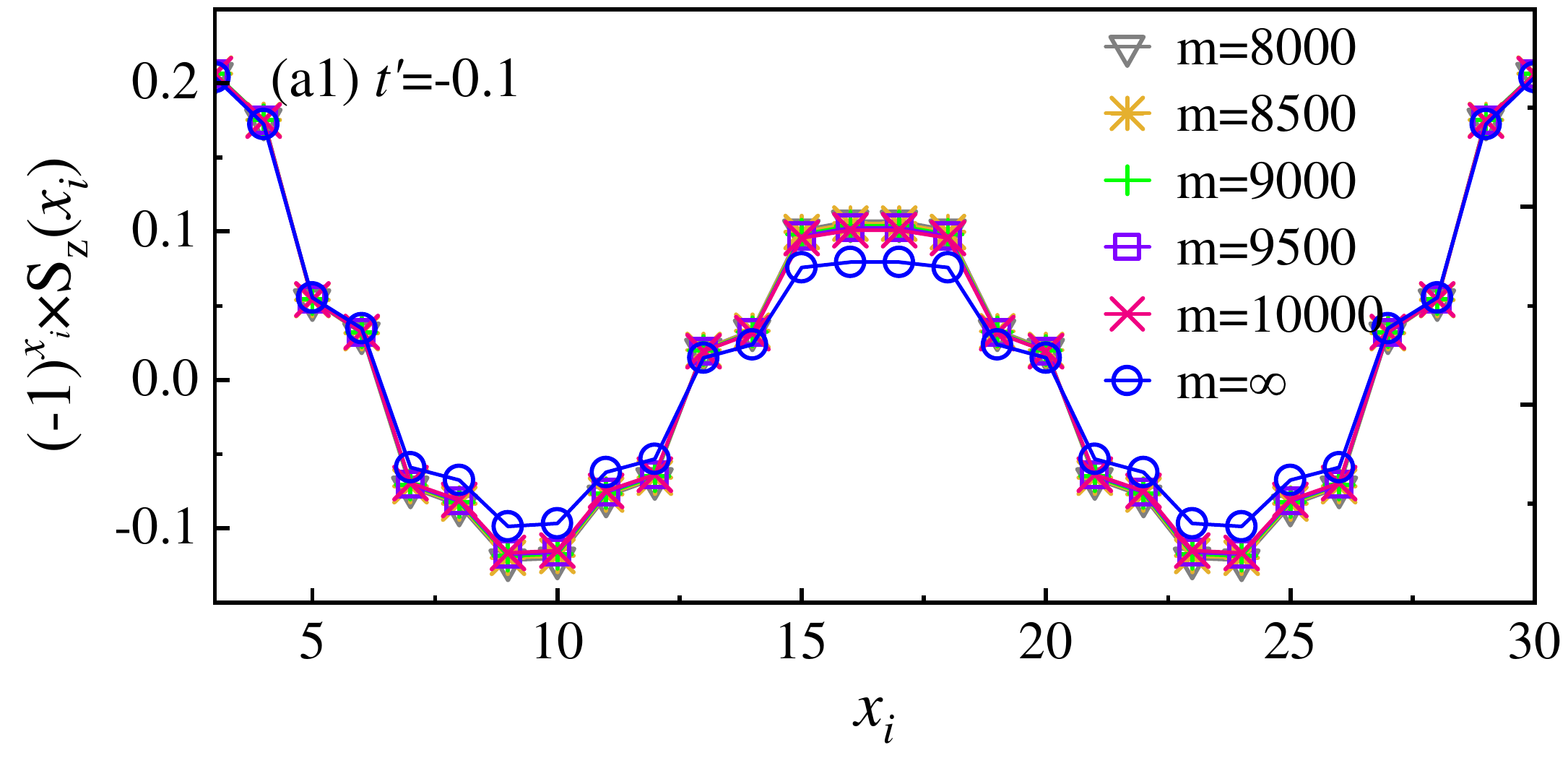}
   \includegraphics[width=0.45\textwidth]{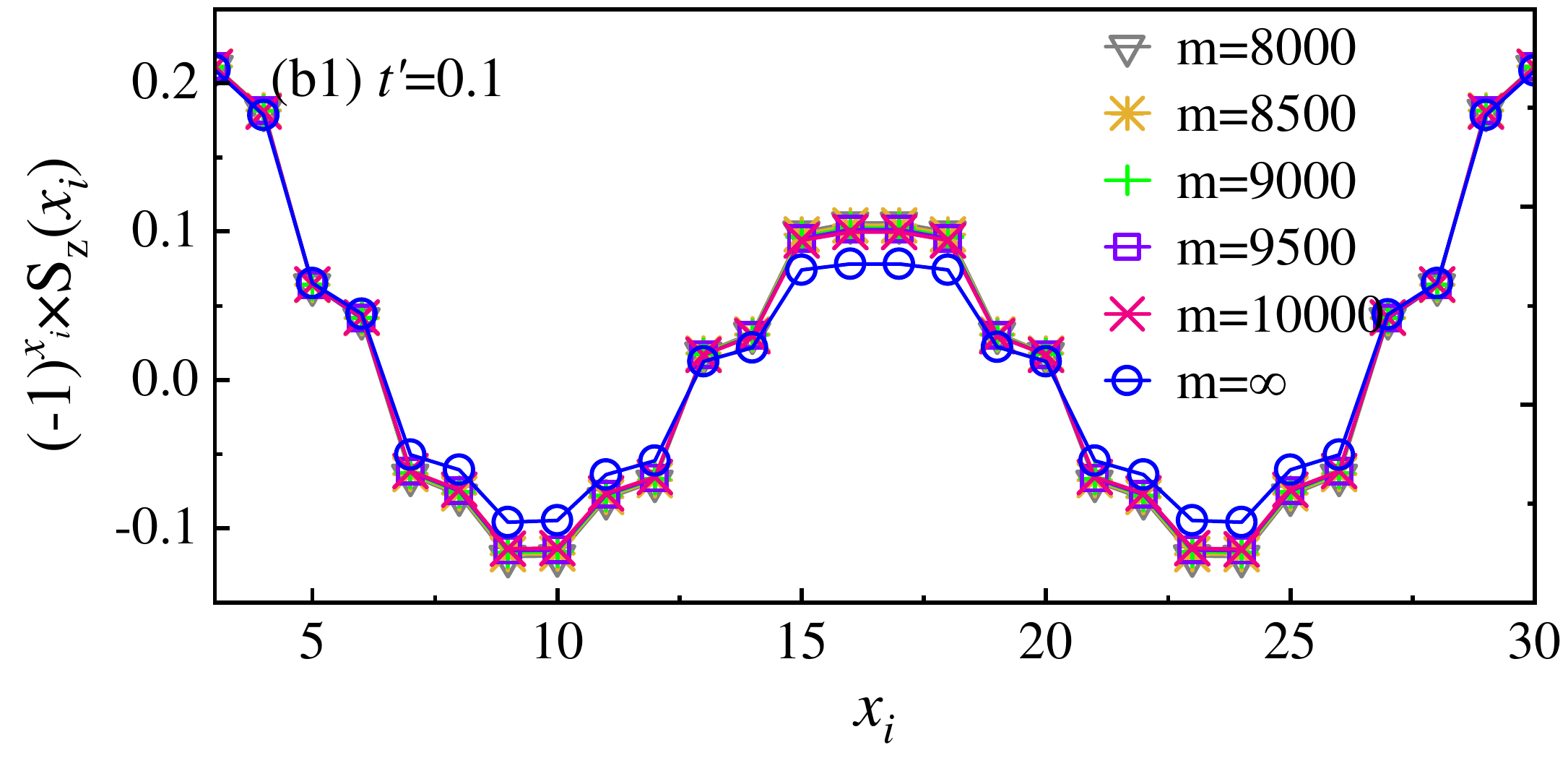}
   \includegraphics[width=0.45\textwidth]{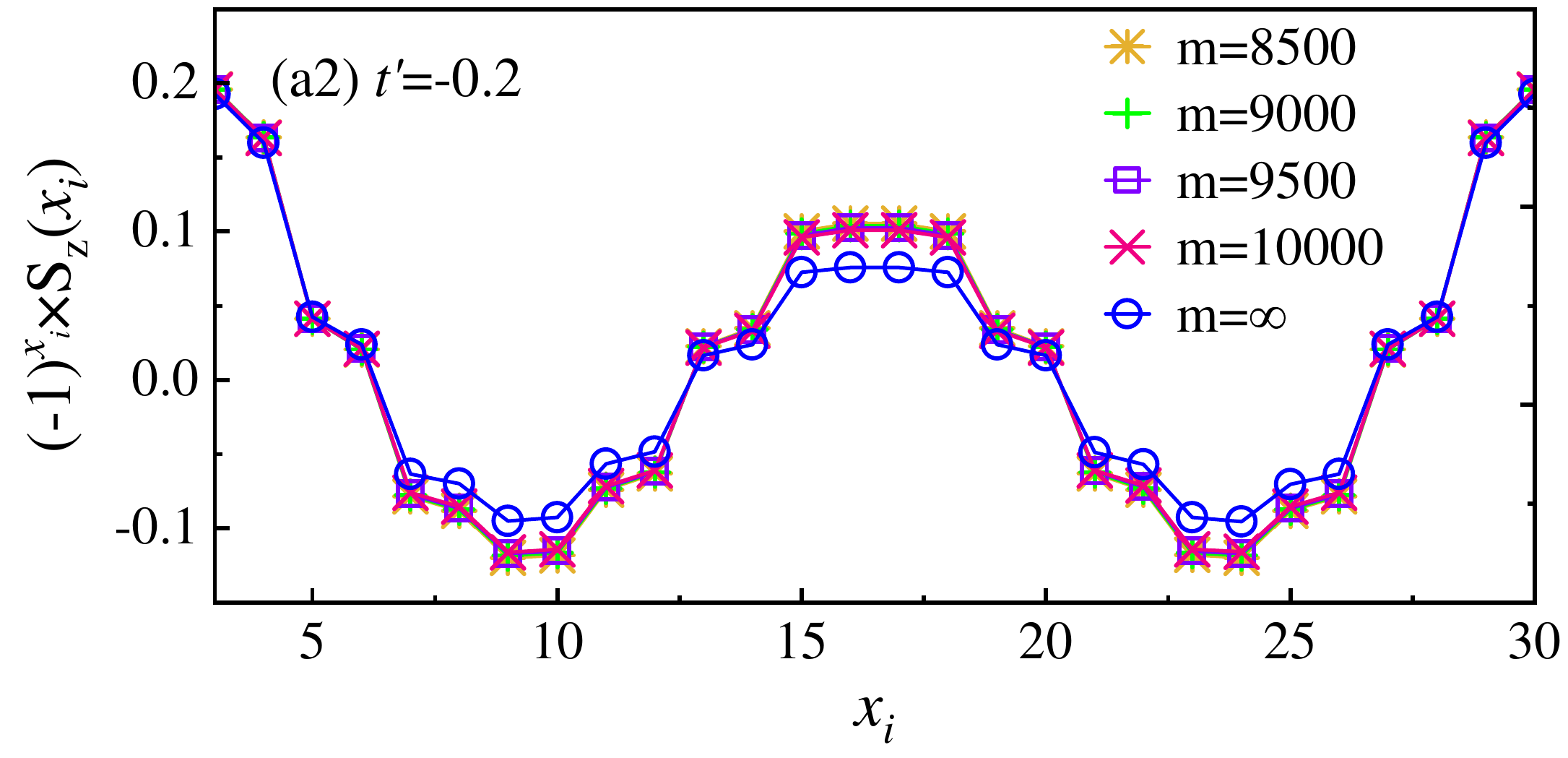}
   \includegraphics[width=0.45\textwidth]{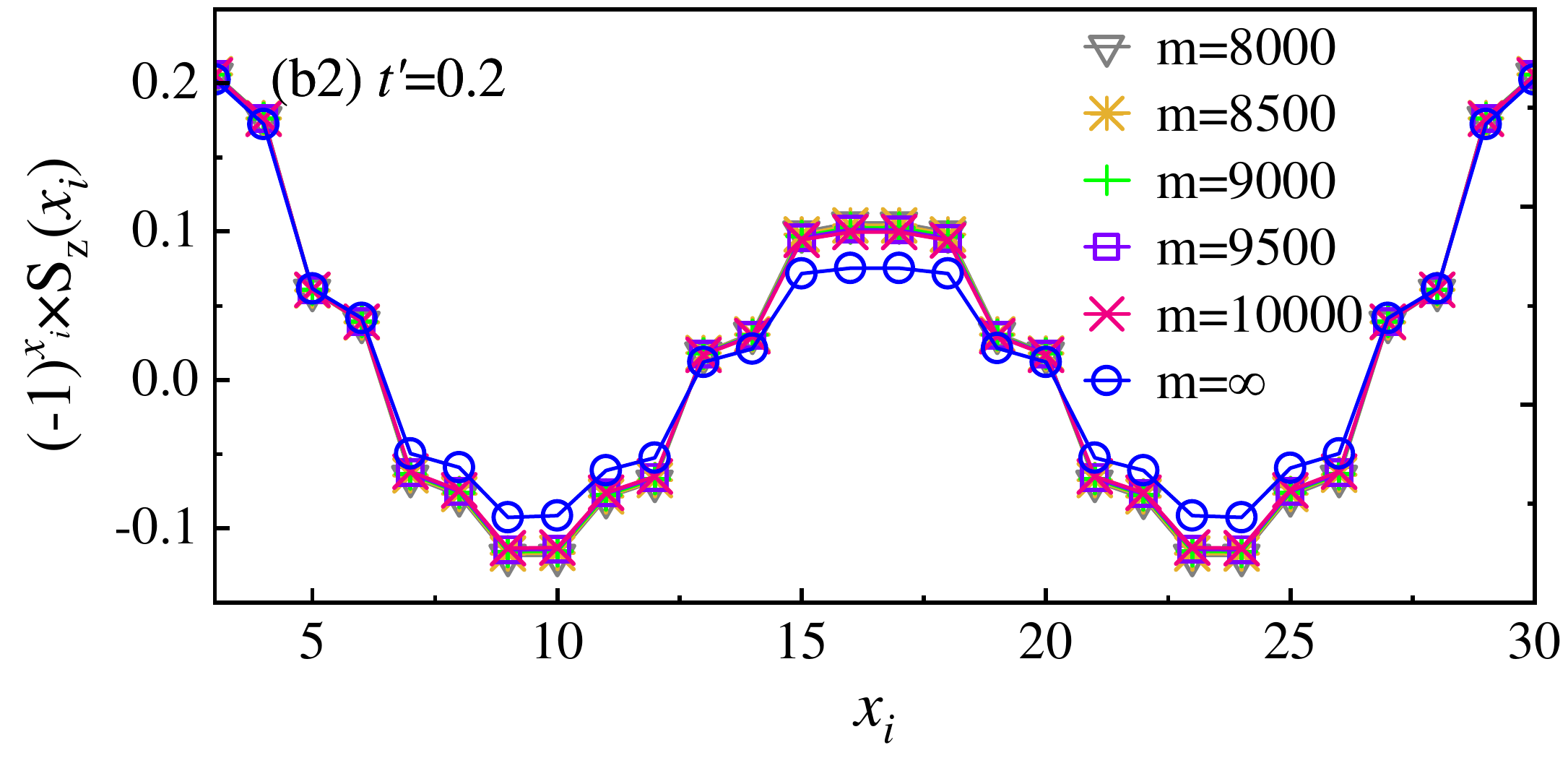}
   \includegraphics[width=0.45\textwidth]{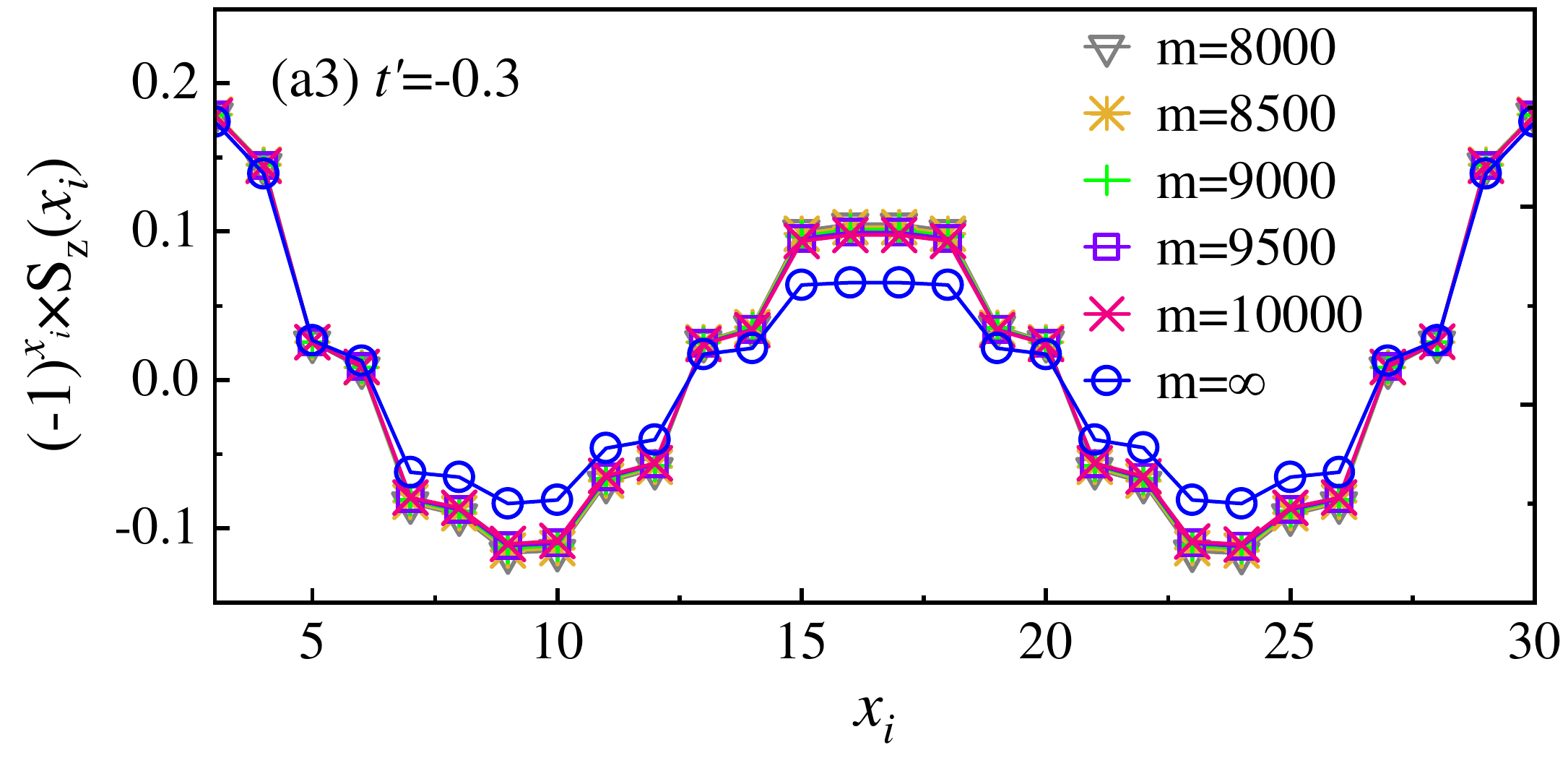}
   \includegraphics[width=0.45\textwidth]{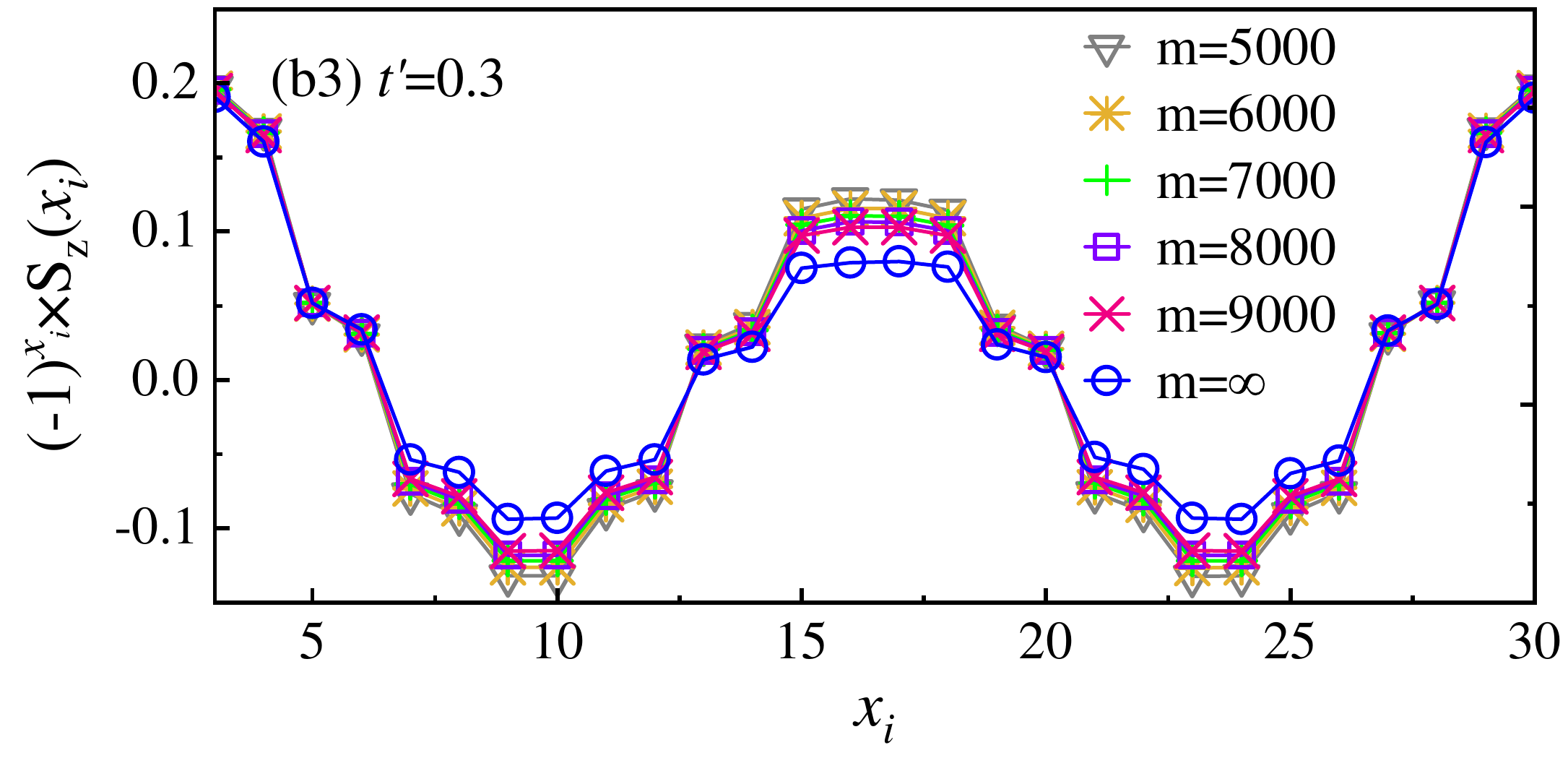}
   \includegraphics[width=0.45\textwidth]{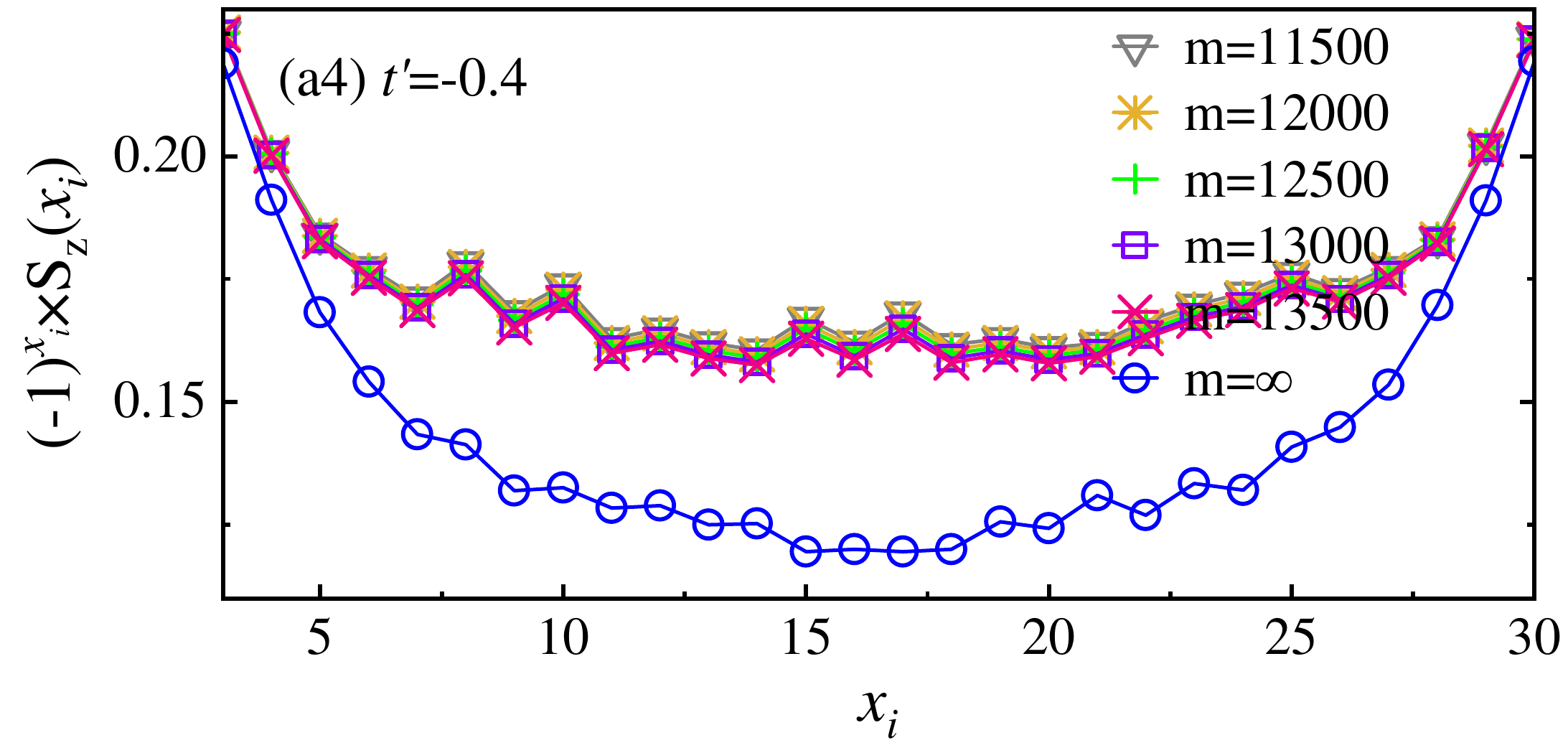}
   \includegraphics[width=0.45\textwidth]{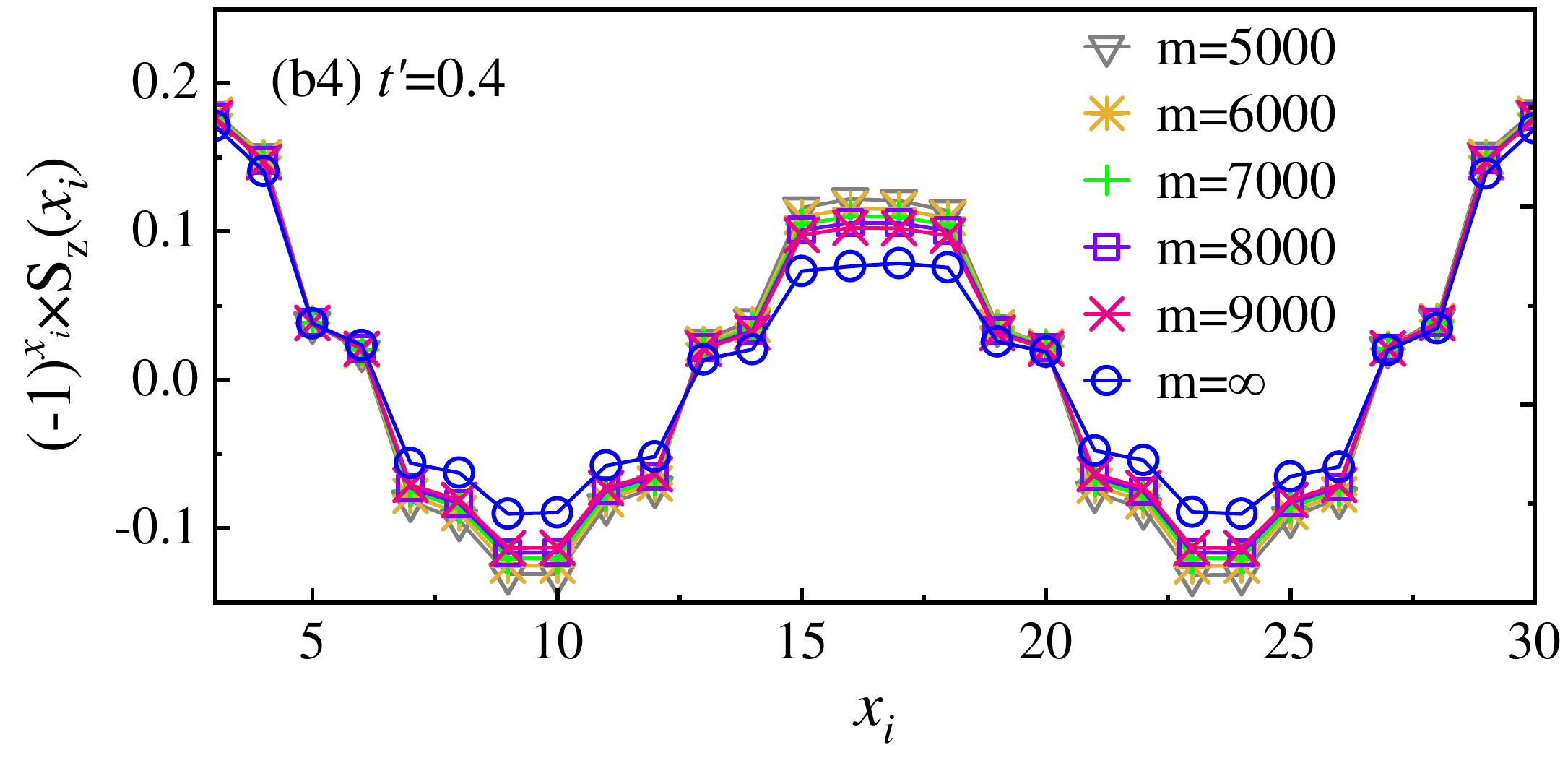}
   \includegraphics[width=0.45\textwidth]{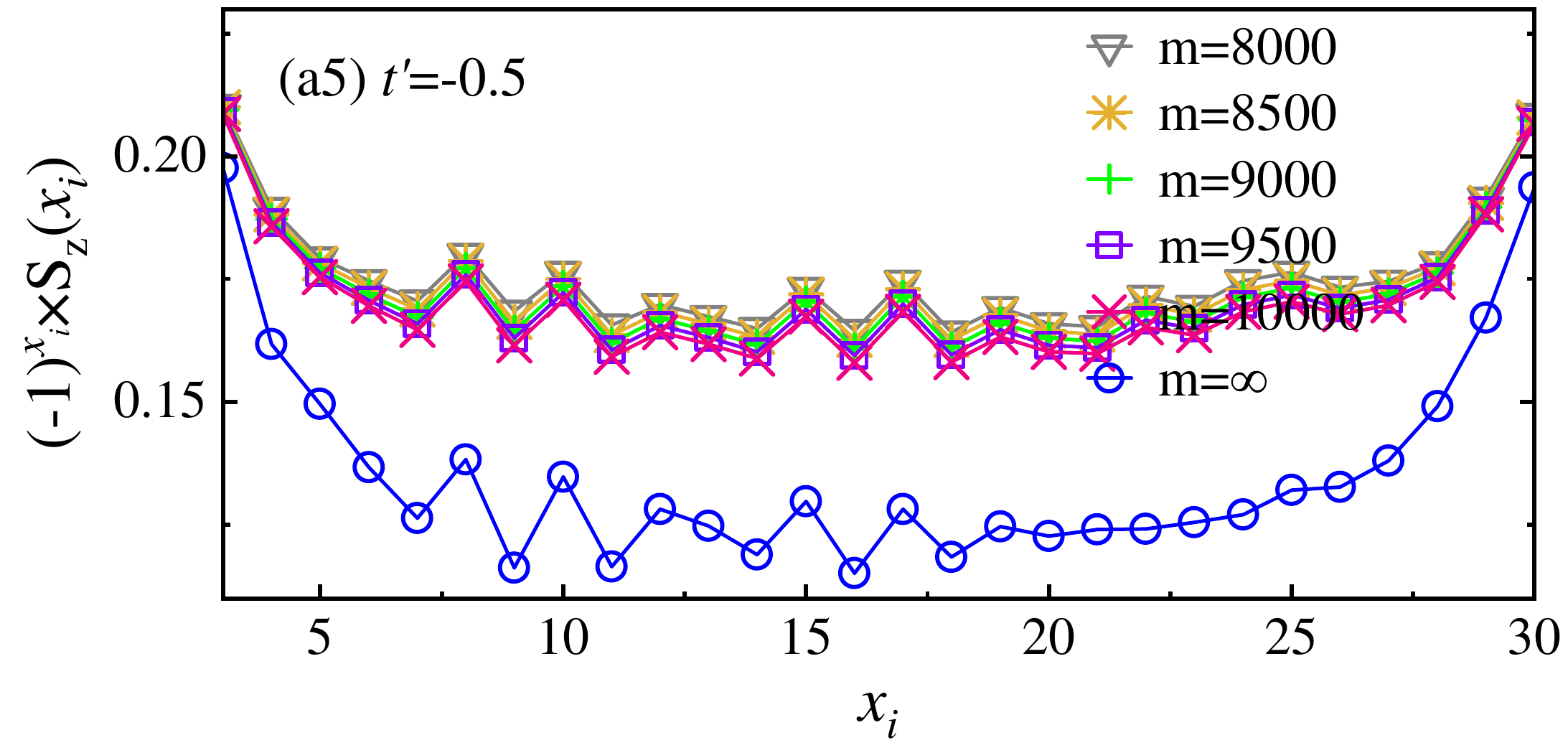} 
   \includegraphics[width=0.45\textwidth]{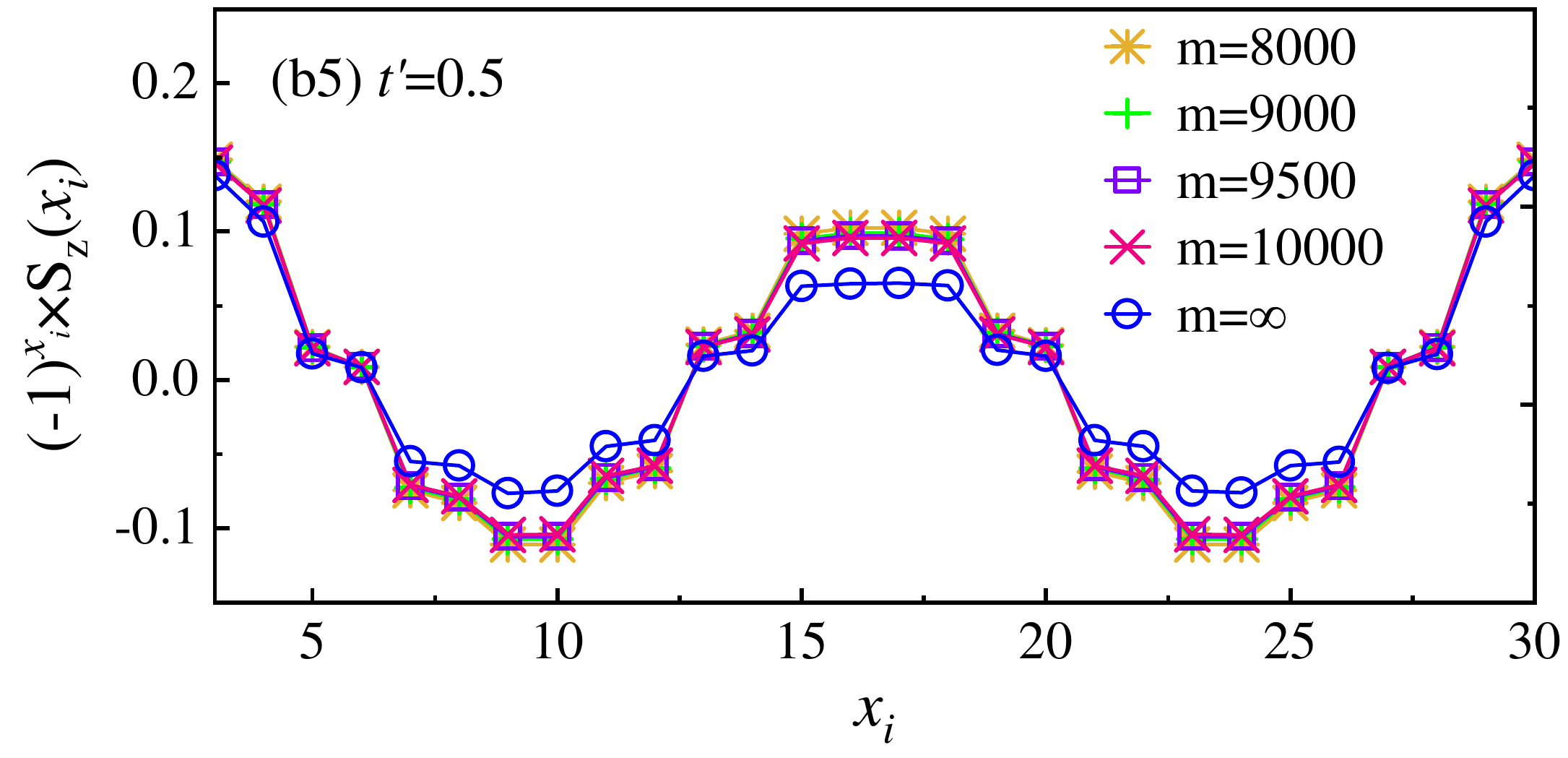}
   \caption{Staggered spin density for a varies of $t'$ values in the stripe and N$\acute e$el antiferromagnetic phases under anti-periodic boundary conditions. Both the finite bond dimension and the extrapolated results are shown.}
   \label{stagSz_APBC}
\end{figure*}

Figs.~\ref{stagSz_PBC} and ~\ref{stagSz_APBC} show the staggered spin density profiles for a varies of $t'$ values in the stripe and N$\acute e$el antiferromagnetic phases under periodic and anti-periodic boundary conditions, respectively. Similar as the local charge density, the results are well converged when $t'$ is away from $-0.4 < t'_c < -0.3$. Near the phase transition point, extrapolations are not well-behaved.

\subsection{C. Extrapolation of charge and staggered spin density with truncation errors in DMRG}

\begin{figure*}[b]
   \includegraphics[width=0.45\textwidth]{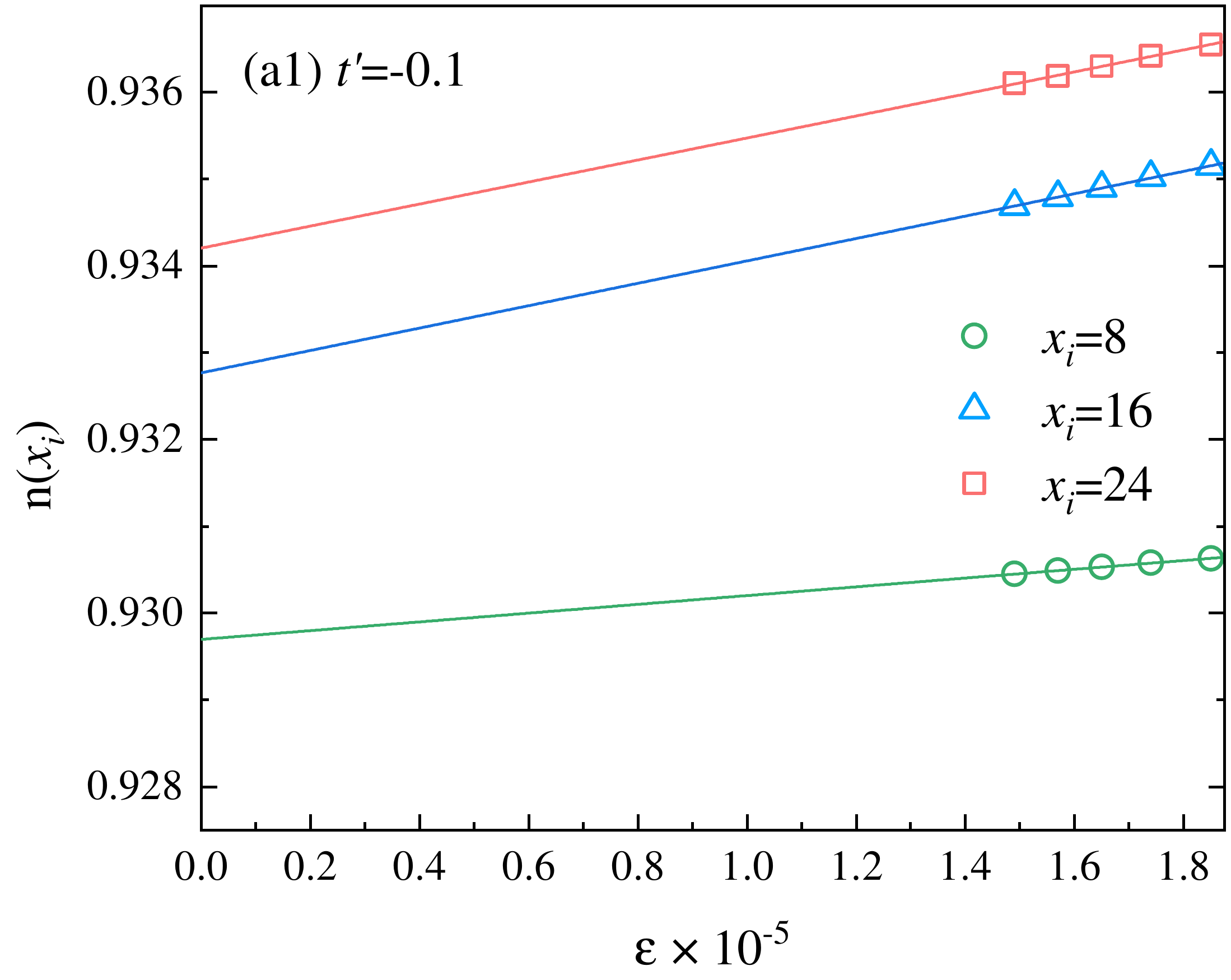}
   \includegraphics[width=0.45\textwidth]{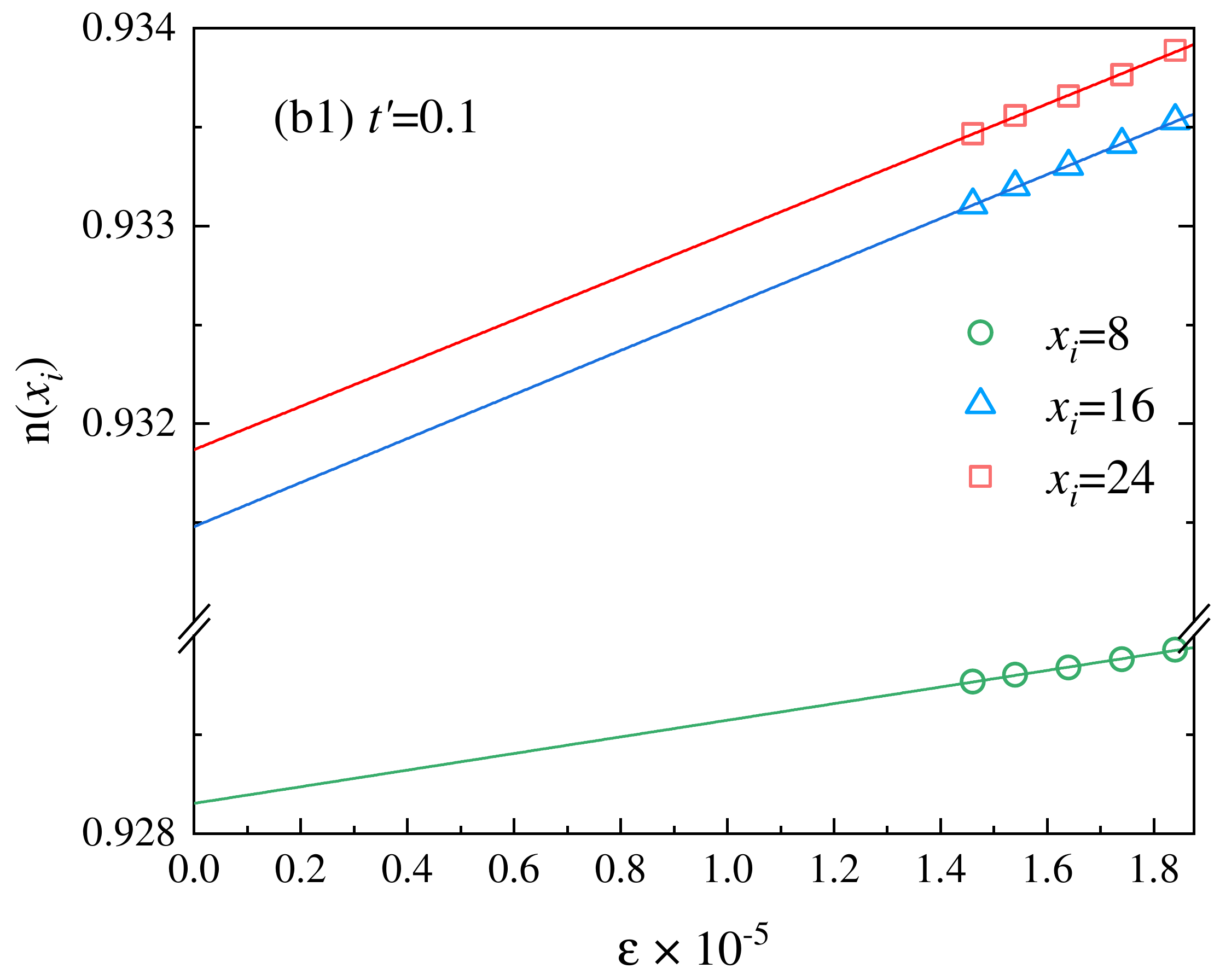}
   \includegraphics[width=0.45\textwidth]{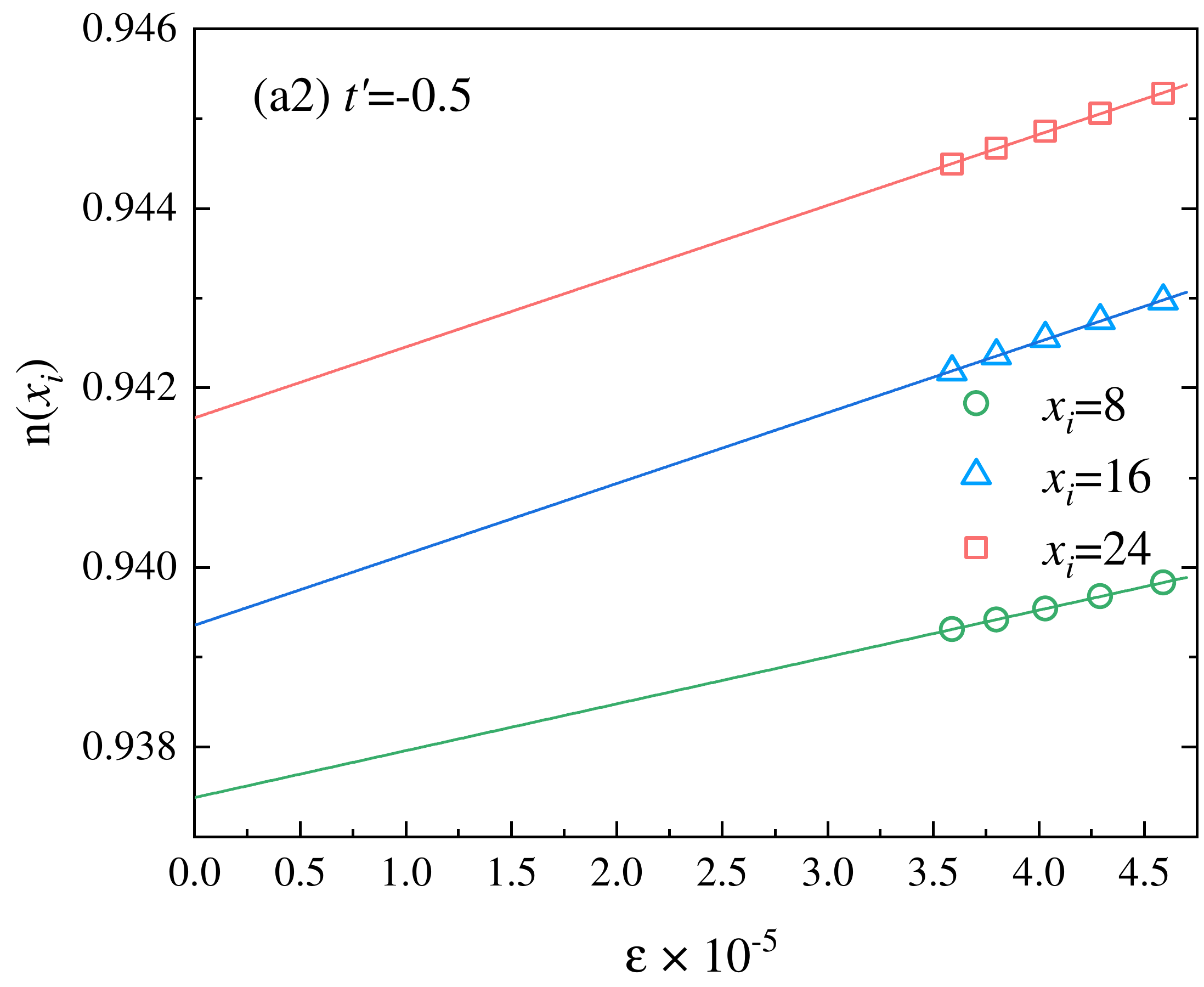}
   \includegraphics[width=0.45\textwidth]{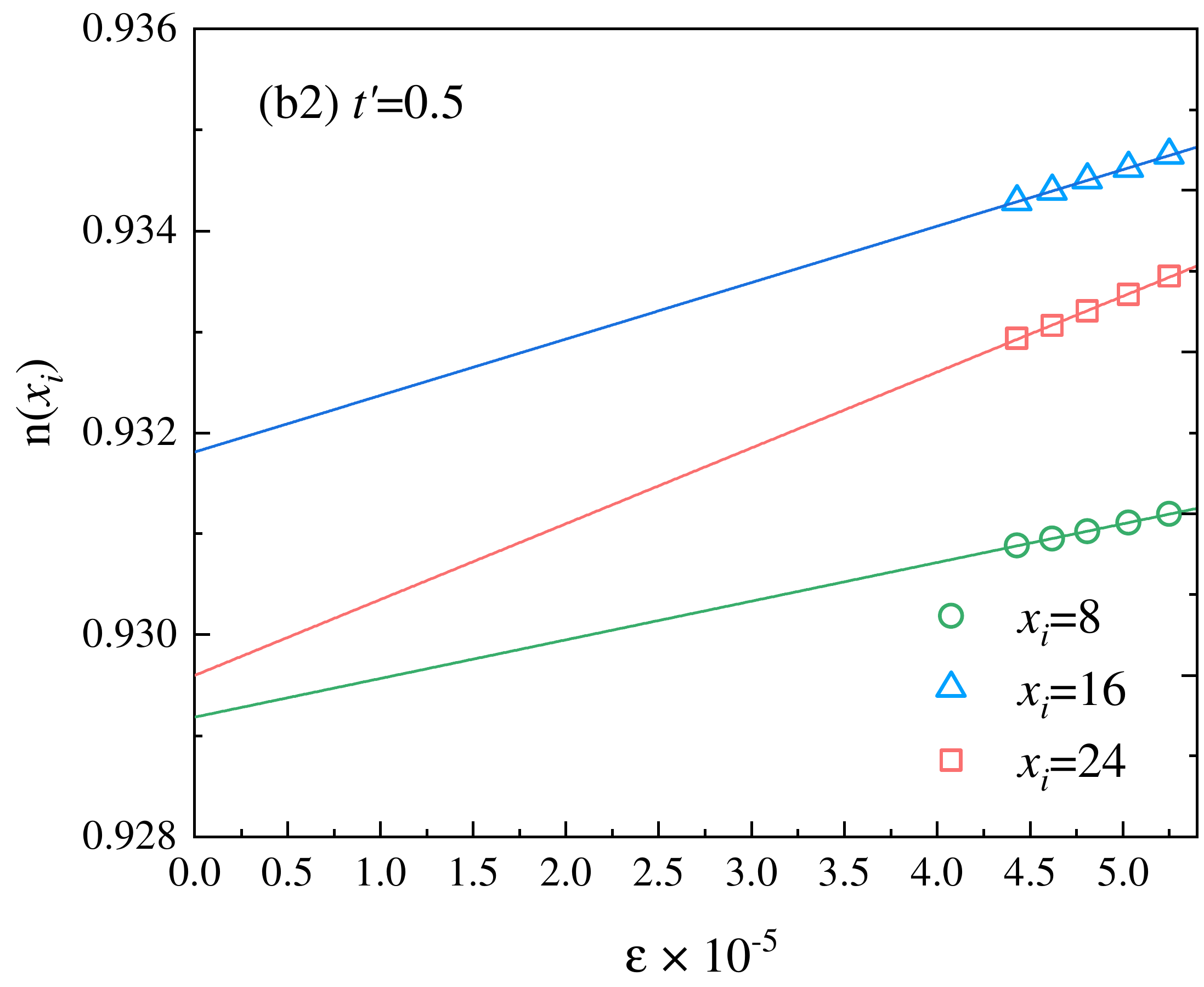}
   \caption{Extrapolation of charge density as a function of the truncated error for different $t'$ under periodic boundary conditions. The local charge density is fitted linearly with truncation error $\epsilon$.}
   \label{ElecPBCfit}
\end{figure*}

\begin{figure*}[b]
   \includegraphics[width=0.45\textwidth]{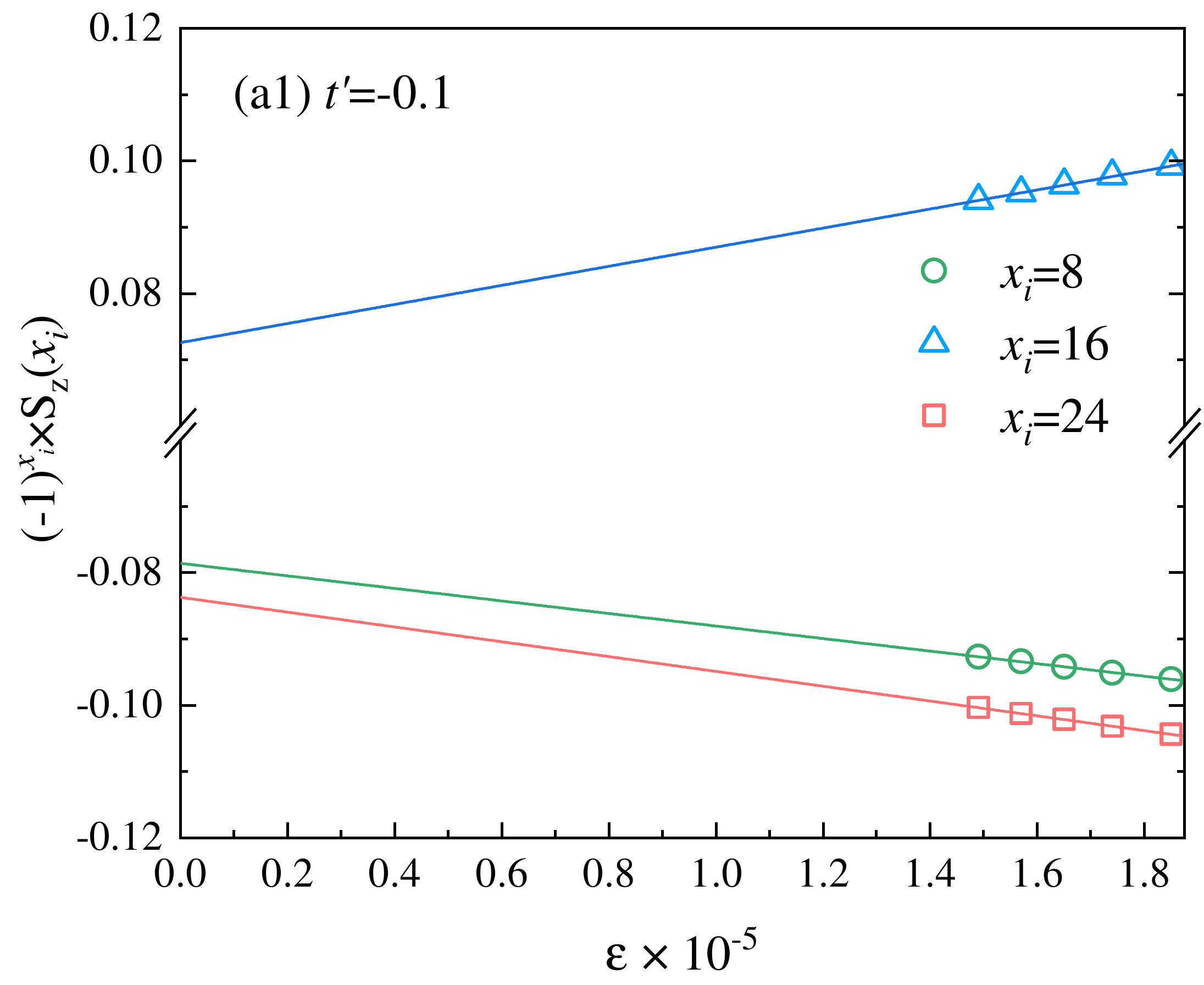}
   \includegraphics[width=0.45\textwidth]{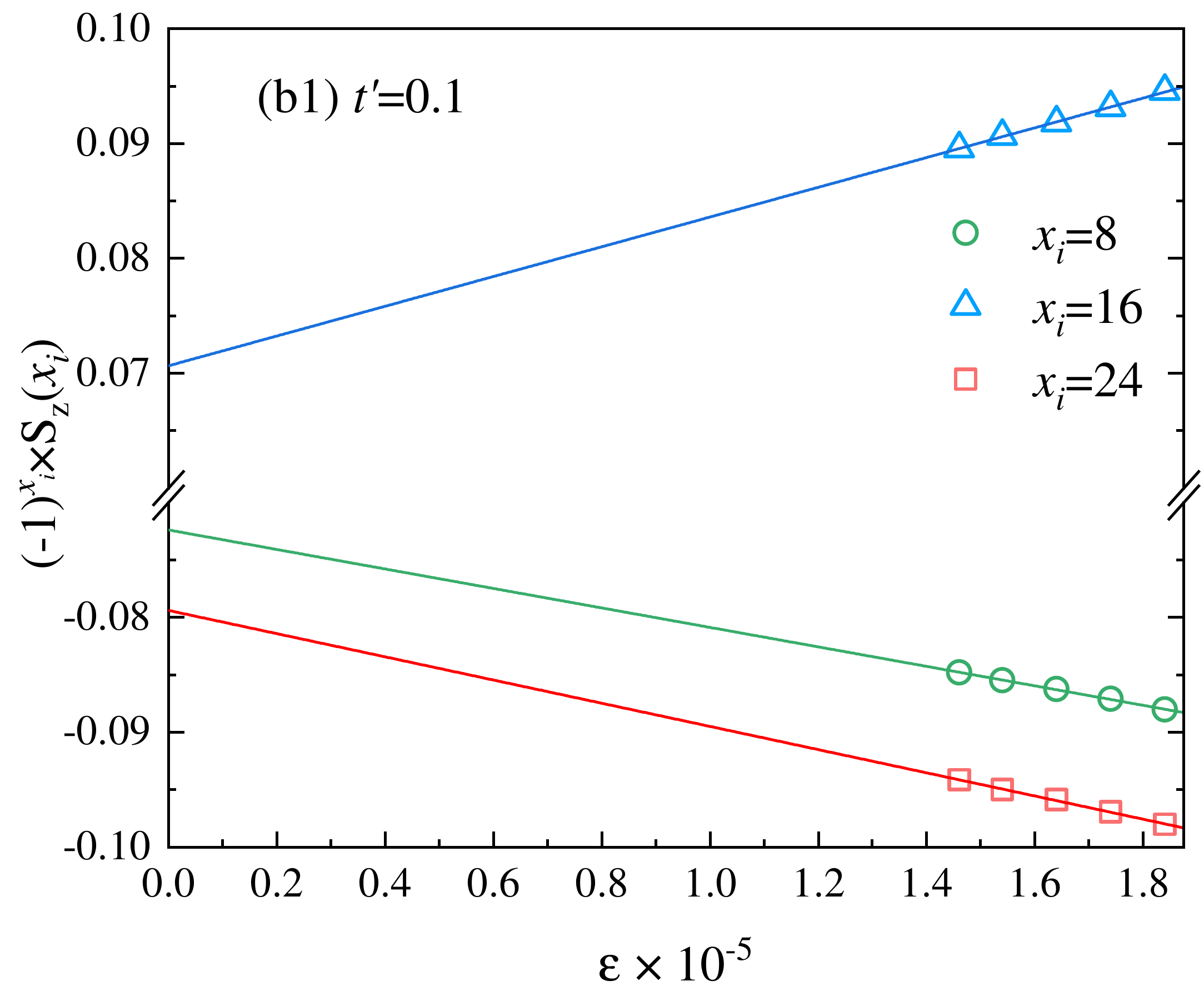}
   \includegraphics[width=0.45\textwidth]{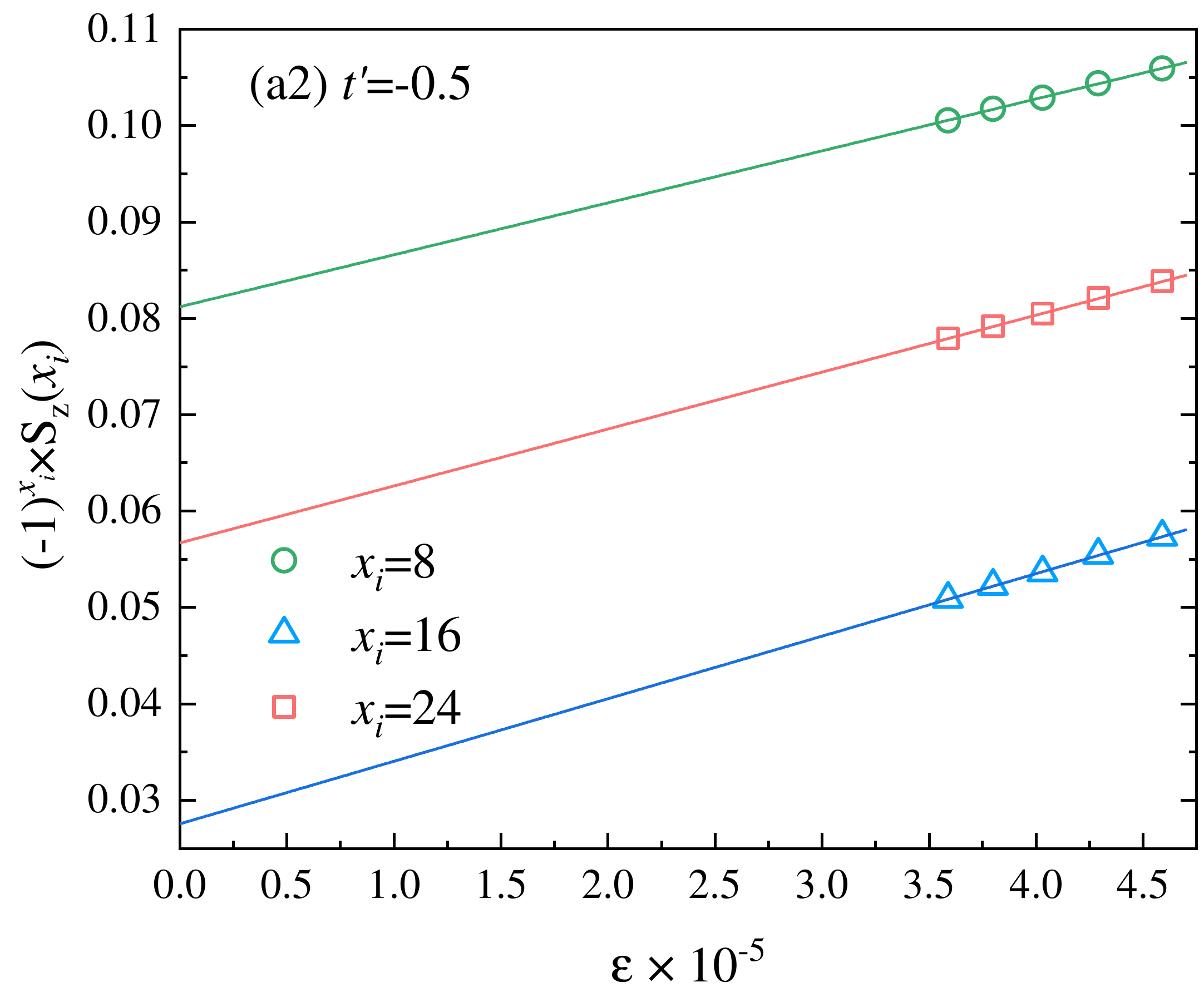}
   \includegraphics[width=0.45\textwidth]{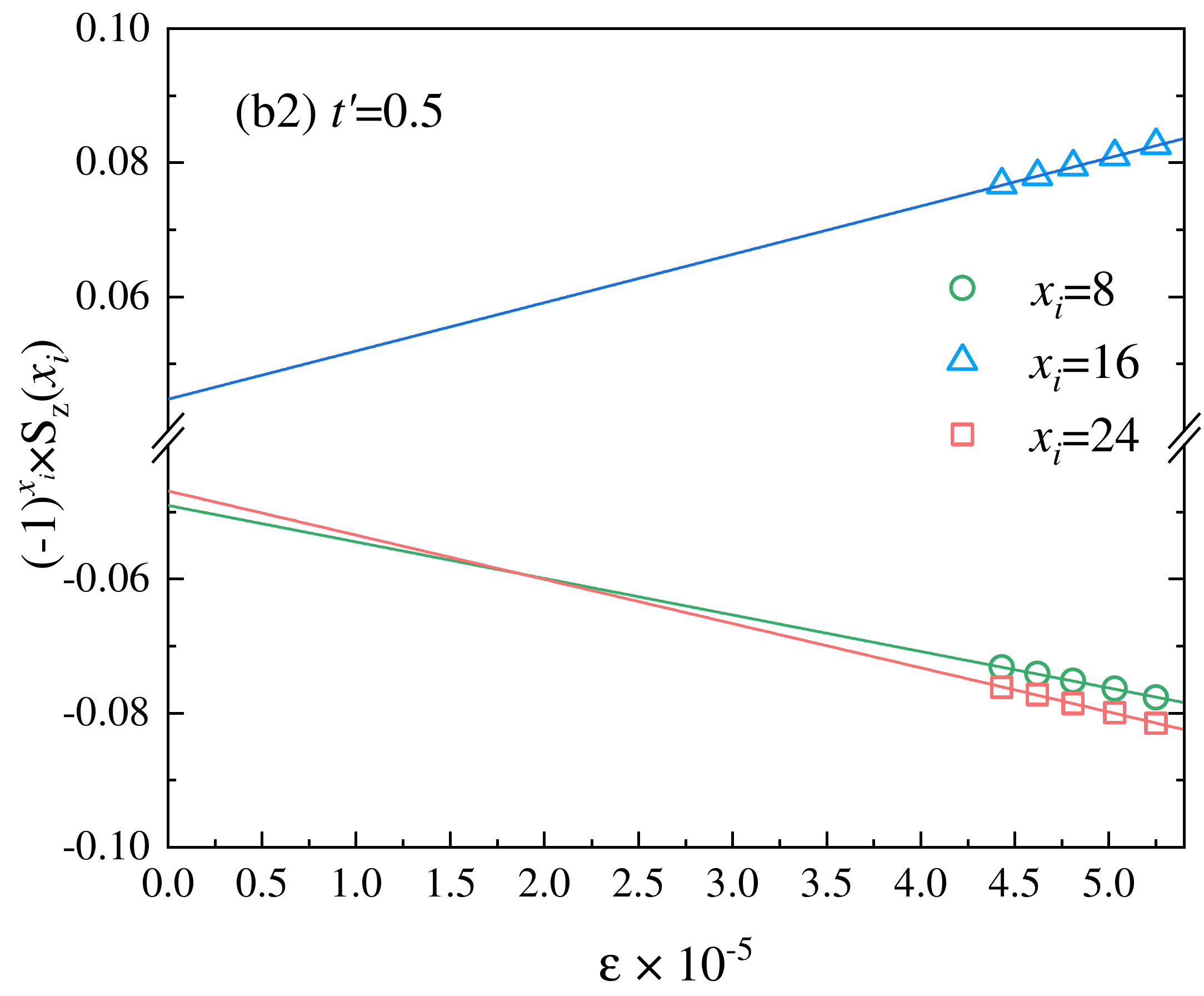}
   \caption{Extrapolation of staggered spin density as a function of the truncated error for different $t'$ under periodic boundary conditions. The local staggered spin density is fitted linearly with truncation error $\epsilon$.}
   \label{stagSzPBCfit}
\end{figure*}

\begin{figure*}[b]
   \includegraphics[width=0.45\textwidth]{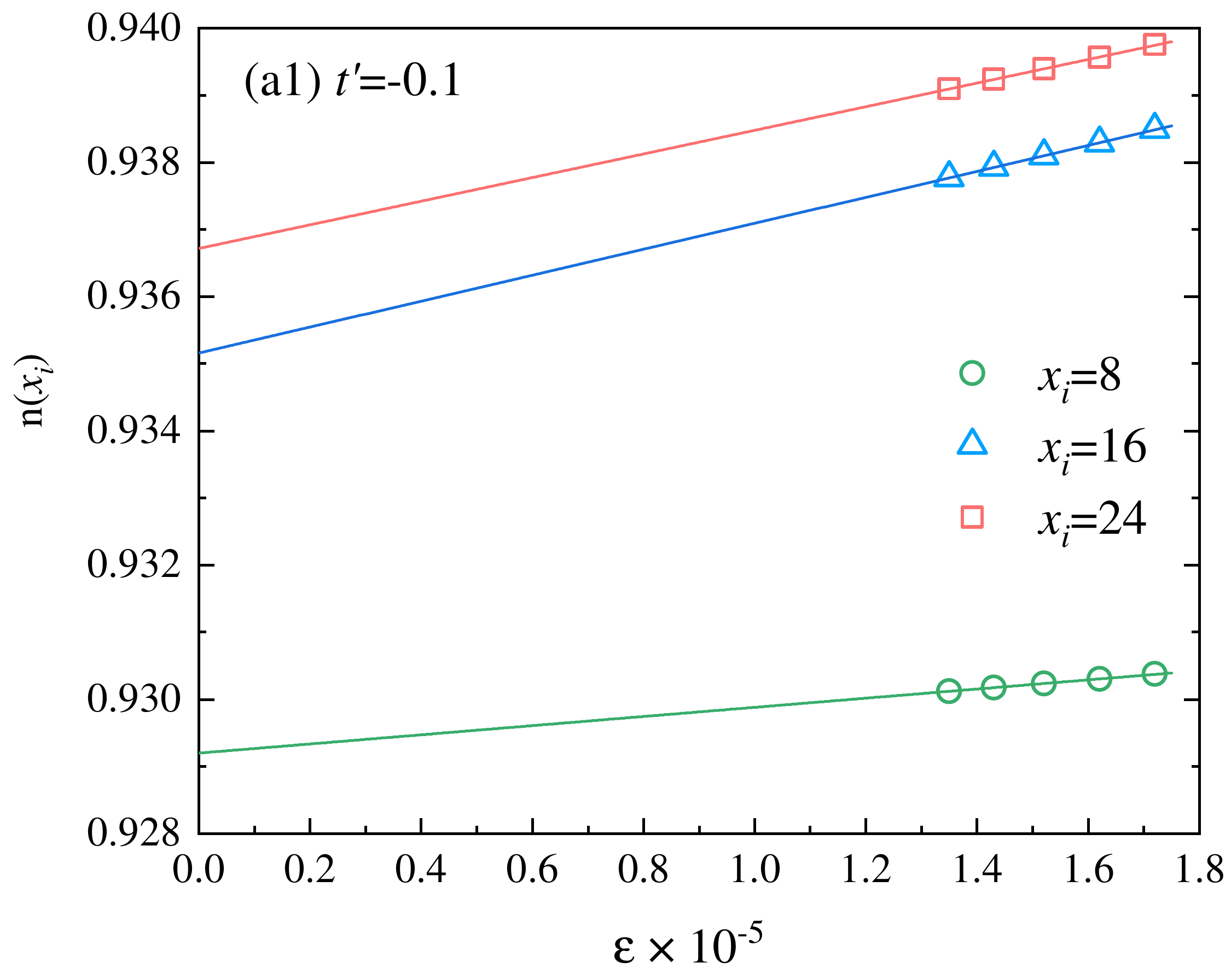}
   \includegraphics[width=0.45\textwidth]{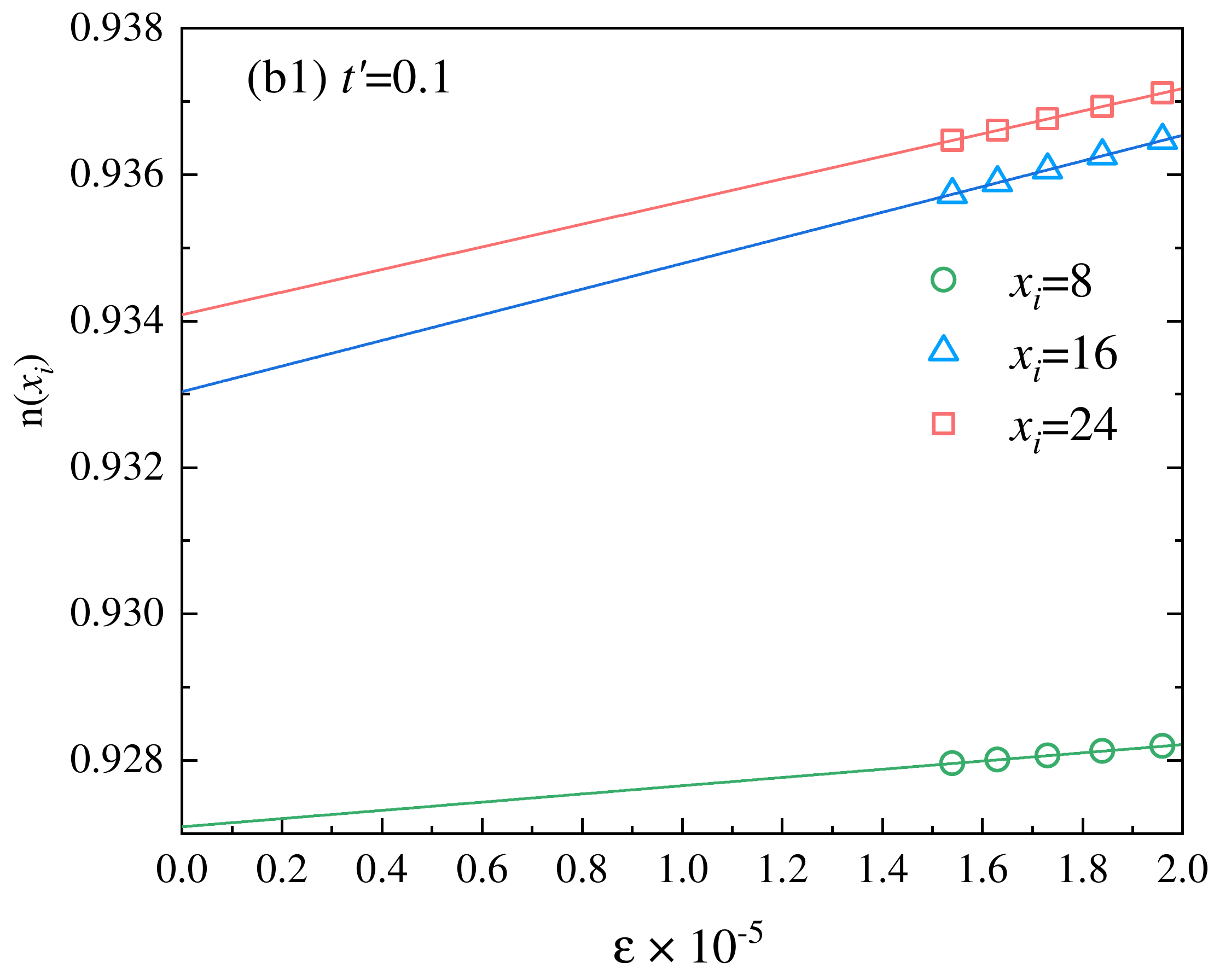}
   \includegraphics[width=0.45\textwidth]{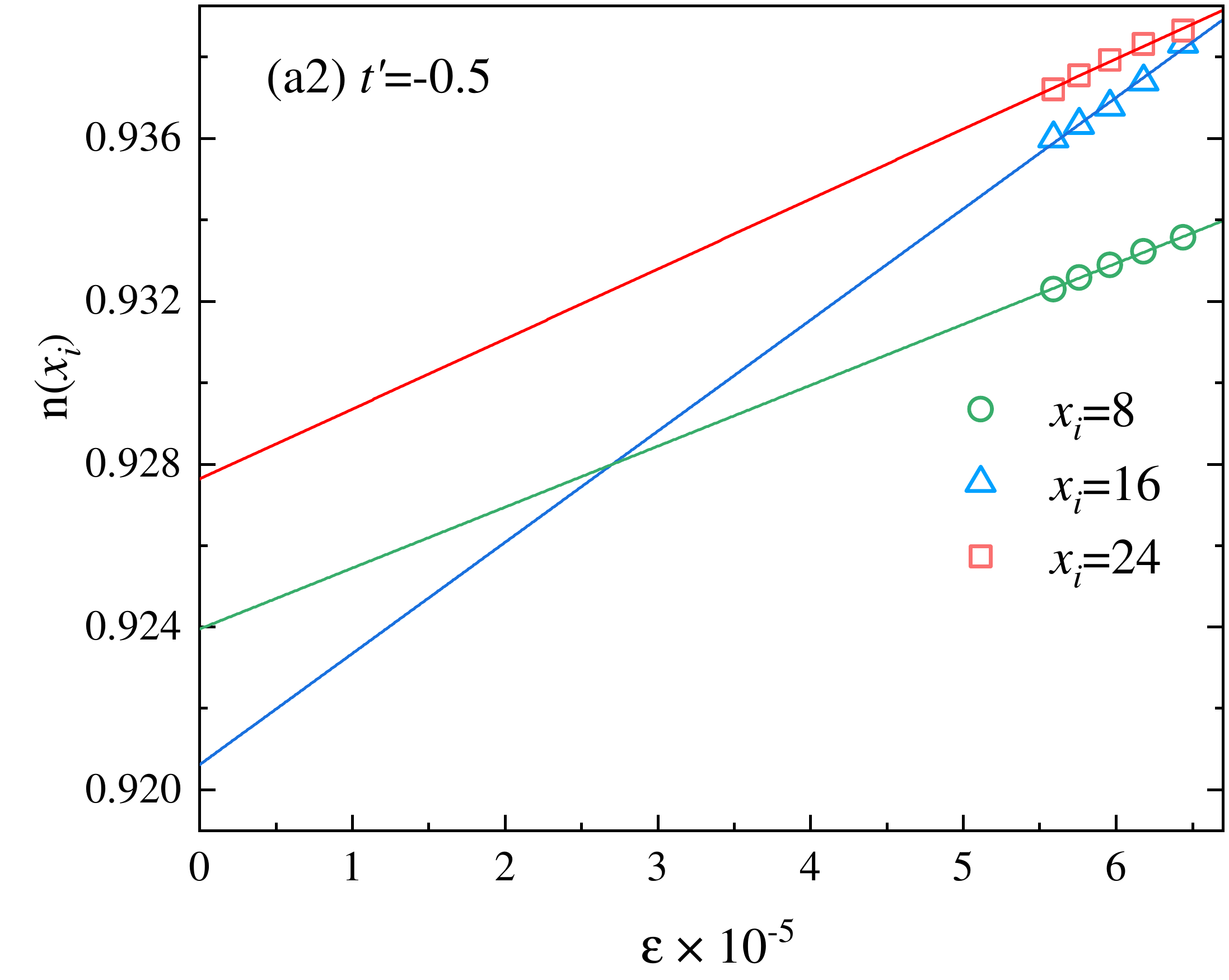}
   \includegraphics[width=0.45\textwidth]{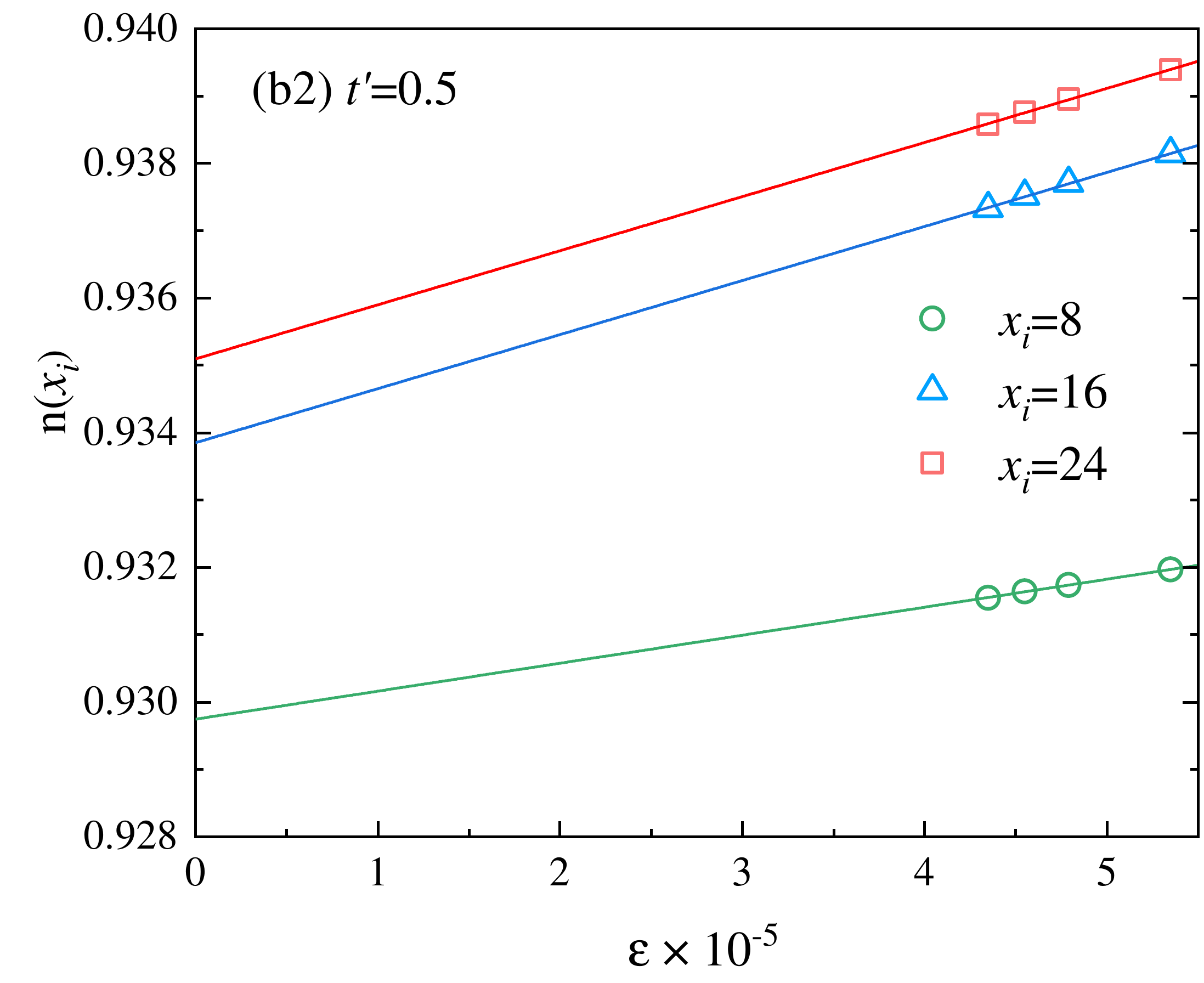}
   \caption{Extrapolation of charge density as a function of the truncated error for different $t'$ under anti-periodic boundary conditions. The local charge density is fitted linearly with truncation error $\epsilon$.}
   \label{ElecAPBCfit}
\end{figure*}

\begin{figure*}[b]
   \includegraphics[width=0.45\textwidth]{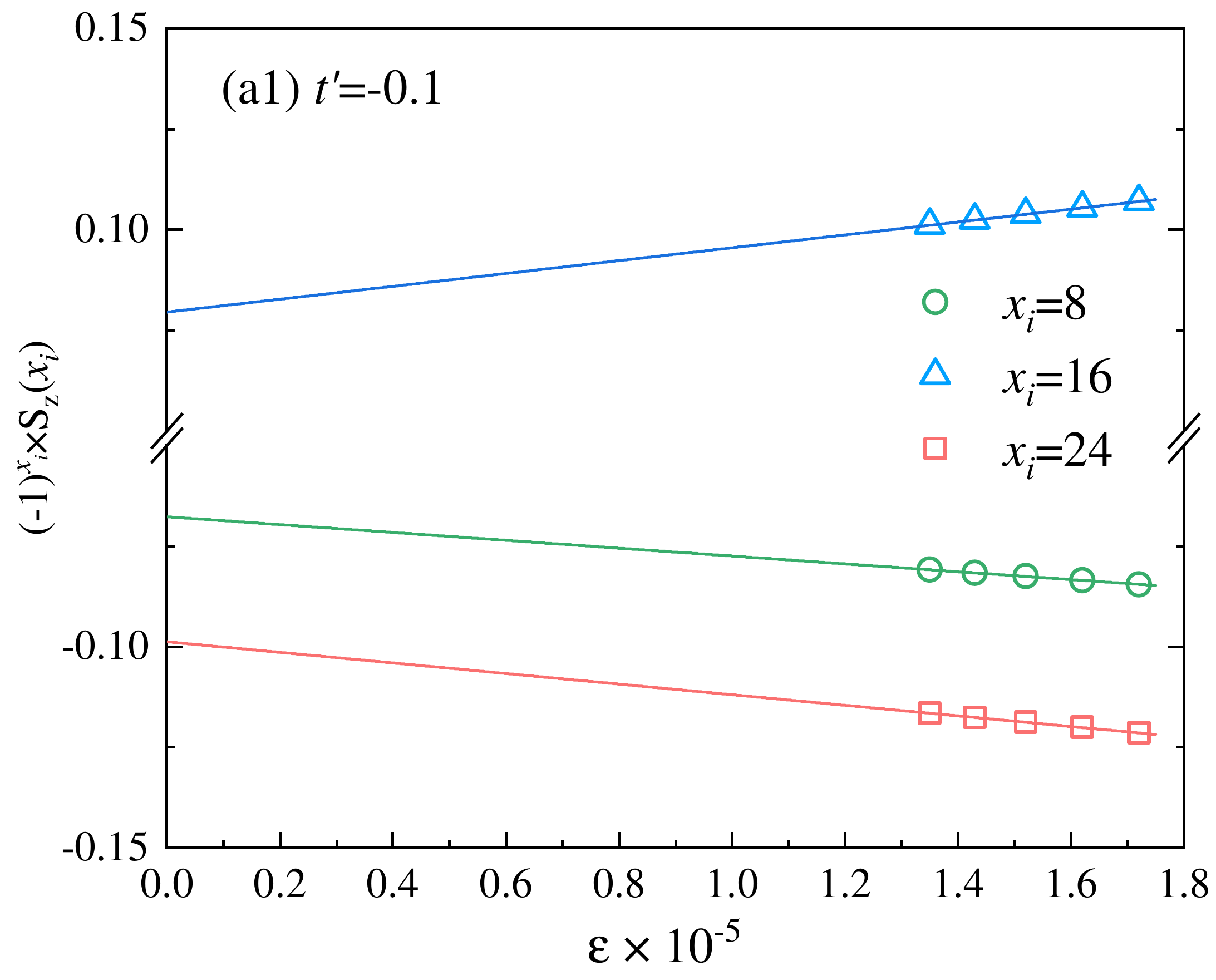}
   \includegraphics[width=0.45\textwidth]{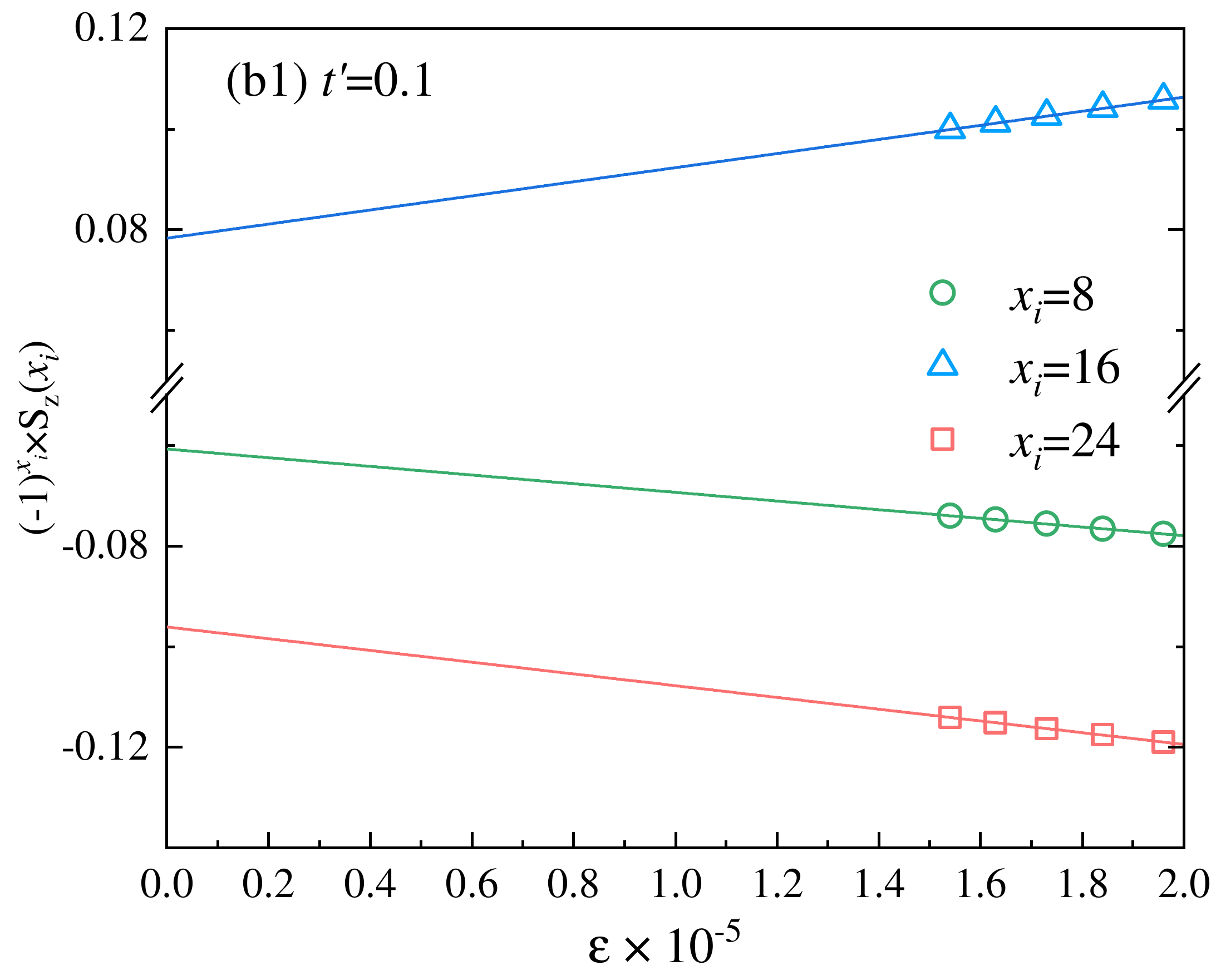}
   \includegraphics[width=0.45\textwidth]{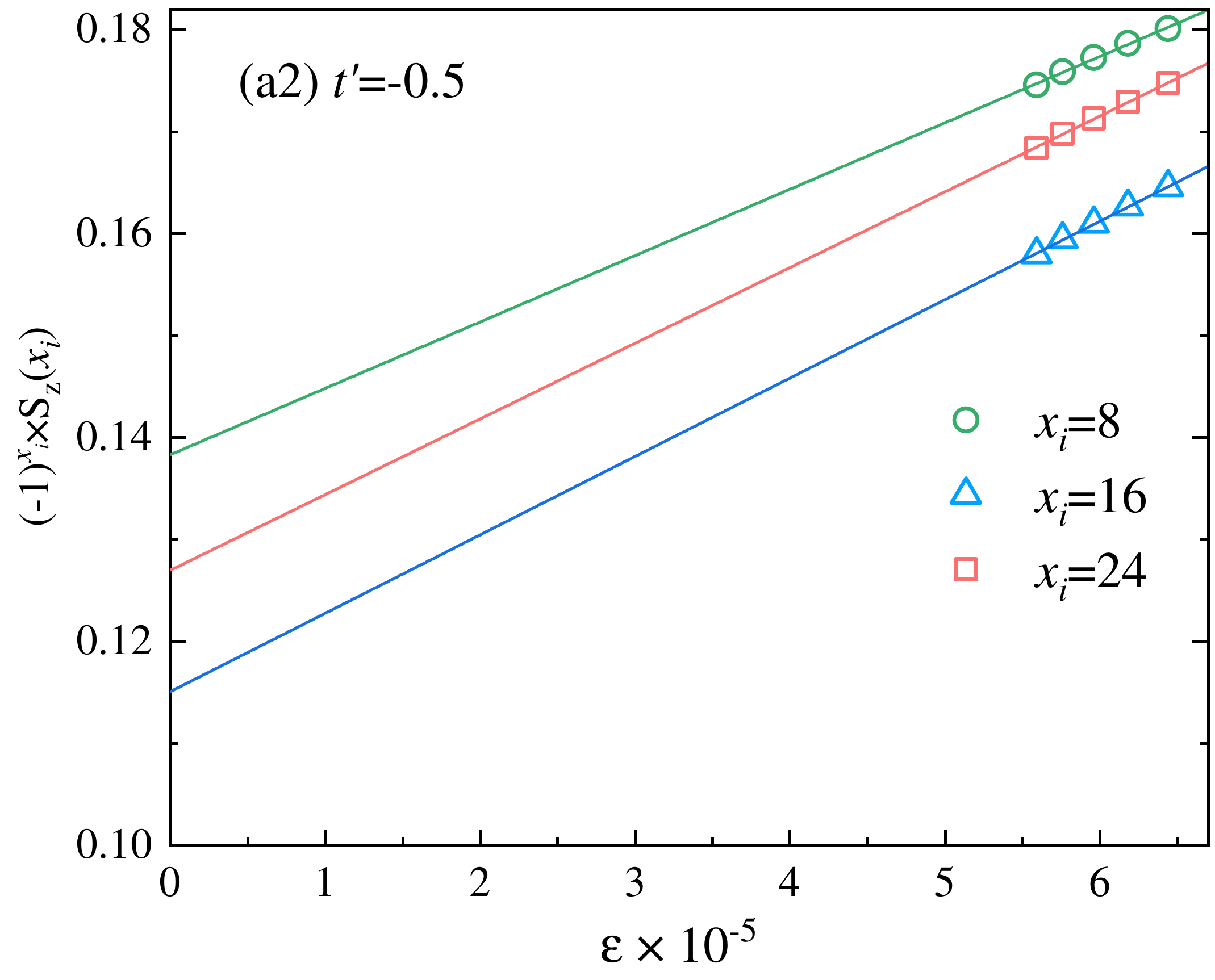}
   \includegraphics[width=0.45\textwidth]{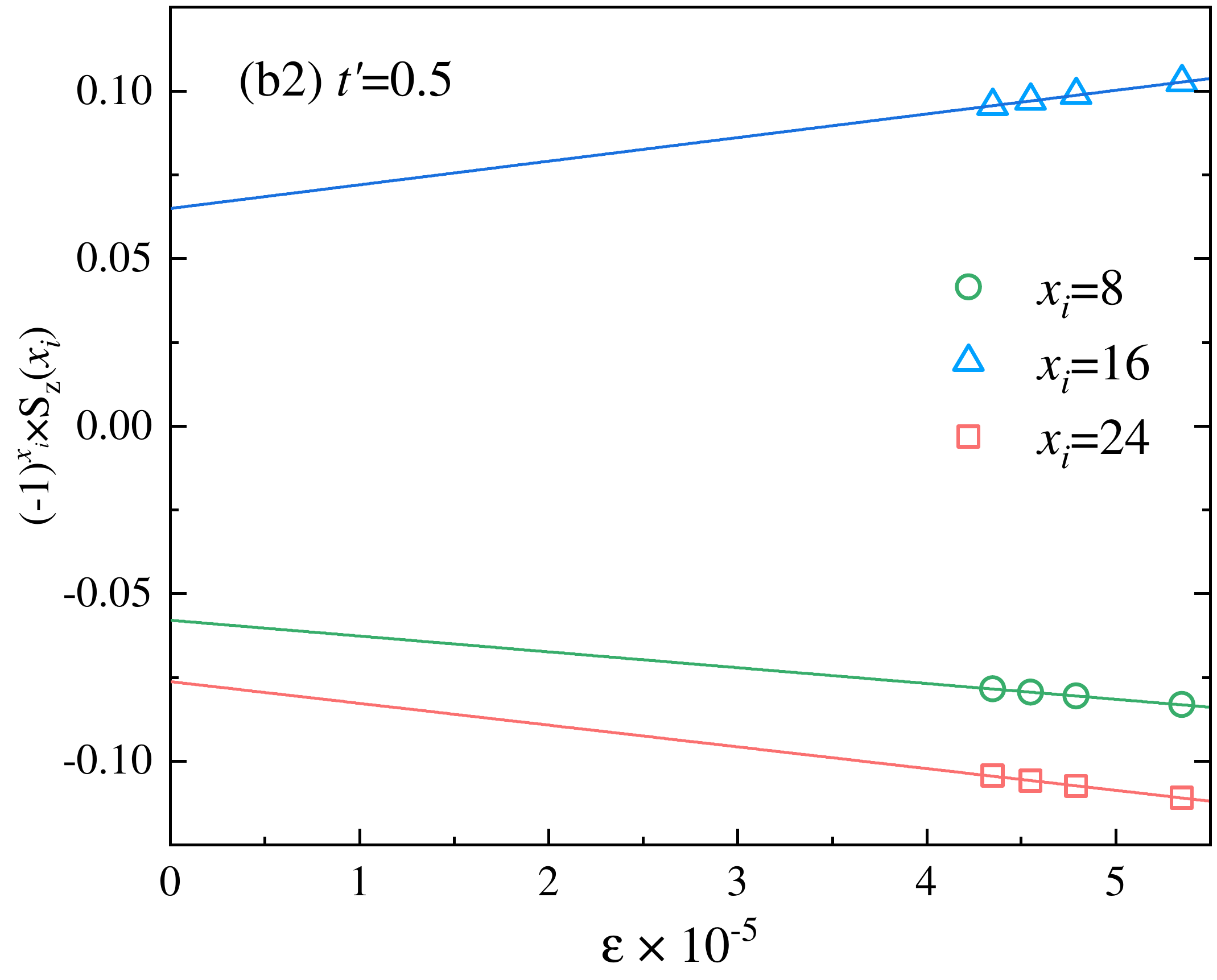}
   \caption{Extrapolation of staggered spin density as a function of the truncated error for different $t'$ under anti-periodic boundary conditions. The local staggered spin density is fitted linearly with truncation error $\epsilon$.}
   \label{stagSzAPBCfit}
\end{figure*}

In Figs.~\ref{ElecPBCfit} and \ref{ElecAPBCfit}, we show the extrapolation of DMRG results for local
charge density with truncation errors under periodic and anti-periodic boundary conditions, respectively. We choose three typical sites to show the extrapolation process. Similar procedures are performed for the staggered spin density in Figs.~\ref{stagSzPBCfit} and \ref{stagSzAPBCfit}. These local quantities are extrapolated linearly with truncation error $\epsilon$. We can see both spin and charge order reach the linear scaling region in our calculation, which ensures the reliability of the extrapolated results.

\subsection{D. Singlet pair-pair correlation}

\begin{figure*}[b]
   \includegraphics[width=0.3\textwidth]{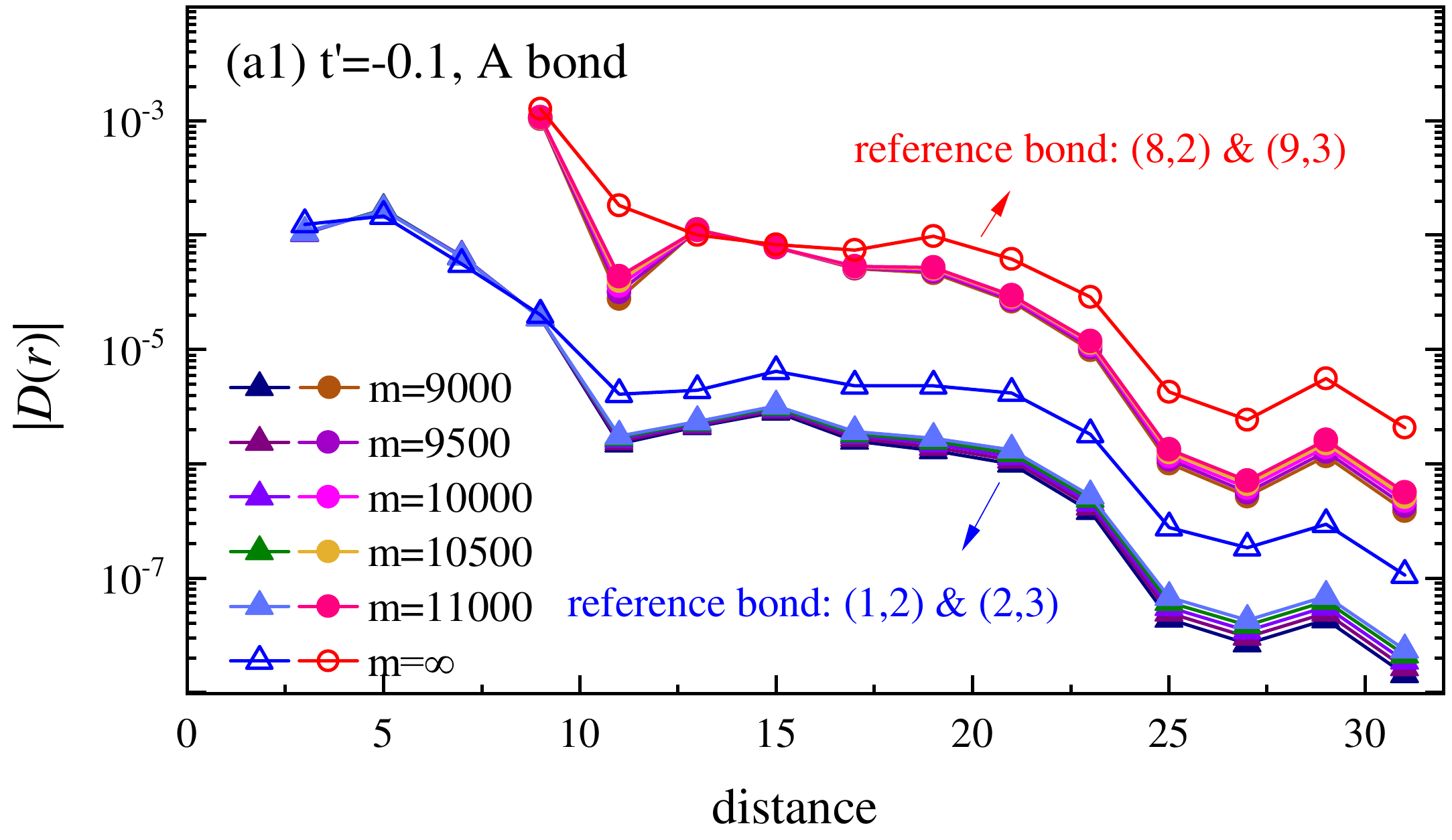}
   \includegraphics[width=0.3\textwidth]{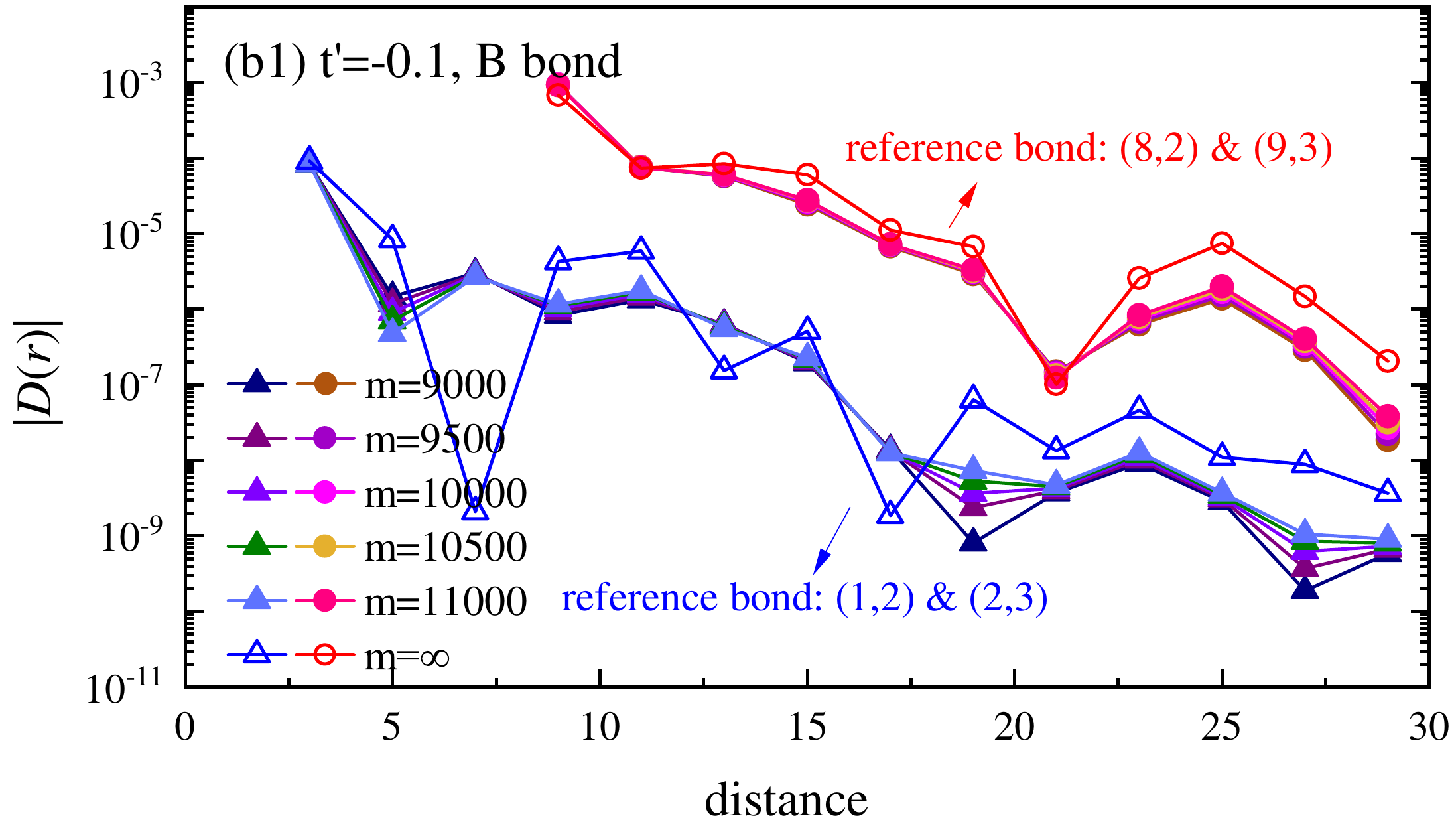}
   \includegraphics[width=0.3\textwidth]{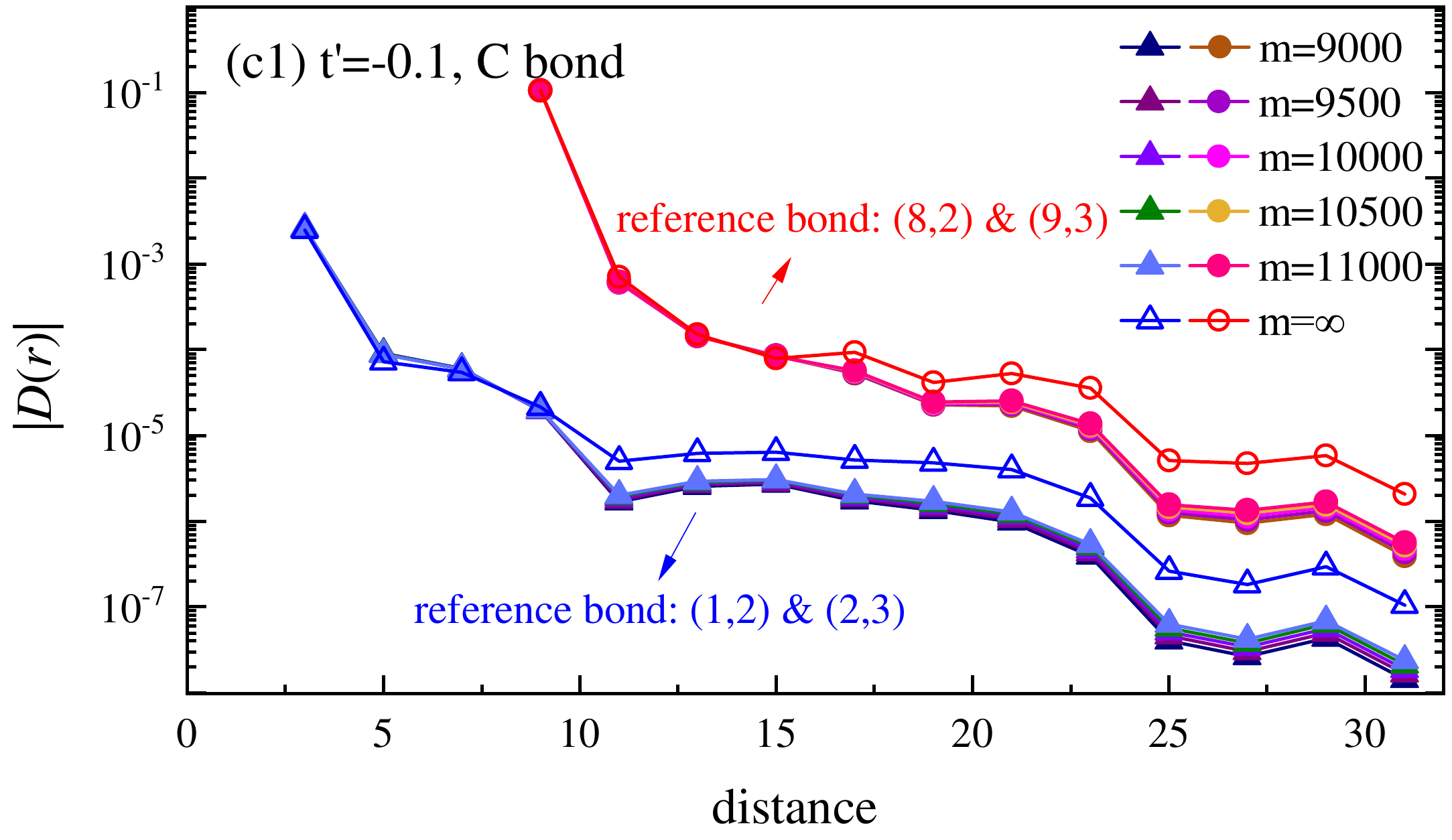}
   \includegraphics[width=0.3\textwidth]{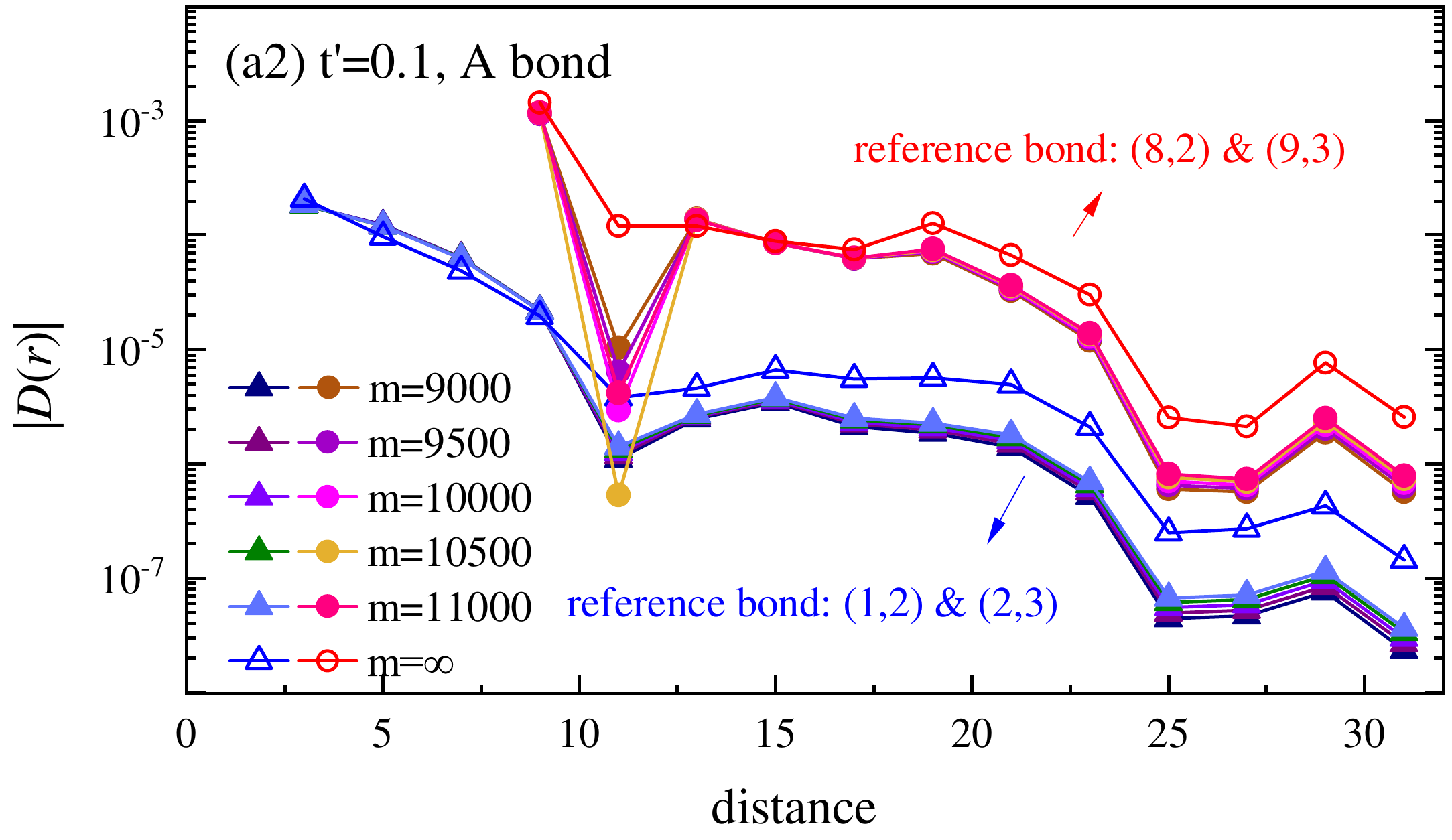}
   \includegraphics[width=0.3\textwidth]{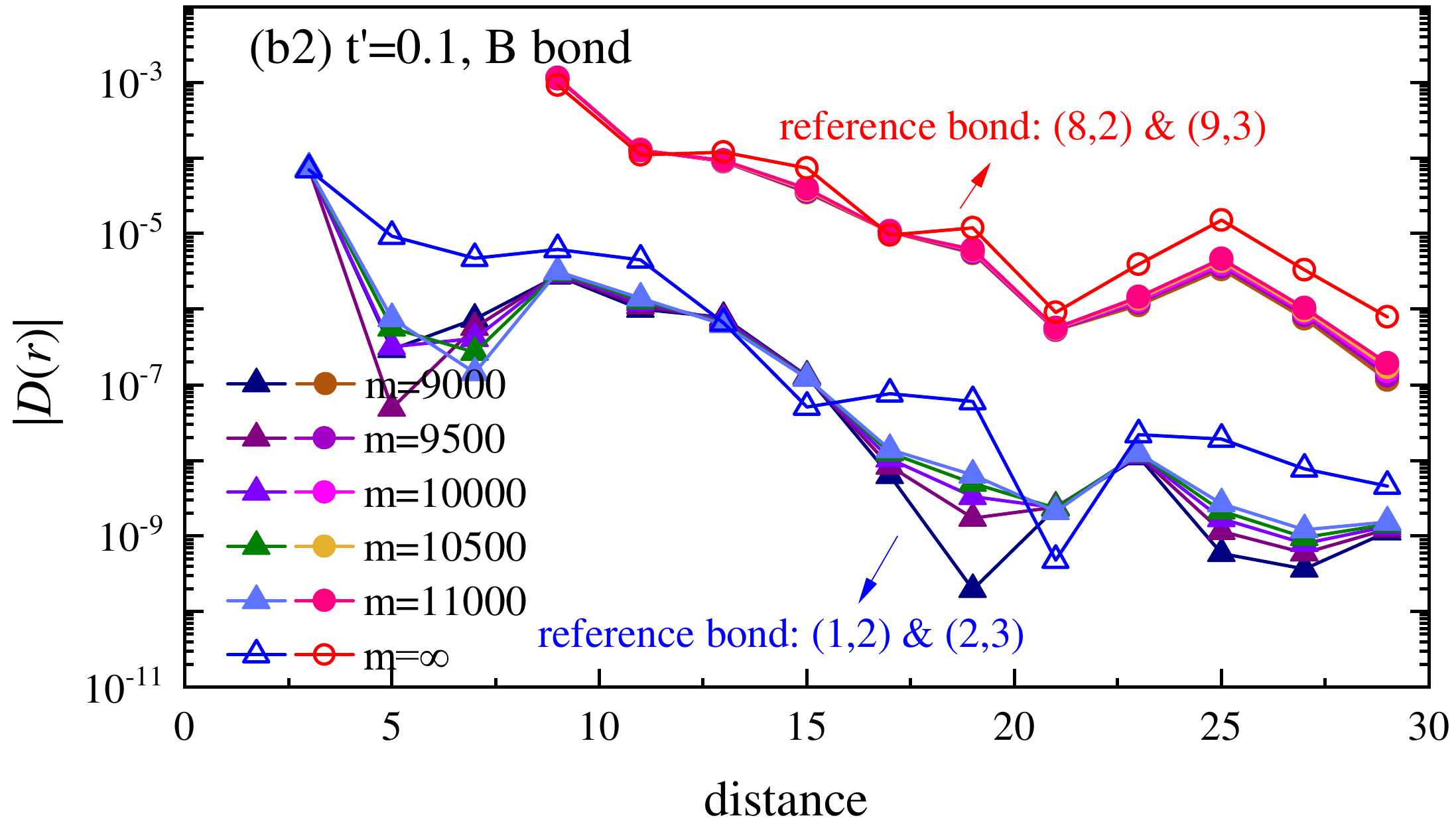}
   \includegraphics[width=0.3\textwidth]{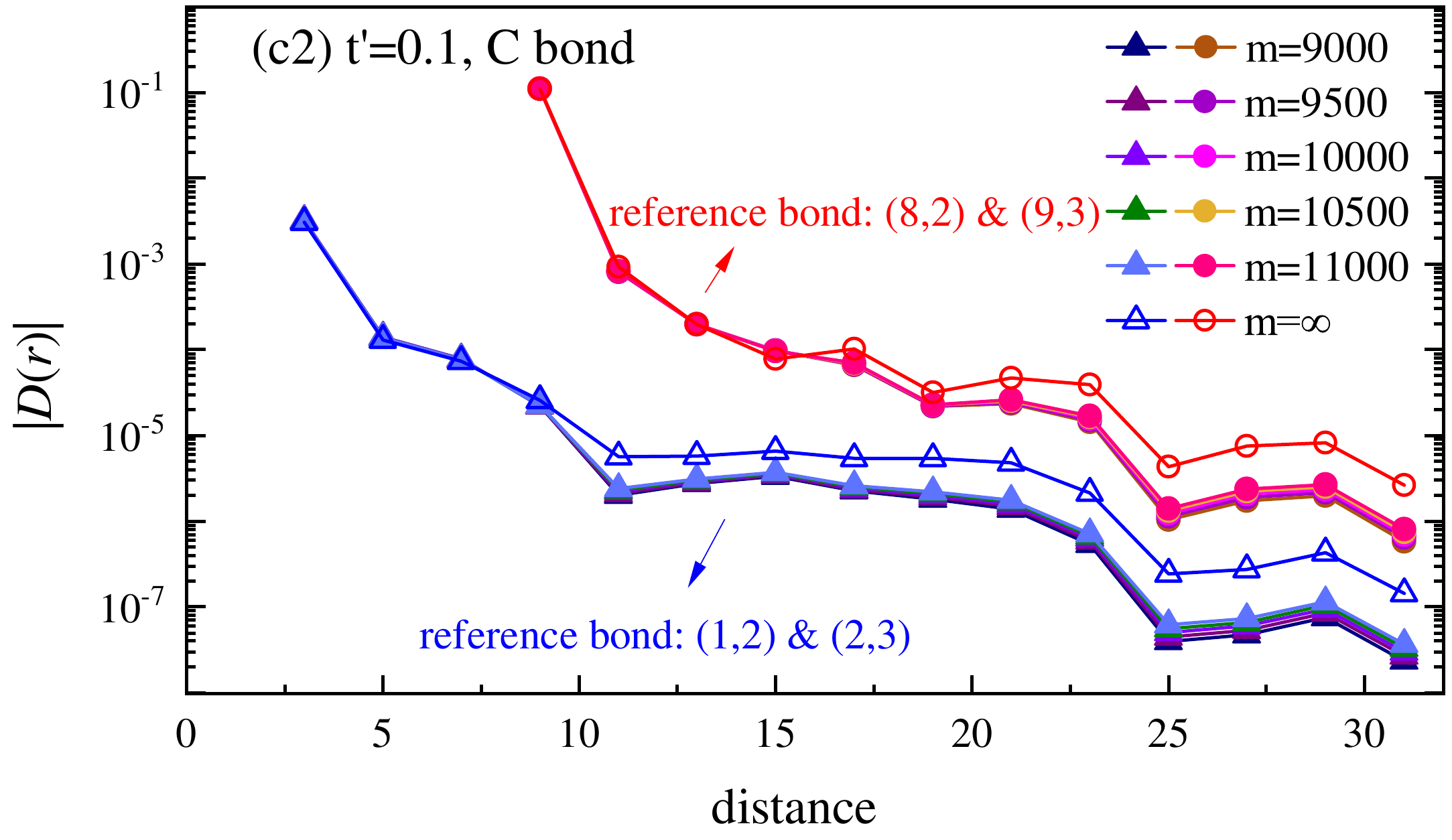}
   \includegraphics[width=0.3\textwidth]{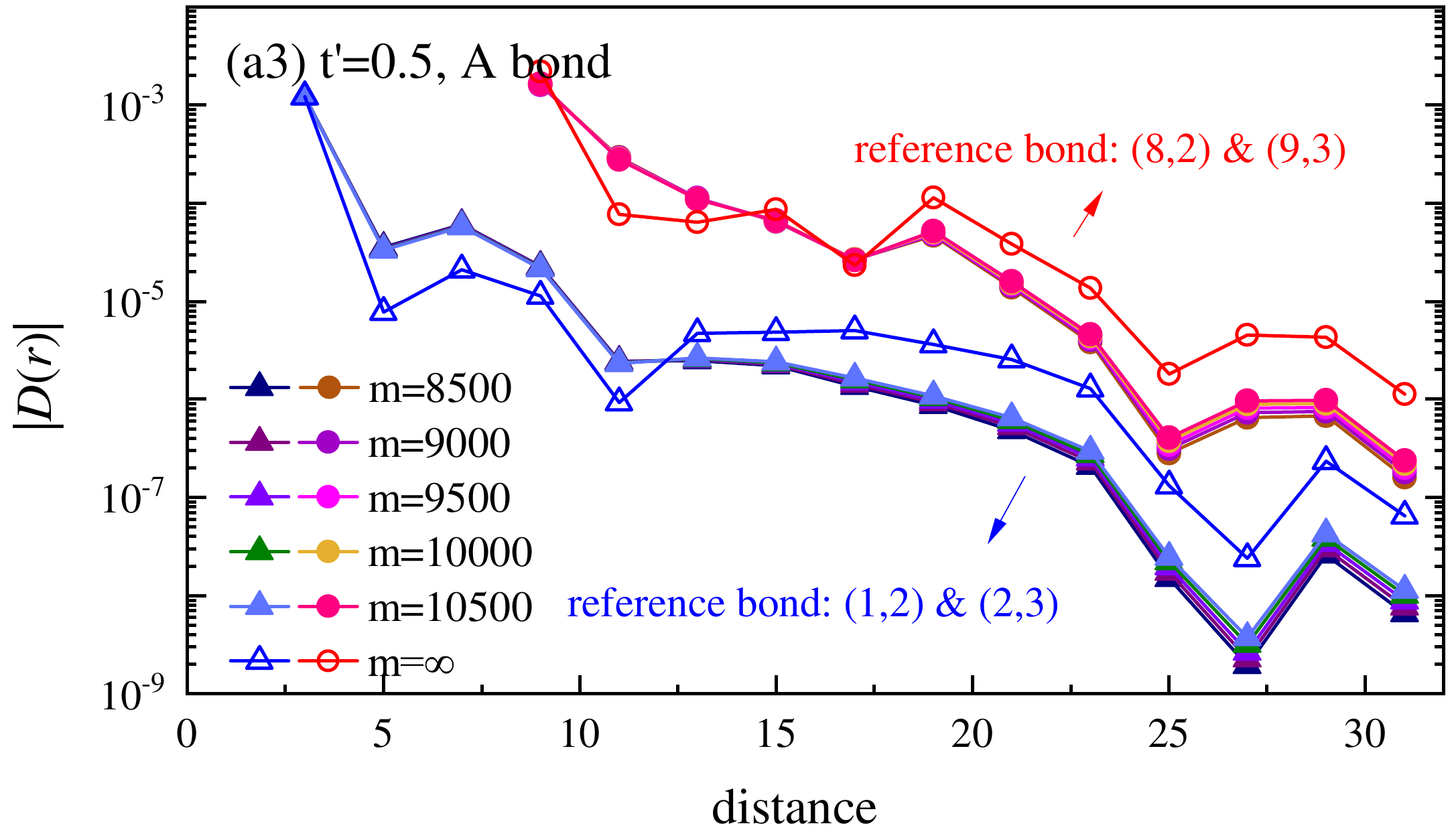}
   \includegraphics[width=0.3\textwidth]{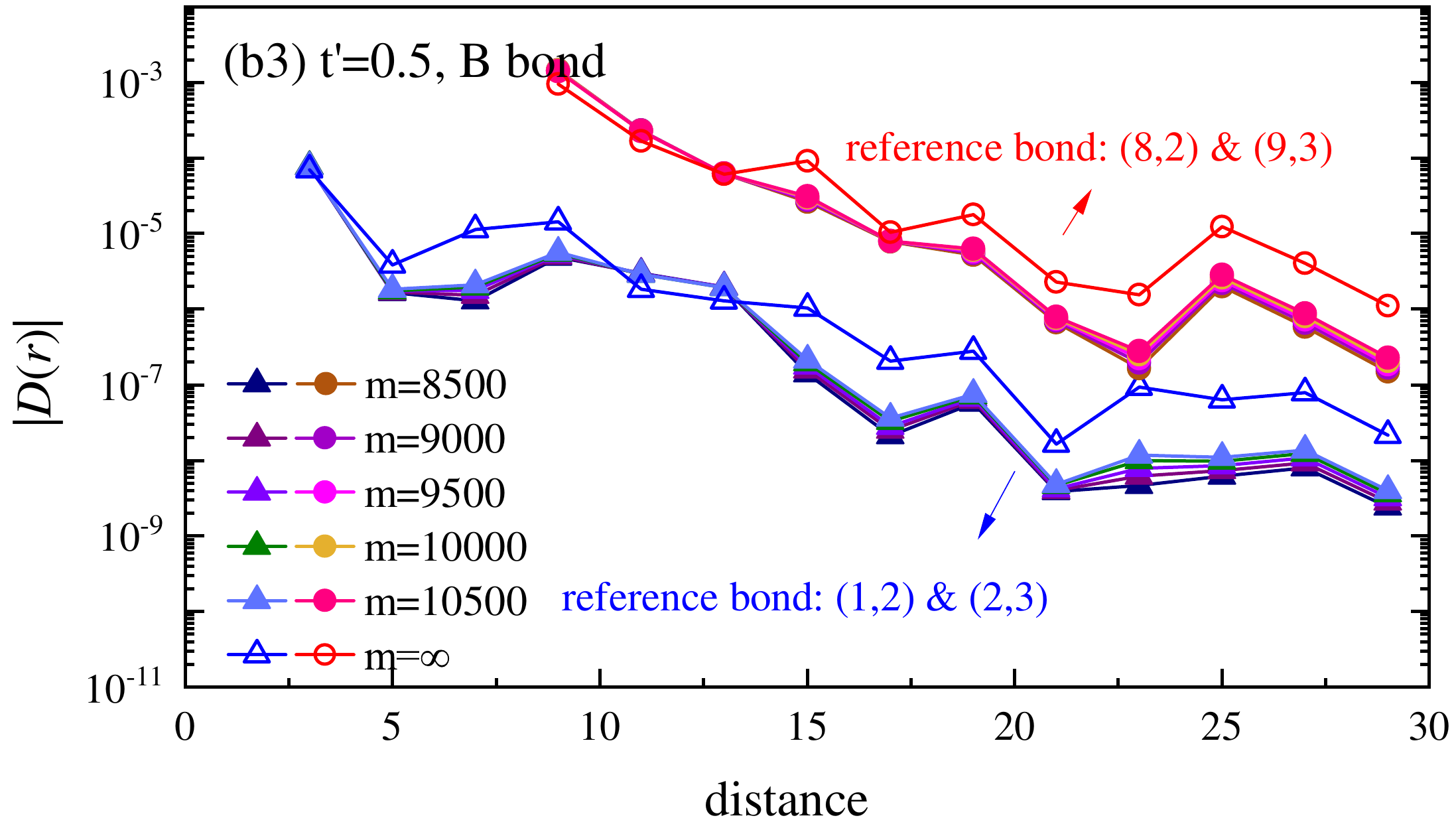}
   \includegraphics[width=0.3\textwidth]{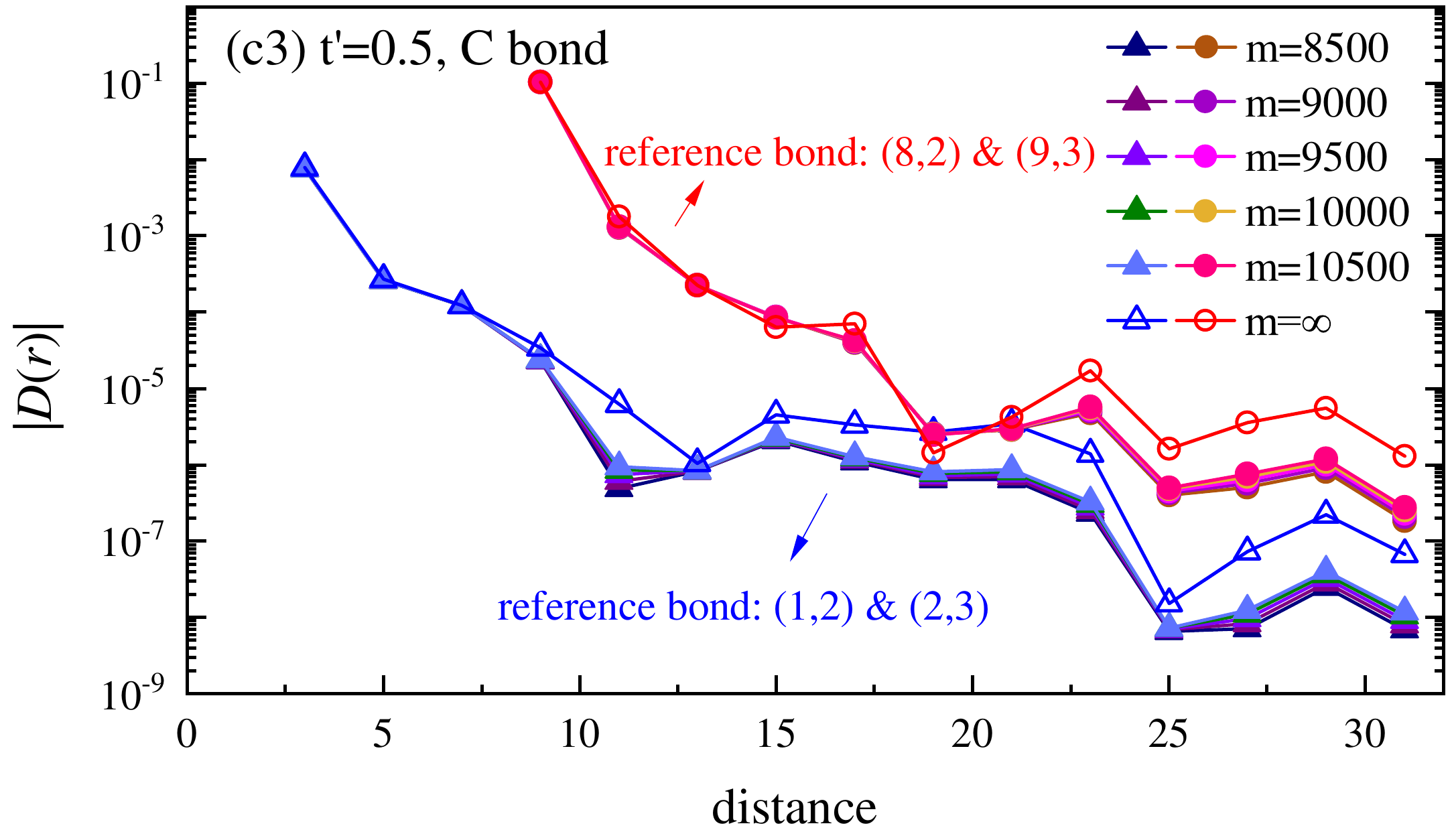}
   \caption{Bond dimension dependency of pairing correlations for A, B and C bonds on the semi-logarithmic scale for different $t'$ under periodic boundary conditions. The red and blue results are the pair-pair correlations from different reference bonds.}
   \label{PairPBC-semilog}
\end{figure*}

\begin{figure*}[b]
   \includegraphics[width=0.3\textwidth]{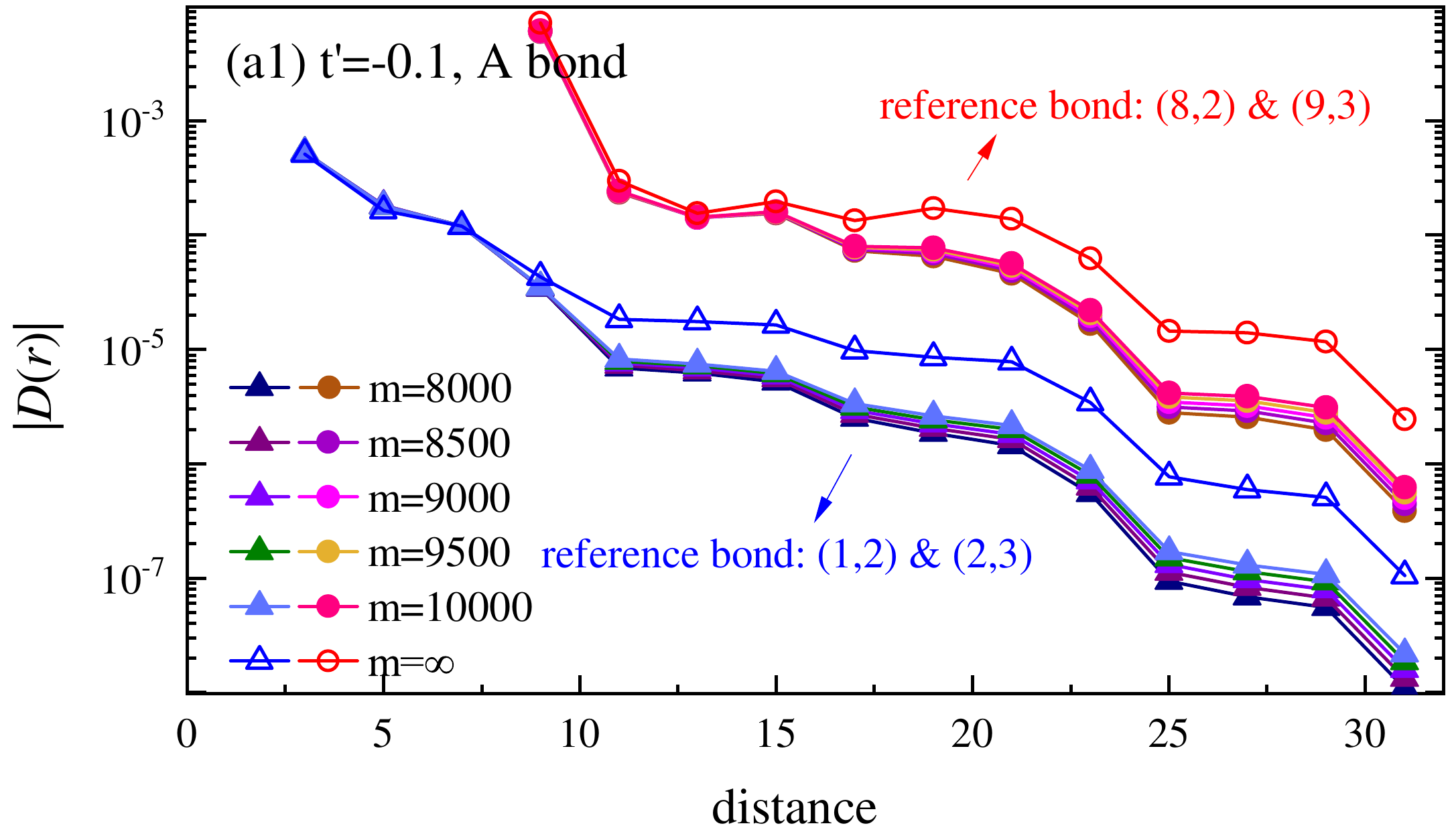}
   \includegraphics[width=0.3\textwidth]{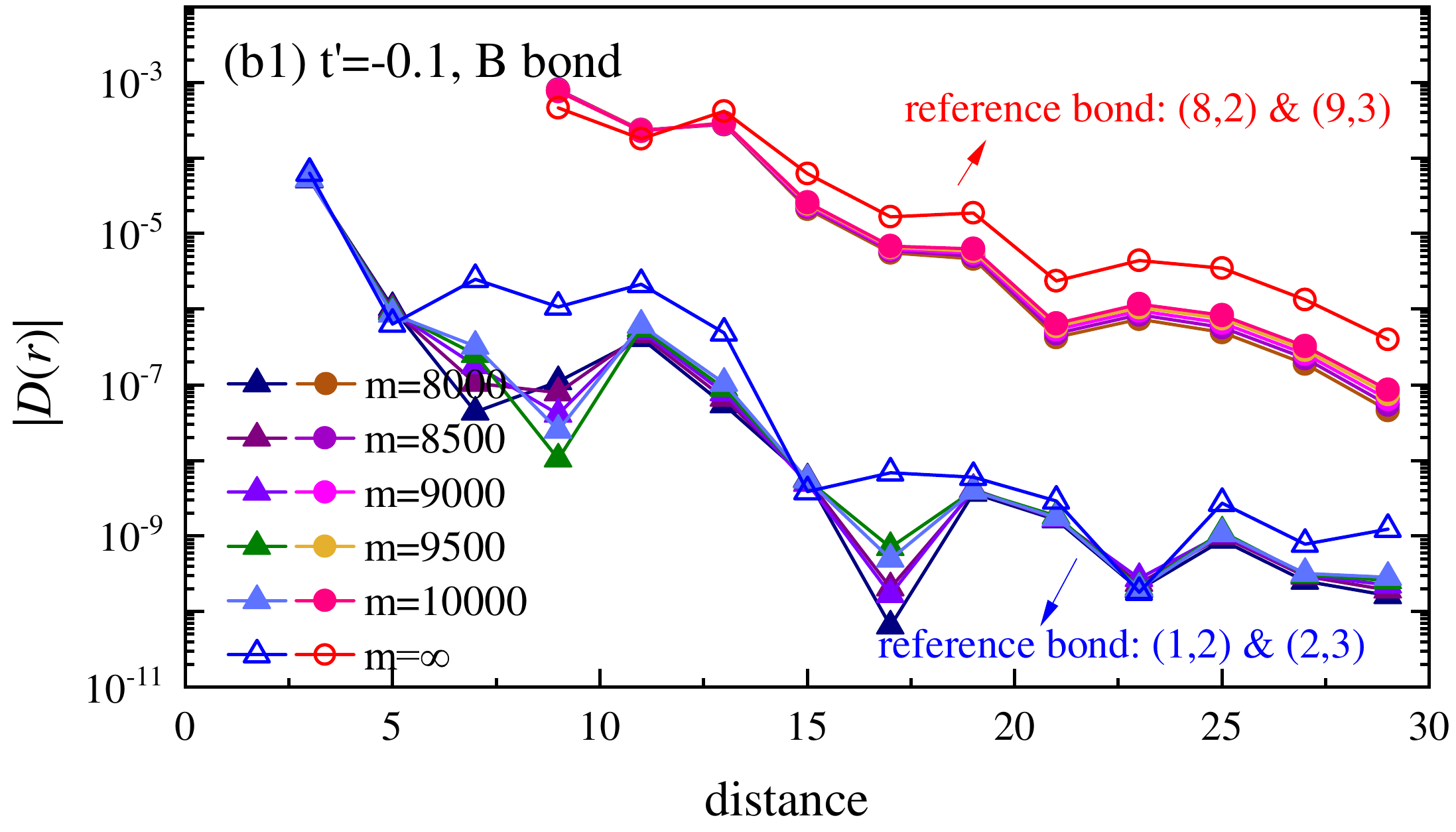}
   \includegraphics[width=0.3\textwidth]{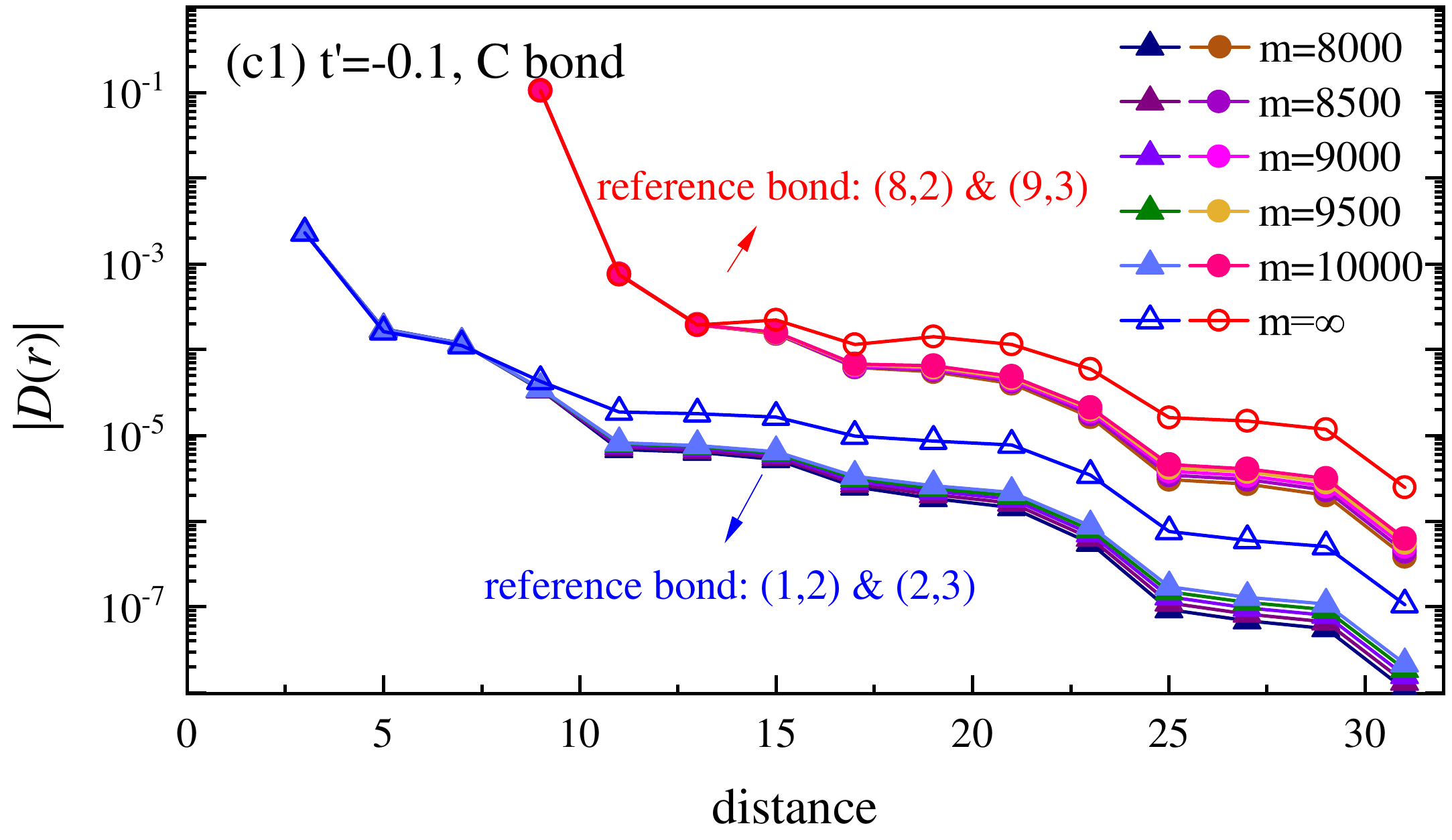}
   \includegraphics[width=0.3\textwidth]{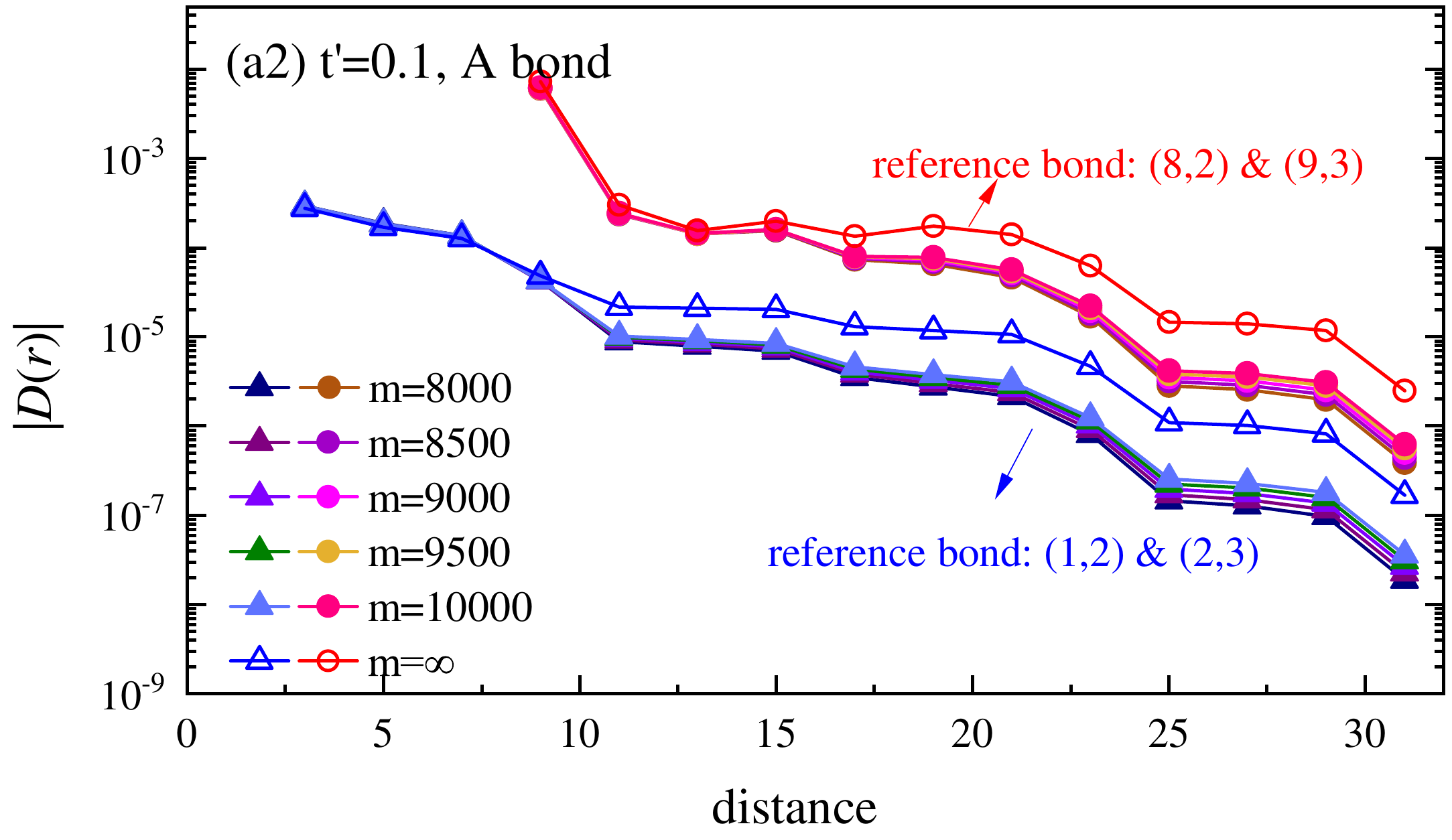}
   \includegraphics[width=0.3\textwidth]{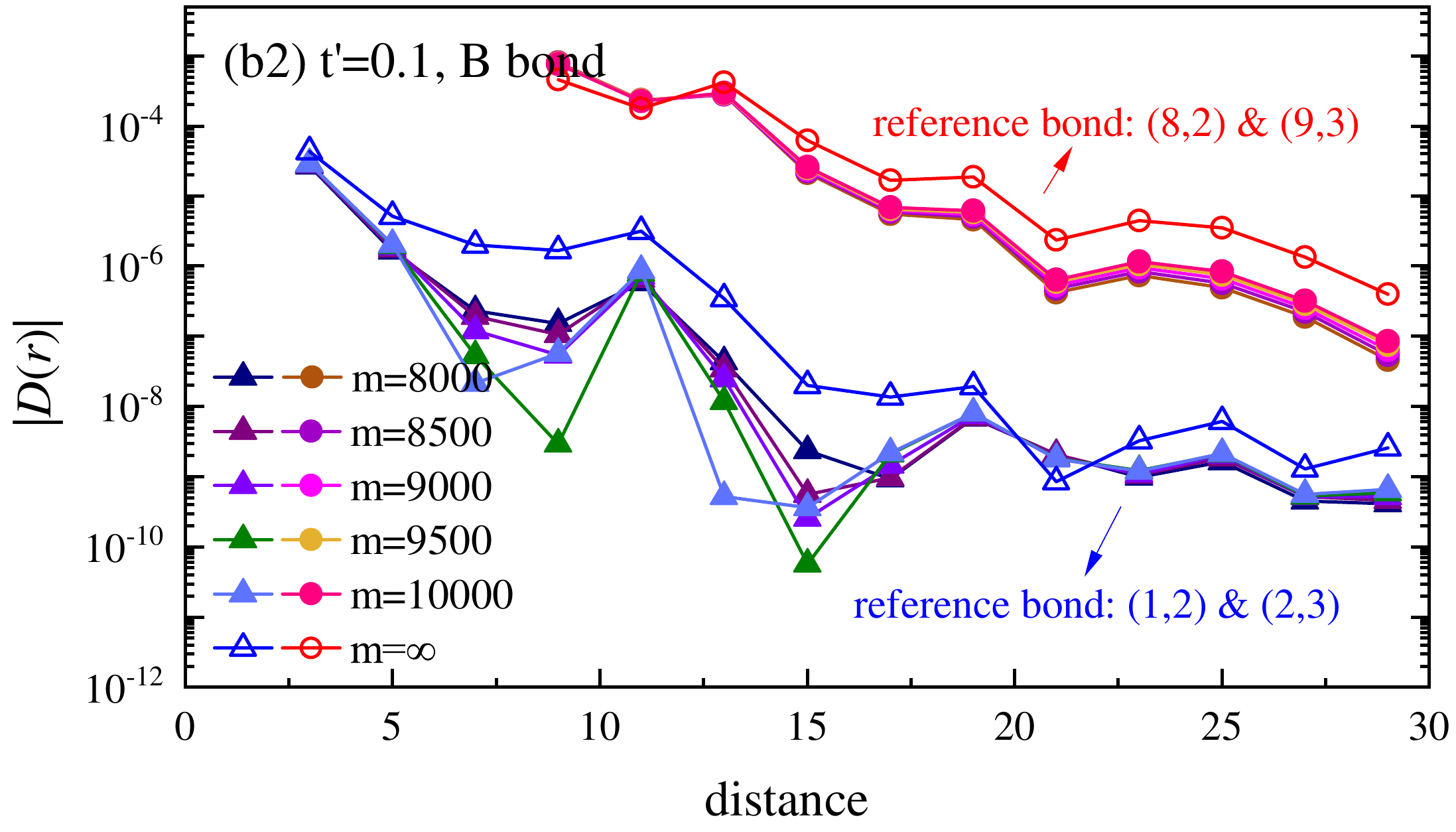}
   \includegraphics[width=0.3\textwidth]{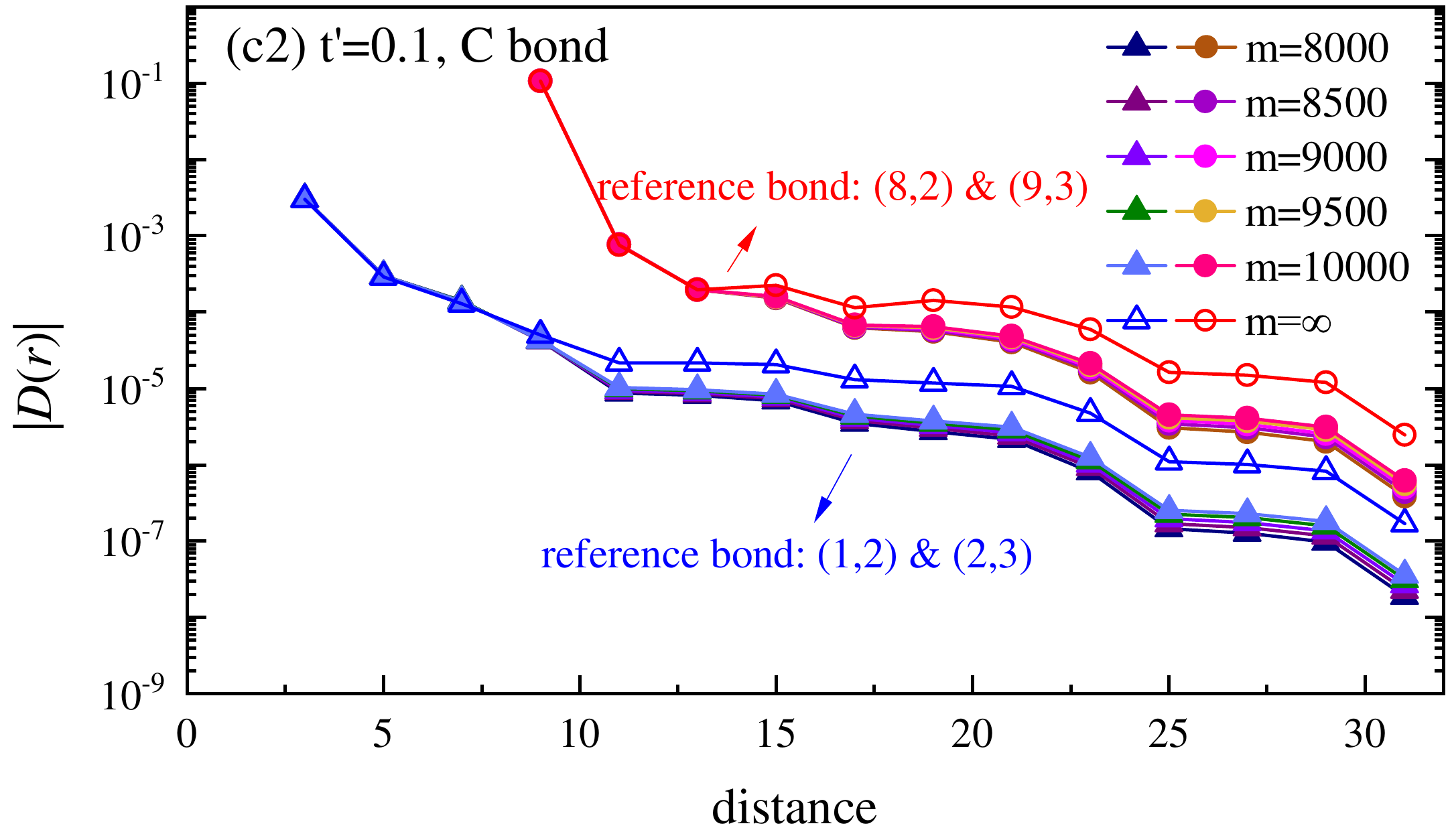}
   \includegraphics[width=0.3\textwidth]{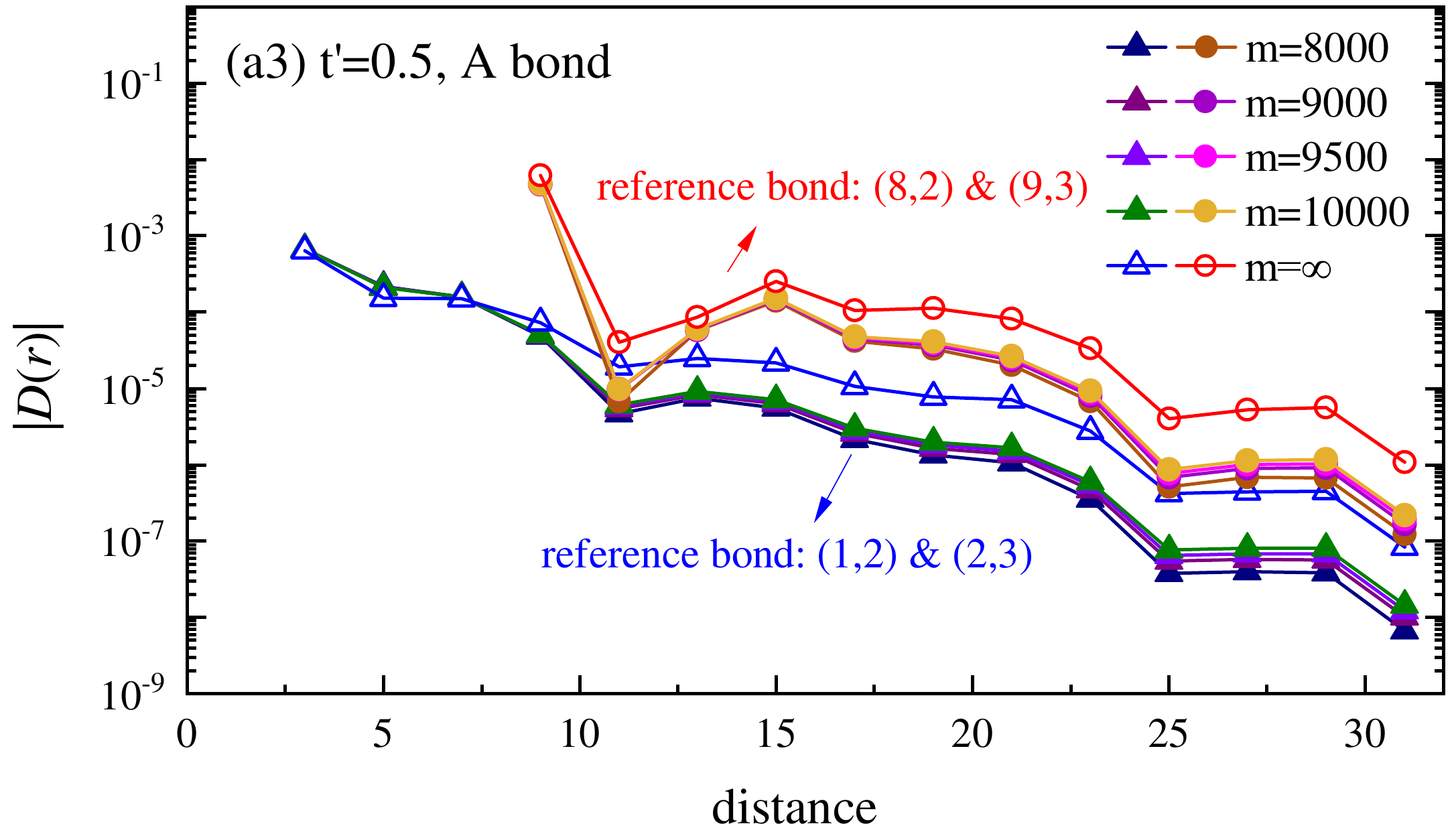}
   \includegraphics[width=0.3\textwidth]{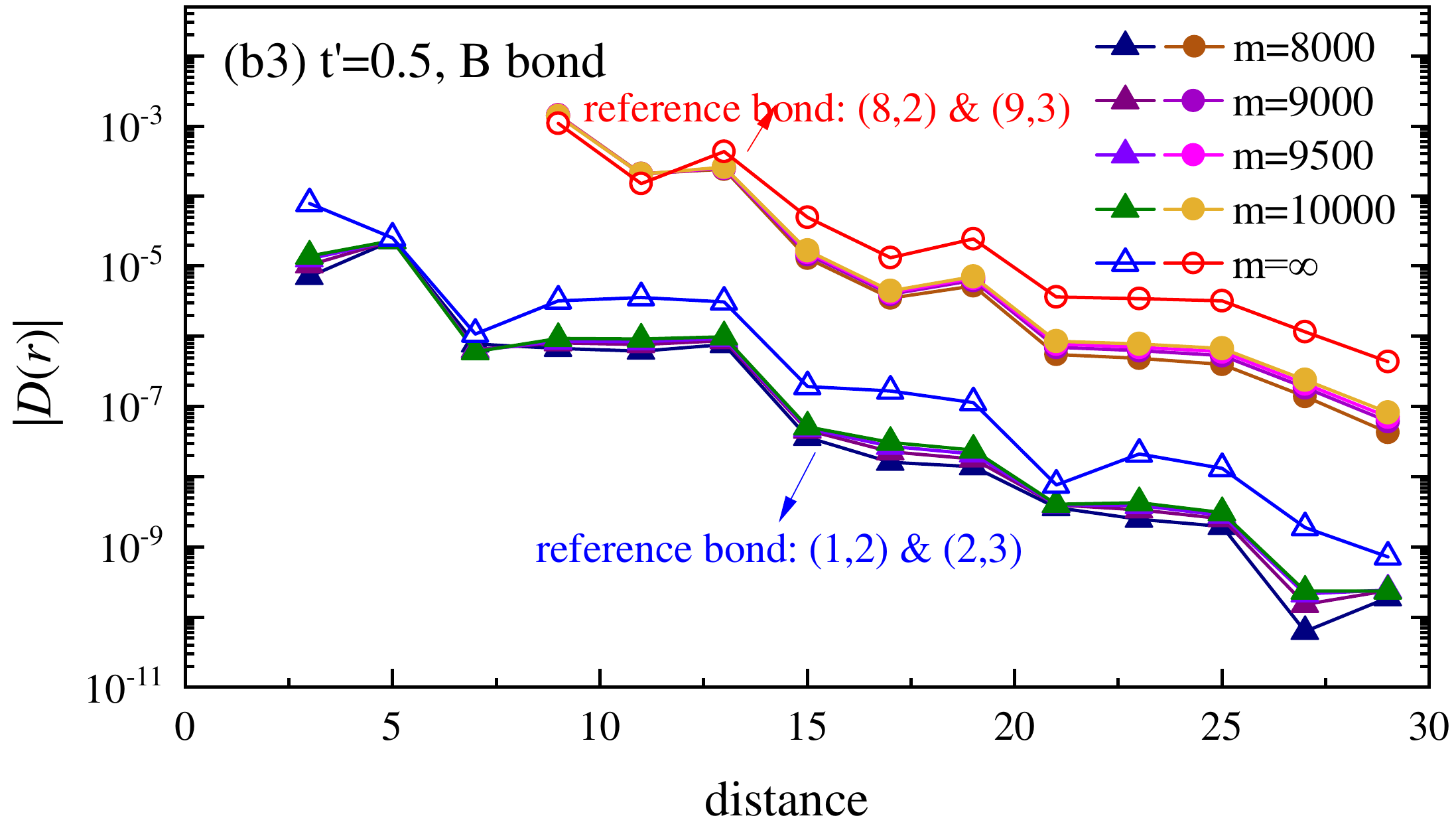}
   \includegraphics[width=0.3\textwidth]{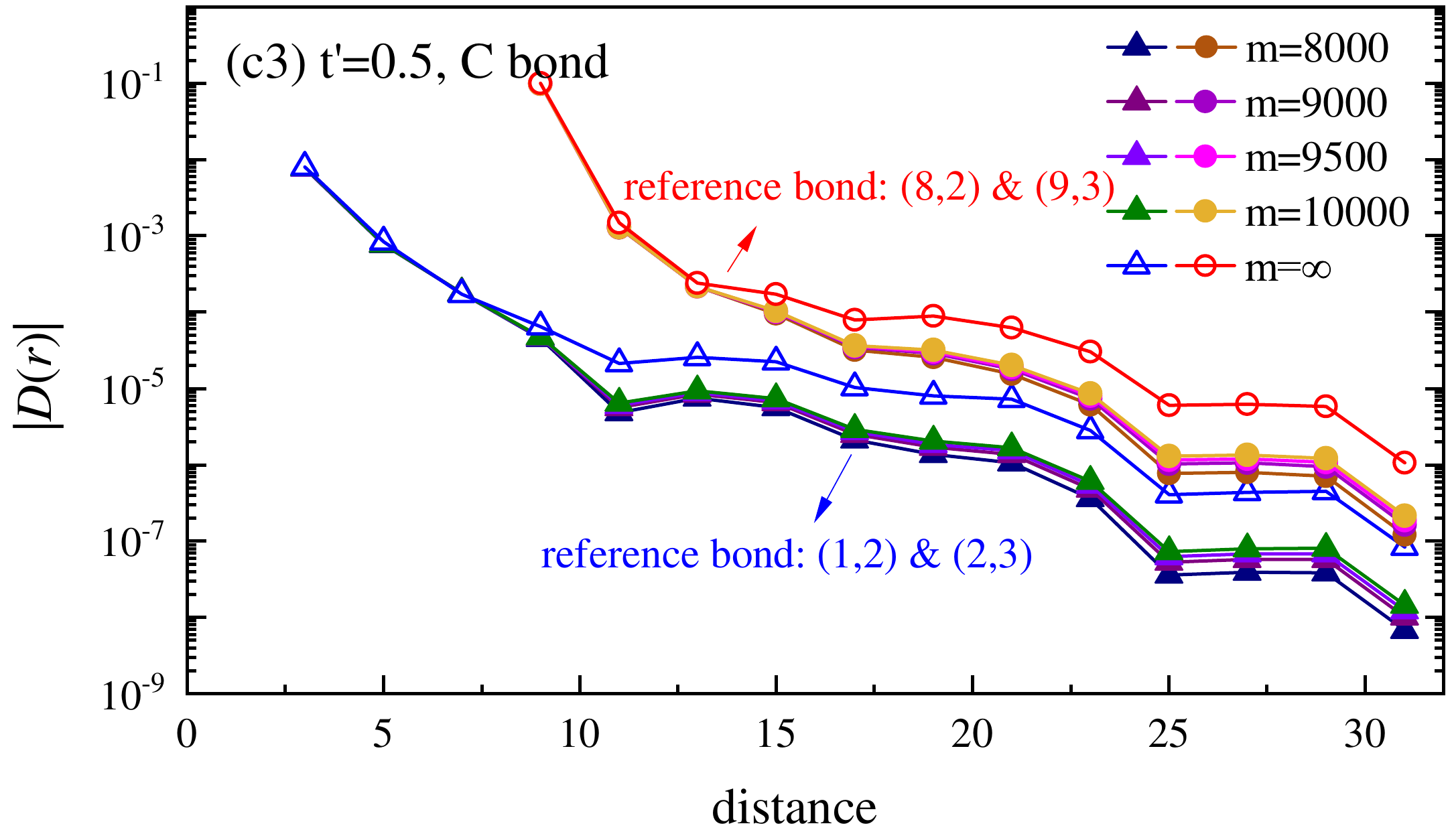}
   \caption{Bond dimension dependency of pairing correlations for A, B and C bonds on the semi-logarithm scale for different $t'$ under anti-periodic boundary conditions. The red and blue results are the pair-pair correlations from different reference bonds.}
   \label{PairAPBC-semilog}
\end{figure*}

\begin{figure*}[b]
   \includegraphics[width=0.3\textwidth]{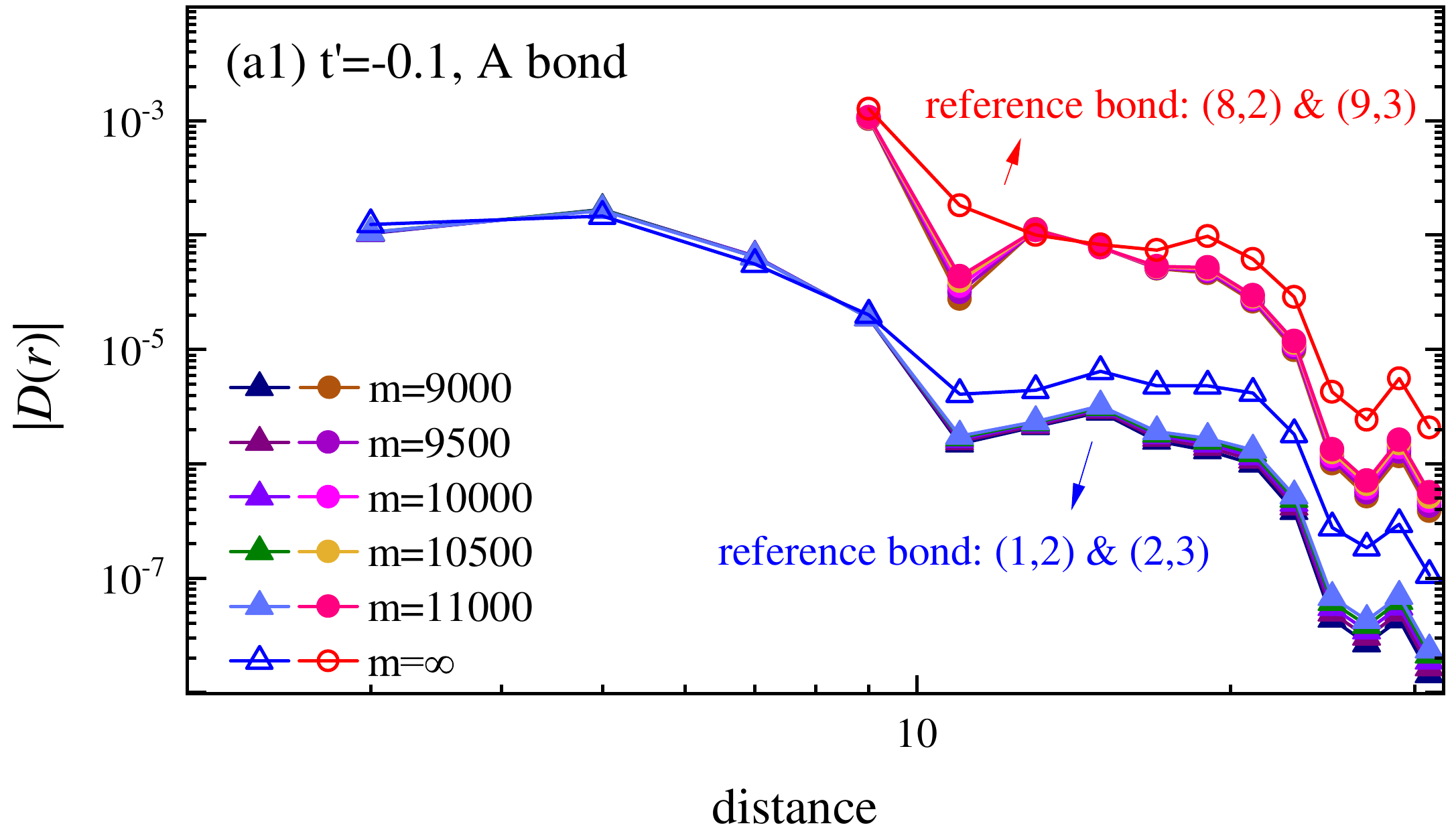}
   \includegraphics[width=0.3\textwidth]{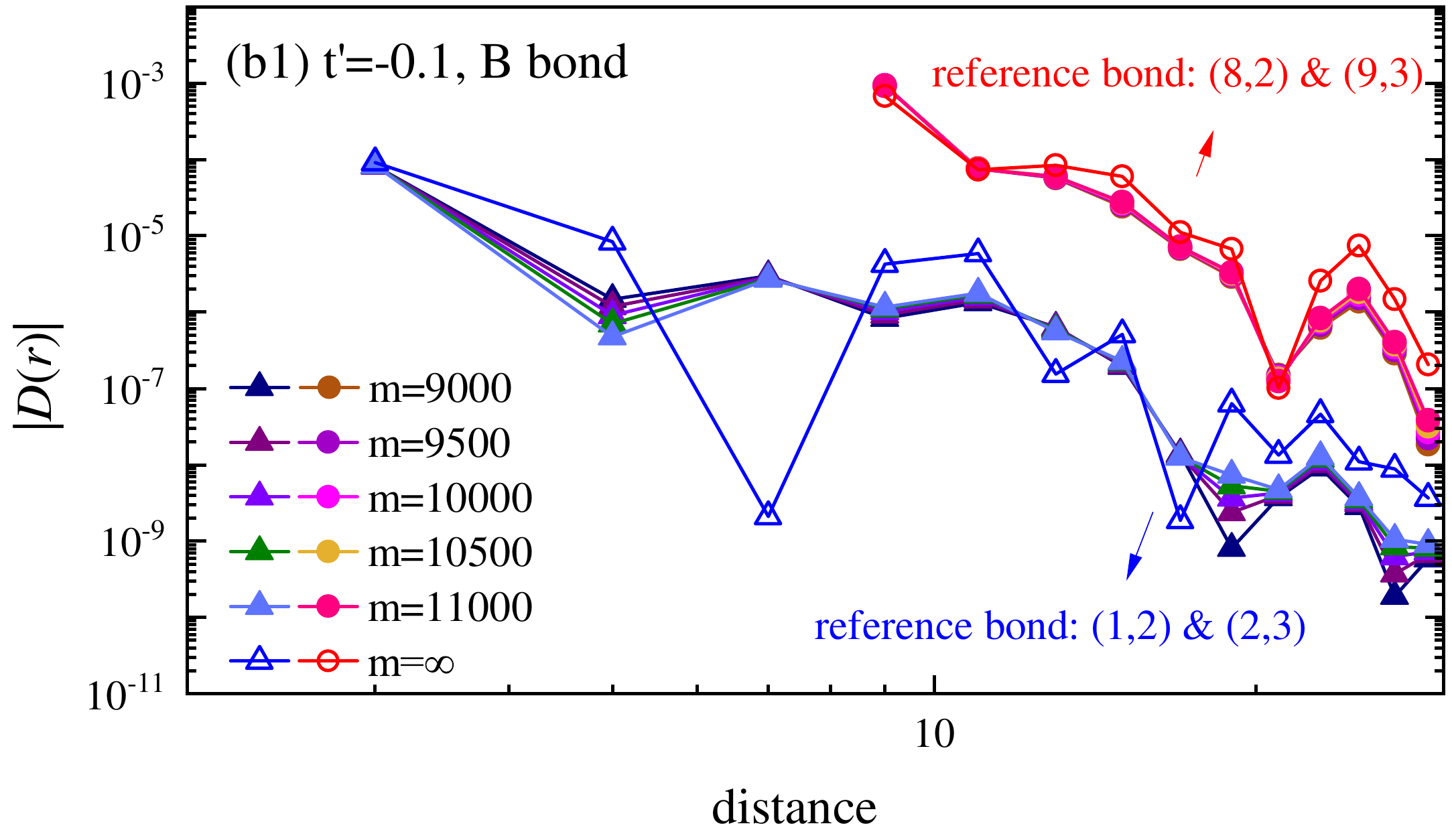}
   \includegraphics[width=0.3\textwidth]{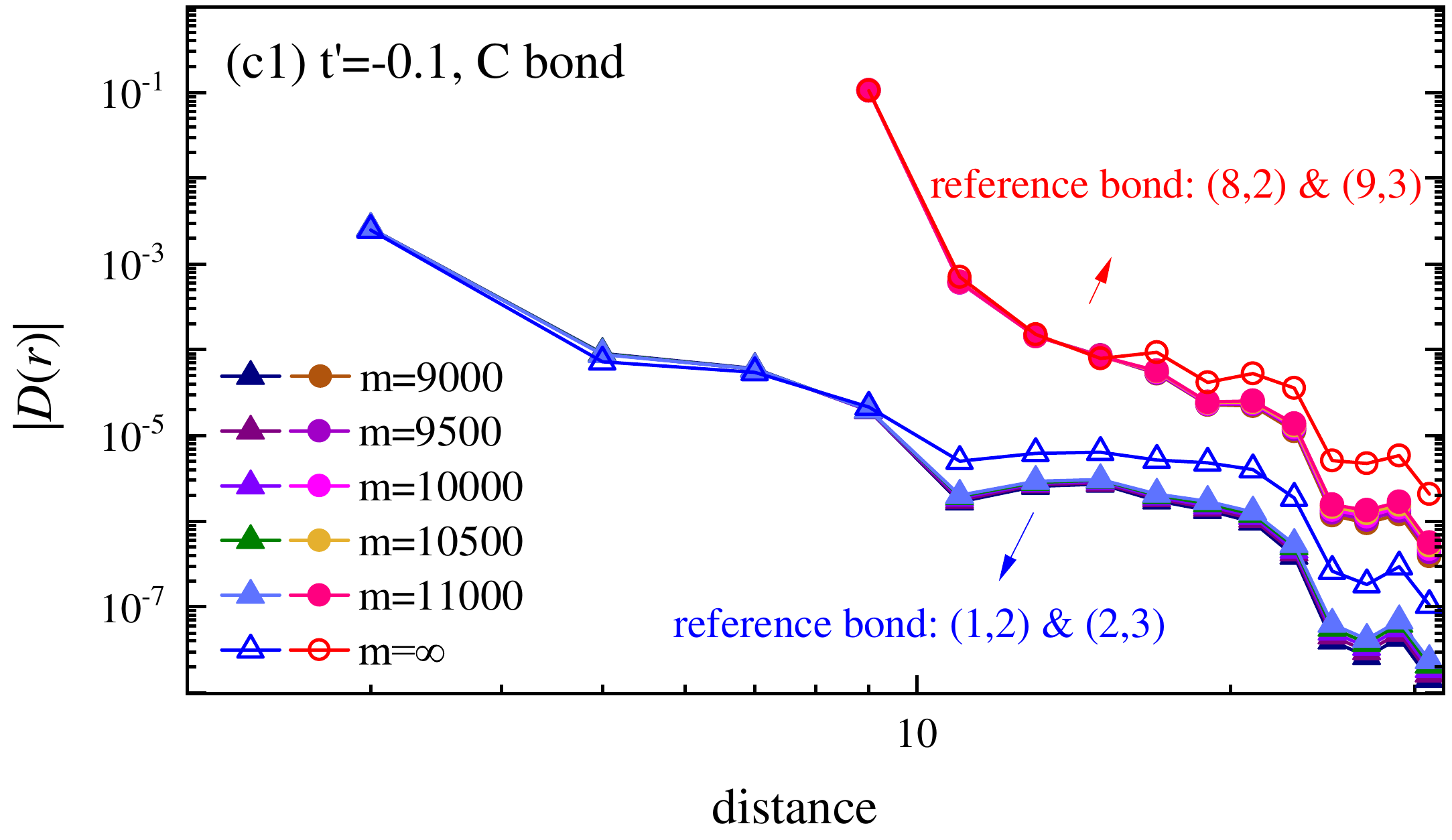}
   \includegraphics[width=0.3\textwidth]{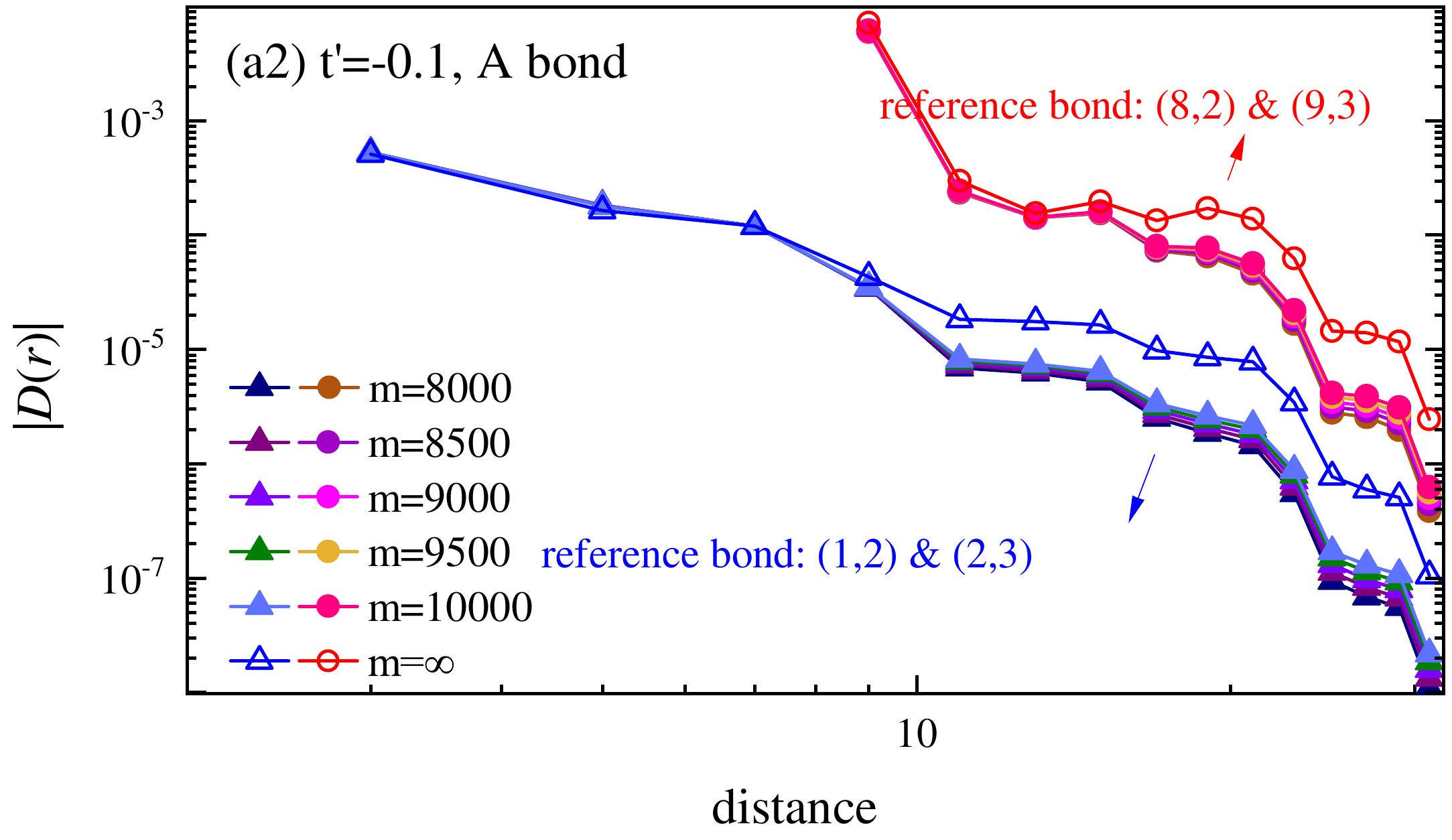}
   \includegraphics[width=0.3\textwidth]{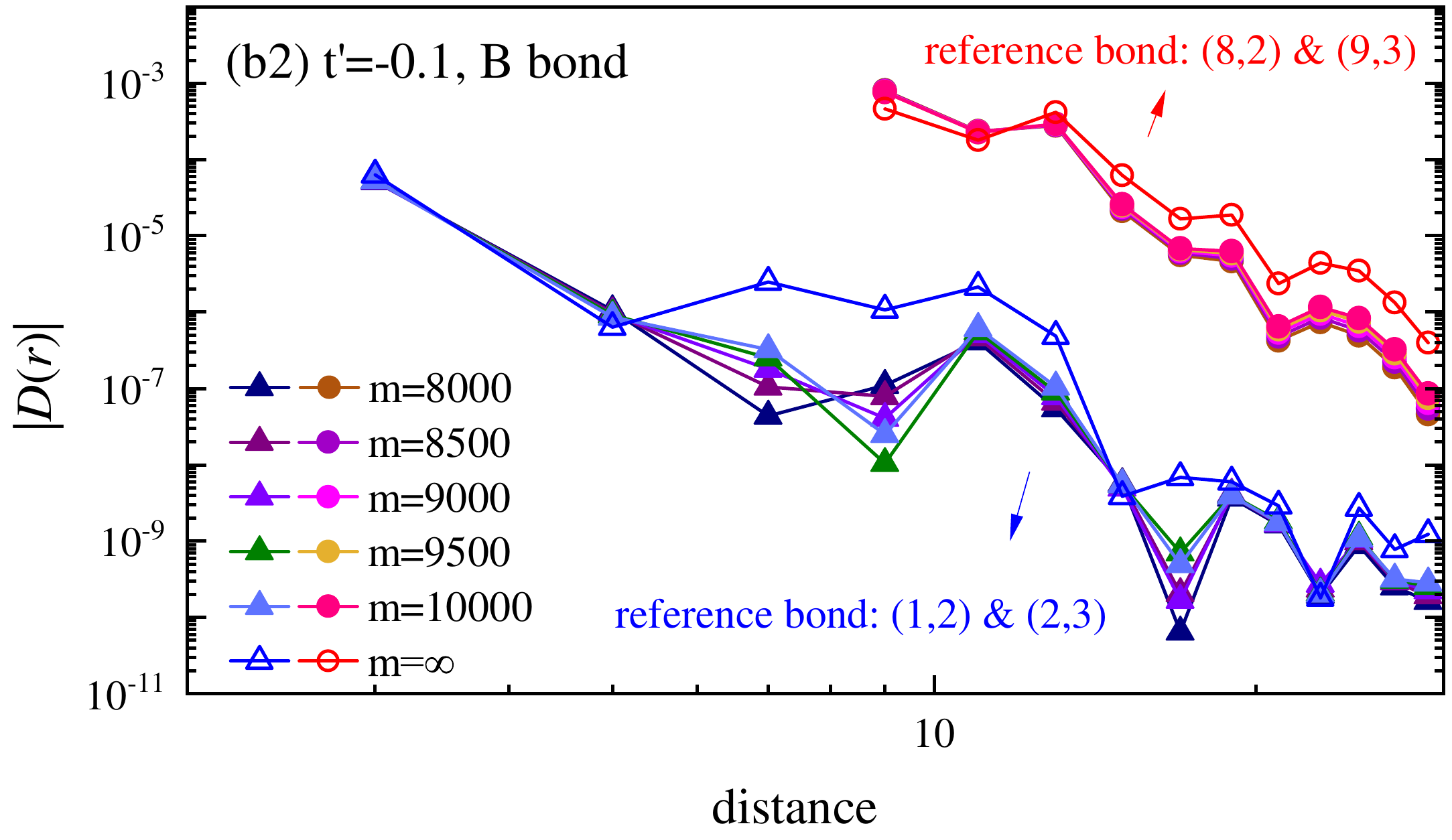}
   \includegraphics[width=0.3\textwidth]{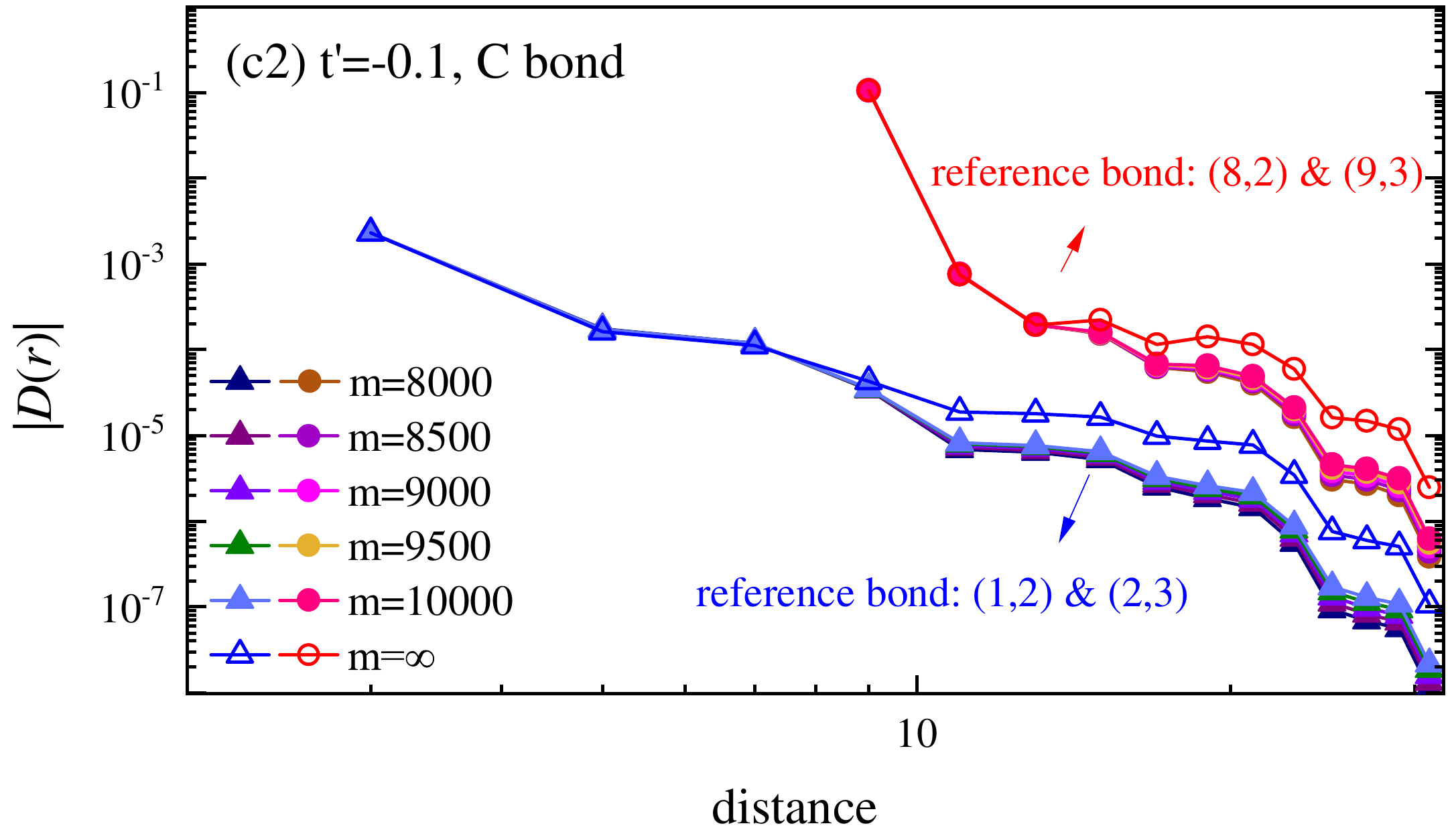}
   \caption{Bond dimension dependency of pairing correlations for A, B and C bonds on the double-logarithm scale for $t'=-0.1$ as a representative under the periodic (upper panel) and anti-periodic (lower panel) boundary conditions. The red and blue results are the pair-pair correlations from different reference bonds.}
   \label{Pair-loglog}
\end{figure*}

In this section, we show the singlet pair-pair correlation function $D(r)$ for a varies of $t'$ values in the stripe and N$\acute e$el antiferromagnetic phases. Figs.~\ref{PairPBC-semilog} and ~\ref{PairAPBC-semilog} represent results in semi-logarithm scale under periodic and anti-periodic boundary conditions, respectively. We also change the position of the reference bond to evaluate its effect on pair-pair correlation function. We can find that the characterization of pairing correlation doesn't change, though the absolute values are different for different reference bonds. These obtained data are also displayed in the double-logarithm scale in Fig.~\ref{Pair-loglog}. It is difficult to tell whether the pairing correlation decays exponentially or algebraically from the data.

\subsection{E. Triplet pair-pair correlation}
\begin{figure*}[b]
   \includegraphics[width=0.4\textwidth]{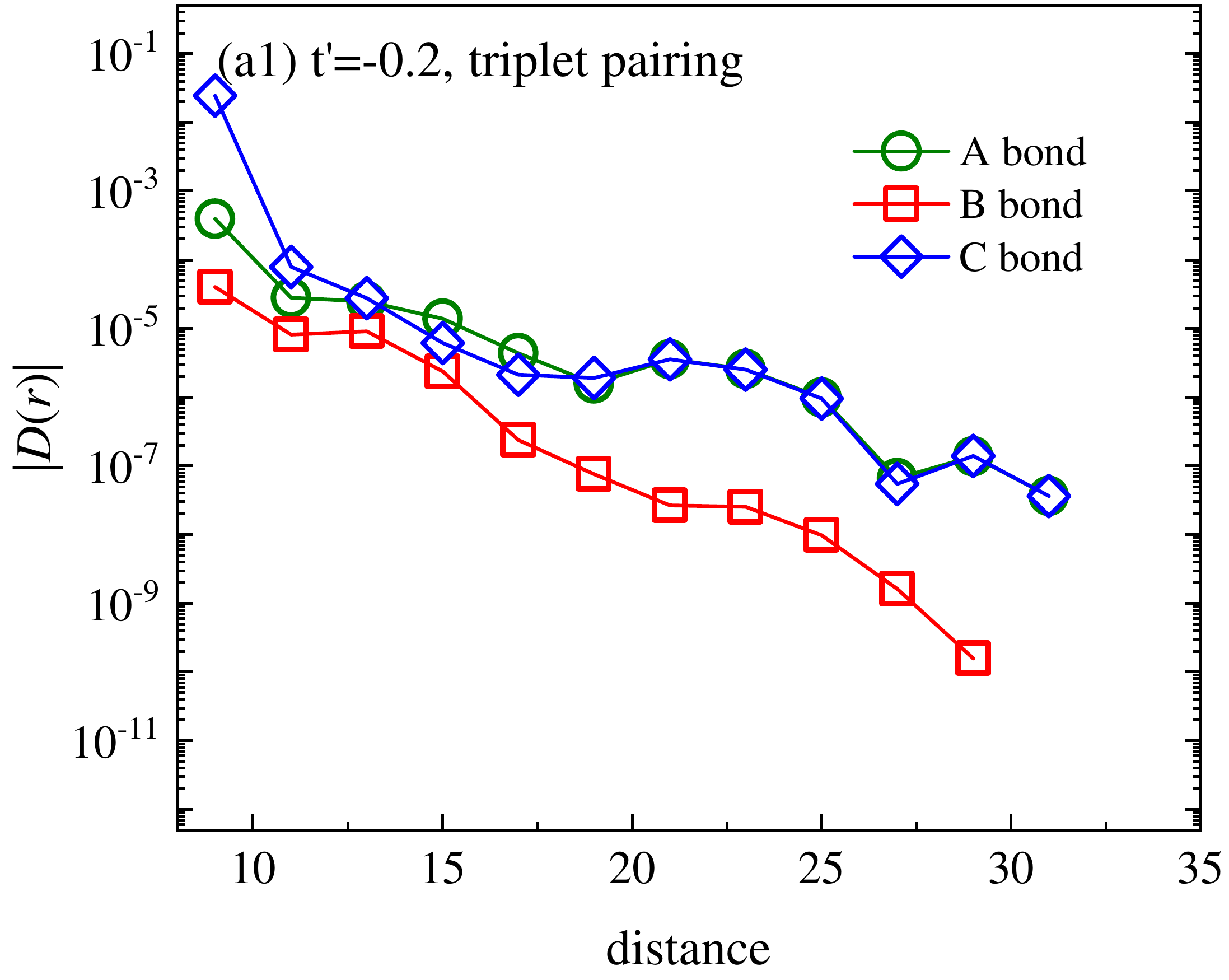}
   \includegraphics[width=0.4\textwidth]{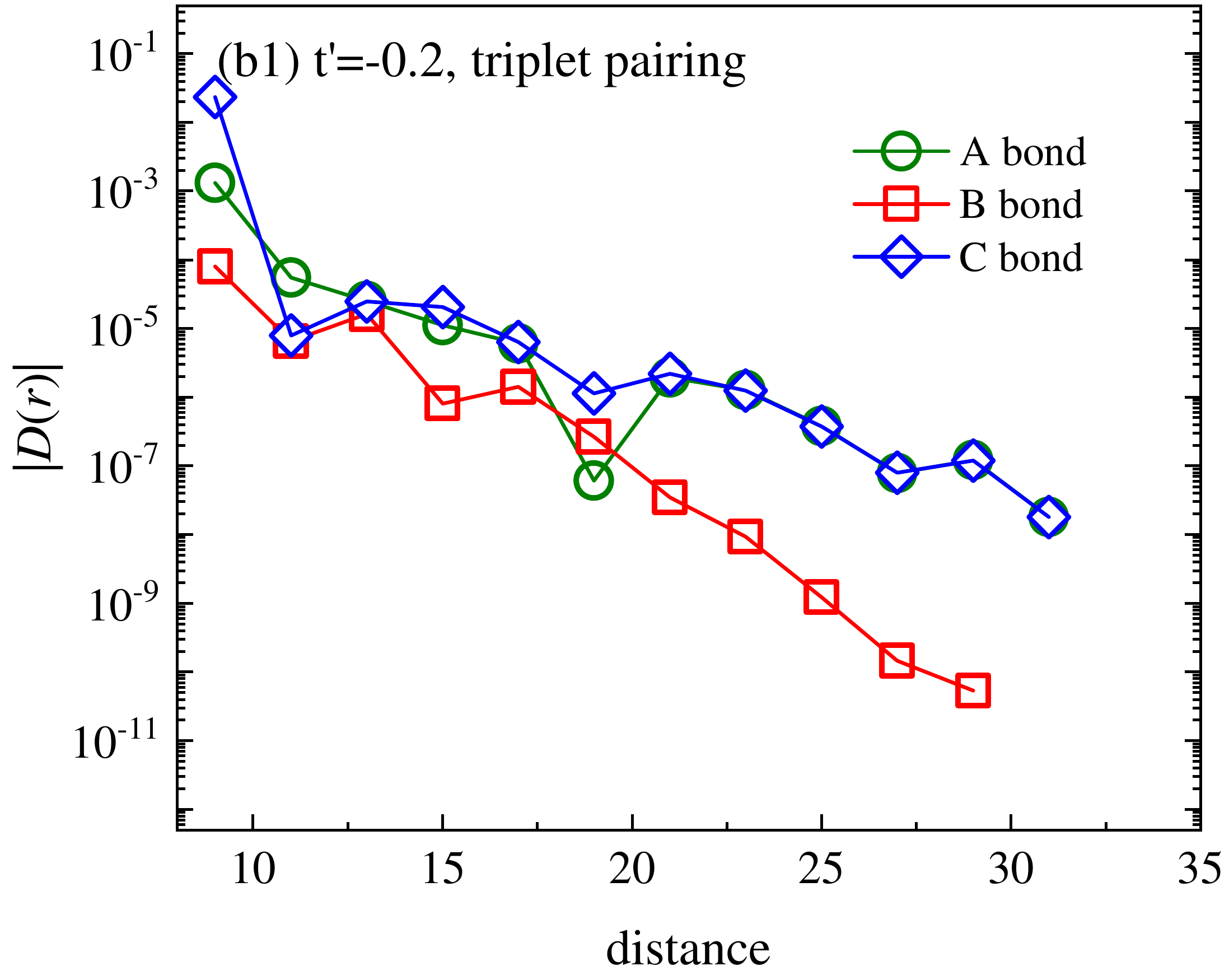}
   \includegraphics[width=0.4\textwidth]{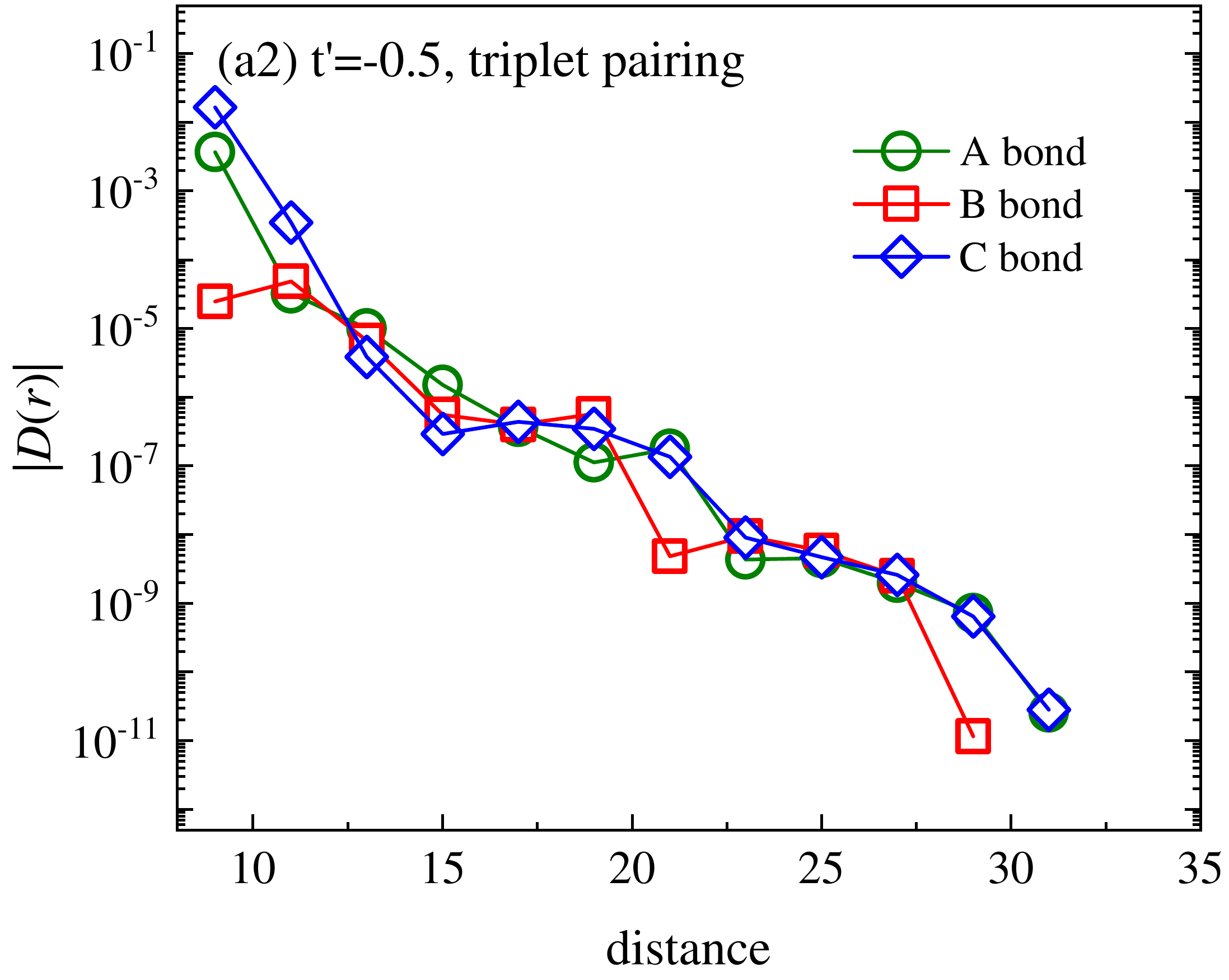}
   \includegraphics[width=0.4\textwidth]{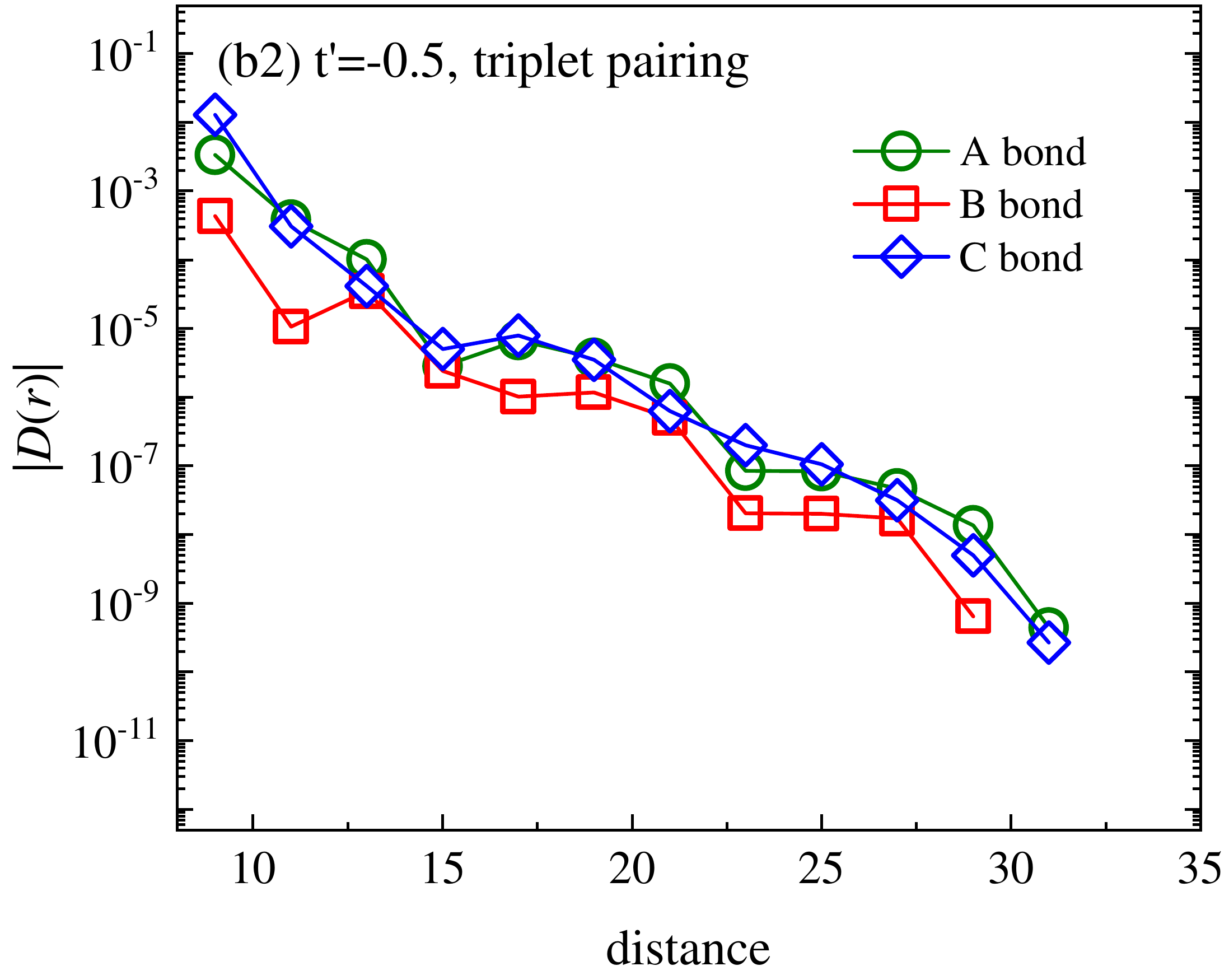}
   \caption{Triplet pairing correlations for A, B and C bonds on the semi-logarithm scale for different $t'$ under the periodic (left) and anti-periodic (right) boundary conditions in DMRG calculations. The reference bond is placed at the edge between sites (8,2) and (9,3) when calculating the triplet pairing correlation functions.}
   \label{tripletPairPBC-semilog}
\end{figure*}

The triplet pair-pair correlations are shown in Fig.~\ref{tripletPairPBC-semilog}. We find the triplet pairings are weaker than the singlet pairings. 

